%% file: Korteweg-BC.tex
 \documentclass[final,3p,times,authoryear]{elsarticle}

\usepackage{leftidx}
\usepackage{amsmath}
\usepackage{color}
\usepackage[normalem]{ulem}
\usepackage{showlabels}

\journal{Int J Eng Sci.}

\begin{document}
\input{macros}

\begin{frontmatter}



\title{On a thermodynamic framework for developing boundary conditions for Korteweg fluids}


\author[label1]{Ond\v{r}ej Sou\v{c}ek\corref{cor1}}
\ead{Ondrej.Soucek@mff.cuni.cz}
\author[label2]{Martin Heida}
\author[label1]{Josef M\'{a}lek}
\address[label1]{Charles University, Faculty of Mathematics and Physics, Mathematical Institute, Sokolovsk\'{a} 83, 186 75, Praha 8, Czech Republic}
\address[label2]{Weierstrass Institute
for Applied Analysis and Stochastics
Mohrenstrasse 39
D-10117 Berlin, Germany}
\cortext[cor1]{Corresponding author}

\begin{abstract}
We provide a derivation of several classes of boundary conditions for fluids of Korteweg-type using a simple and transparent thermodynamic approach that automatically guarentees that the derived boundary conditions are compatible with the second law of thermodynamics. The starting assumption of our approach is to describe the boundary of the domain as the membrane separating two different continua, one inside the domain, and the other outside the domain. With this viewpoint one may employ the framework of continuum thermodynamics involving singular surfaces. This approach allows us to identify, for various classes of surface Helmholtz free energies, the corresponding surface entropy production mechanisms. By establishing the constitutive relations that guarantee that the surface entropy production is non-negative, we identify a new class of boundary conditions, which on one hand generalizes in a nontrivial manner the Navier's slip boundary conditions, and on the other hand describes dynamic and static contact angle conditions. We explore the general model in detail for a particular case of Korteweg fluid where the Helmholtz free energy in the bulk is that of a van der Waals fluid. 
We perform a series of numerical experiments to document the basic qualitative features of the novel boundary conditions and their practical applicability to model phenomena such as the contact angle hysteresis.

\end{abstract}

\begin{keyword}
Continuum  Thermodynamics \sep Korteweg fluid \sep van der Waals fluid \sep Boundary conditions \sep Diffuse interface \sep  Contact angle hysteresis



\end{keyword}

\end{frontmatter}



\input{Korteweg-soucek_invariant}




\end{document}

%% file: macros.tex
\newcommand{\om}{_V}

\newcommand{\tzrg}[1]{\mbox{\boldmath $#1$}}
\newcommand{\velky}[1]{\mbox{\Large $#1$}}
\newcommand{\vt}[1]{\mbox{\Large \boldmath $#1$}}

\newcommand{\bea}{\begin{eqnarray}}
\newcommand{\eea}{\end{eqnarray}}
\newcommand{\be}{\begin{equation}}
\newcommand{\ee}{\end{equation}}
\newcommand{\pa}{\partial}
\renewcommand{\SS}{\tzrg{\sigma}}
\newcommand{\stress}{\mathbb{T}}
\newcommand{\grad}[1]{\nabla #1}
\renewcommand{\div}[1]{\hspace{0.3mm}\mathrm{div}#1}
\newcommand{\heatflux}{{\bf q}}
\newcommand{\DD}{\mathbb{D}}
\newcommand{\vv}{{\bf v}}
\newcommand{\ww}{{\bf w}}

\newcommand{\uu}{{\bf u}}

\newcommand{\cv}{V}

\newcommand{\lb}{\llbracket}
\newcommand{\rb}{\rrbracket}

\newcommand{\bb}{{\bf b}}
\newcommand{\Surface}{\Gamma}
\newcommand{\ga}{_{\mbox{\scriptsize{$\Gamma$}}}}
\newcommand{\gan}{_{\mbox{\scriptsize{$\Gamma$}},\mathrm{n}}}
\newcommand{\gat}{_{\mbox{\scriptsize{$\Gamma$}},\tau}}

\newcommand{\nn}{\mathrm{{\bf n}}}
\newcommand{\ienergy}{e}
\newcommand{\eqdef}{\stackrel{\mathrm{def}}{=}}

\renewcommand{\tt}{{\bf t}_\rho}

\font\zpd=pzdr at 16pt
\def\ding#1{{\zpd \char#1}}
\def\arrow{\ding{226}}
\def\handpen{\ding{45}}
\def\hand{\ding{43}}
\def\dx{dx}
\def\ds{dS}
\def\dl{dl}
\def\tflux{\mathcal{F}}
\def\tprodu{\mathcal{P}}
\def\tsupply{\mathcal{S}}
\def\flux{\tzrg{\Phi}}
\def\produ{\Pi}
\def\supply{\Sigma}
\def\helm{\psi}

\def\genericflux{\tflux^{\mbox{\tiny{$\Psi$}}}}

\def\genericbulkfluxdensity{\flux\om^{\mbox{\tiny{$\Psi$}}}}

\def\entropybulkfluxdensity{\flux^{\mbox{\tiny{$\eta$}}}}

\def\genericsurffluxdensity{\flux\ga^{\mbox{\tiny{$\Psi$}}}}

\def\entropysurffluxdensity{\flux\ga^{\mbox{\tiny{$\eta$}}}}

\def\genericproduction{\tprodu^{\mbox{\tiny{$\Psi$}}}}

\def\genericbulkproductiondensity{\produ\om^{\mbox{\tiny{$\Psi$}}}}

\def\entropybulkproductiondensity{\produ^{\mbox{\tiny{$\eta$}}}}

\def\genericsurfproductiondensity
{\produ\ga^{\mbox{\tiny{$\Psi$}}}}

\def\entropysurfproductiondensity
{\produ\ga^{\mbox{\tiny{$\eta$}}}}

\def\genericsupply{\tsupply^{\mbox{\tiny{$\Psi$}}}}

\def\genericbulksupplydensity{\supply\om^{\mbox{\tiny{$\Psi$}}}}

\def\entropybulksupplydensity{\supply^{\mbox{\tiny{$\eta$}}}}

\def\genericsurfsupplydensity{\supply\ga^{\mbox{\tiny{$\Psi$}}}}

\def\entropysurfsupplydensity{\supply\ga^{\mbox{\tiny{$\eta$}}}}

\def\Id{\mathbb{I}}

\def\tens#1{\boldsymbol{\mathsf{#1}}}
\def\pp#1#2{\frac{\partial #1}{\partial #2}}
\def\energy{\ensuremath{e}}
\def\entropyflux{\tzrg{\Phi}}

\def\rhotest{\varphi_\rho}
\def\vvtest{\varphi_\vv}
\def\pitest{\varphi_z}
\def\uutest{\varphi_\uu}

\def\rhotesth{\varphi^h_\rho}
\def\vvtesth{\varphi^h_\vv}
\def\pitesth{\varphi_z^h}
\def\uutesth{\varphi_\uu^h}

\newcommand{\sigmalw}{\sigma_{{lw}}}
\newcommand{\sigmavw}{\sigma_{{vw}}}
\newcommand{\sigmalv}{\sigma_{{lv}}}

\newcommand{\COL}[1]{{\color{blue}{#1}}}

\newcommand{\COLg}[1]{{\color{green}{#1}}}

\newcommand{\REPLACE}[2]{{\color{red}\sout{#1}}{\ \color{black}\uline{#2}}}

\newcommand{\MAYBE}[2]{{\color{blue}{#1}}{{\color{blue}{/#2}}}}

\newcommand{\INSERT}[1]{{\color{blue}\uuline{#1}\color{black}}}

\newcommand{\DELETE}[1]{{\color{red}\sout{#1}\color{black}}\ }

\newcommand{\CHECK}[1]{\color{red}\uwave{#1}\color{black}\ }

\newcommand{\COLOR}[2]{{\color{#1}{#2}}}

\newcommand{\REM}[1]{\marginpar{{\color{red}#1}}}

%% file: Korteweg-soucek_invariant.tex
\section{Introduction}
\label{sec:introduction} 
The seminal papers by Dutch scientists Johannes Diederik van der Waals and Diederik Johannes Korteweg at the turn of the 19th century \citep{van-der-Waals-1983,korteweg1901} provided the first thermodynamic insight into the physics of capilarity. In their theory, interaction phenomena at the interfaces between liquid and vapor phases of one substance are described in terms of properties of an interfacial zone of finite thickness where density changes continuously albeit with a very steep gradient. A cornerstone of their theory can be formulated as the assumption that the Helmholtz free energy of such a two-phase system is composed of two contributions - a (local) double well part with two minima related to the two coexisting phases and a gradient term penalizing the volume of the interfacial regions, the latter term being related to the notion of surface energy and surface tension. A considerable effort has been spent in an attempt to incorporate these ideas consistently into the framework of continuum mechanics and thermodynamics and to couple these models of capillarity with equations of flow \citep[e.g.][]{dunn-serrin-1986, anderson1998,heida.m.malek.j:on}. 

Korteweg-type models have gained great popularity 
in the modeling of granular materials and also in the modeling of two-phase flows; see survey papers \cite{hutter1994, Rohde2018}. A key feature of Korteweg-type models is  their ability to naturally deal with complex changes of domain topology in contrast with the sharp interface counterparts of these models. On the other hand, their apparent disadvantage is due to the presence of Korteweg stress in the balance of linear momentum that calls for additional boundary conditions which are very difficult to specify in an ad-hoc manner.  This issue is clearly not just a mathematical subtlety. In the discussed class of models the boundary conditions describe real physical phenomena such as the motion of the contact line, i.e., dynamics of advancement or retreat of the vapor-fluid interface attached to the solid surface, see for instance \cite{heida2013} and references therein, in particular \cite{bonn2009}. Another observable real-world phenomenon most likely related to boundary conditions is the so-called contact angle hysteresis, that is, the difference in the measured contact angles of sliding droplets on the advancing and receding parts of the contact line \citep[see e.g.][]{bormashenko2013}. Furthermore, one expects that a formulation of the boundary conditions based on solid physical grounds would result in formulations of the problems that might be robust from the point of view of computer simulations and amenable from the point of view of mathematical analysis.

The general aim of this paper is to address the question of the identification of appropriate boundary conditions for problems in continuum thermodynamics. Towards this goal, we use a transparent thermodynamic approach that has been successful in identification of the constitutive equations in the bulk for various complex materials and that stems from specification of the energy storage and dissipation mechanisms. Here, we follow a similar methodology, but we extend it also to surface phenomena.
A crucial viewpoint adopted here is that the outer boundary of a liquid-vapor body may be viewed as an  interface between this body and its exterior. This viewpoint provides a framework for considering a rather general class of boundary processes and admits a natural coupling between the processes on the surface and in the bulk. This in turn leads to a relatively straightforward procedure for deriving the constitutive relations on the surface delimiting the boundary, i.e., the boundary conditions. This approach is illustrated on the derivation of boundary conditions for a Korteweg-type fluid, for which we can explicitly characterize both the bulk and the surface Helmholtz free energies - in the bulk using the standard thermodynamic relations for van der Waals fluid and at the surface by exploiting the idea of wall-interaction energy for diffuse-interface models \citep{Jacqmin2000}. Let us, however, note that the methodology developed in this paper can be extended in a relatively straightforward manner to other diffuse interface (or order parameter) models, such as Cahn-Hilliard or Allen-Cahn models; see the concluding remarks in the final section.

The structure of the paper is as follows. In Section 2, we first formulate a general integral form of the balance equation for a quantity comprising bulk and interfacial contributions and provide a corresponding local form of the balances in the bulk and at the interfaces. We explicitly list the local forms of balance equations for mass, linear momenta, energy and entropy for a single-component body. In Section 3, we recapitulate the thermodynamic derivation of a constitutive model for Korteweg-type fluid in the bulk following and slightly modifying the approach of \cite{heida.m.malek.j:on}. In Section 4, we extend this approach to surface phenomena, and by mutual coupling between the bulk and the interface processes, we identify the surface entropy production and the surface entropy flux. The surface entropy production is then rearranged into the form of a sum of the products of mutually related quantities (sometimes called thermodynamic fluxes and thermodynamic affinities) where the individual terms represent different physical mechanisms. Requiring that these mutually related quantities are linearly related\footnote{To our understanding, it means that we have employed the framework of linear irreversible thermodynamics \citep{groot.sr.mazur.p:non-equilibrium}.} (with positive coefficient of proportionality), we not only specify the convex quadratic form for the entropy production, but we also obtain linear constitutive relations on the boundary that automatically comply with the second law of thermodynamics. These constitutive relations (i.e., the boundary conditions) involve a novel type of static and dynamic contact angle boundary conditions as well as a non trivial generalization of the Navier slip boundary condition for the Korteweg model. In Section 5, we discuss a particular variant of the model obtained by considering the bulk Helmholtz free-energy of a van der Waals fluid under isothermal conditions. In Section 6, we present several numerical experiments which demonstrate in a simplified two-dimensional setting the effects of the obtained contact angle and  generalized Navier slip boundary conditions for the Korteweg - van der Waals model and show its potential to model dynamic contact angle phenomena and in particular the contact angle hysteresis.

\section{General local form of the balance equations in a body with singular surface}
\label{sec:General form}

\begin{figure}[h!]
\begin{center}
\includegraphics[scale=0.15]{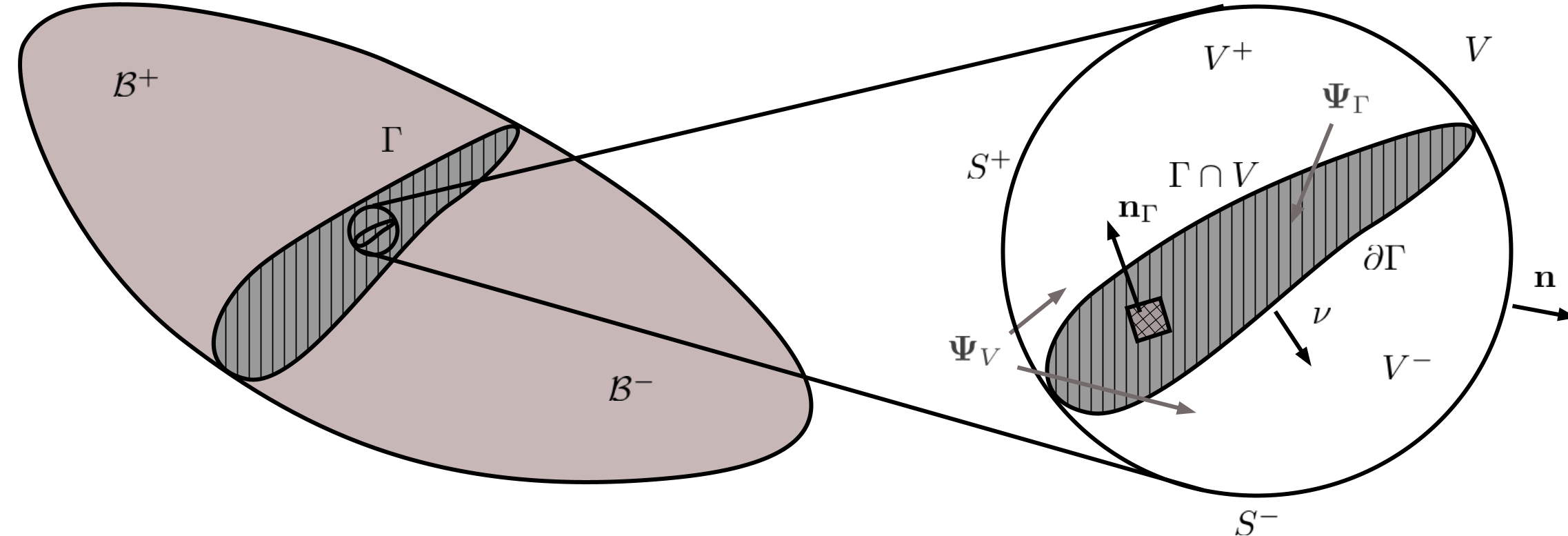} 
\caption{A body $\mathcal{B}$ separated by a singular surface into two subregions 
$\mathcal{B}^+$ and $\mathcal{B}^-$ and associated control volume $\cv$. The quantity $\Psi(\cv)$ is assumed to be composed of a bulk part with volumetric density $\Psi\om$ defined in $\cv^+\cup\cv^-$ and an interfacial contribution with surface density $\Psi_\Gamma$ defined at $\Gamma$.   
}
\label{fig-mueller}
\end{center}
\end{figure}

Let us consider a material body $\mathcal{B}$ 
in the current configuration which contains a singular surface $\Gamma$. The singular surface is understood as a mathematical model for a thin wall or membrane which separates one part of the body from another. On the surface $\Gamma$, counterparts of bulk properties and processes may take place. Let us consider arbitrary control volume $\cv$ and let $\tzrg{\Psi}(\cv)$ denote a generic additive quantity (such as mass, momentum, energy, etc.) contained in $\cv$. Let us consider an integral form of a general balance equation for such a quantity, evaluating the rate of change of $\tzrg{\Psi}(\cv)$, as a result of three independent processes: (i) a flux $\genericflux$ of the quantity $\tzrg{\Psi}$ through the boundaries $\pa\cv$ of the control volume $V$, (ii) an internal production $\genericproduction$ of the quantity $\tzrg{\Psi}$ within the control volume $V$ and (iii)  an outer supply $\genericsupply$ of the quantity $\tzrg{\Psi}$ to the control volume $V$. The general balance equation is thus postulated in the form
            \bea
                \label{eq:general_balance}
                \nonumber
                \frac{d}{dt} \tzrg{\Psi}(\cv) = -\genericflux(\cv) + \genericproduction(\cv) + \genericsupply(\cv)\ .
            \eea
Although this general description is rather formal, we feel it is useful to see all the balance equations of continuum thermodynamics under a unifying frame.            
The quantity $\tzrg{\Psi}$ is assumed to be composed of a bulk contribution and a surface contribution localized at a singular surface $\Gamma$ (see Fig.~\ref{fig-mueller}) and both the bulk and the surface contributions are assumed to be representable by corresponding densities $\Psi\om$ and $\Psi_\Gamma$, respectively. Here, we implicitly follow the standard notion of Gibbs' surface excess when discussing the surface quantities \citep{		gibbs-selected-vol1}. 

Considering a control volume $\cv$ as in Fig.~\ref{fig-mueller}, we thereby assume the following representation:
    \begin{itemize}
        \item Quantity $\tzrg{\Psi}$:
            \bea
                \nonumber
                \tzrg{\Psi}(\cv) = \int_{\cv^+\cup\cv^-}\Psi\om\,\dx + \int_{\Gamma} \Psi_\Gamma\,\ds\ ,
            \eea
        \item Flux $\genericflux$:
            \bea
                \nonumber
                \genericflux(\cv) = \int_{S^+\cup S^-}\genericbulkfluxdensity\cdot\nn\,\ds
                + \int_{\pa\Gamma}\genericsurffluxdensity\cdot{\tzrg{\nu}}\,\dl\ ,
            \eea
            where $\nn$ is the outer unit normal to the  boundary $\pa\cv$ and $\tzrg{\nu}$ is the outer unit normal to the line $\pa\Gamma$ (lying in $\Gamma$), and $S^+$, $S^-$ are the ``outer'' parts of $\pa\cv^+$, $\pa\cv^-$ in the sense that $S^+{\eqdef}\pa\cv^+{\cap}\pa\cv$ and $S^-{\eqdef}\pa\cv^-{\cap}\pa\cv$.
        \item Production $\genericproduction$:
            \bea
                \nonumber
                \genericproduction(\cv) = \int_{\cv^+\cup\cv^-}\genericbulkproductiondensity\,\dx
                + \int\ga\genericsurfproductiondensity\,\ds\ ,
            \eea
        \item Supply $\genericsupply$:
            \bea
                \nonumber\genericsupply(\cv) = \int_{\cv^+\cup\cv^-}\genericbulksupplydensity\dx
                + \int_{\Gamma}\genericsurfsupplydensity\,\ds\ .
            \eea
    \end{itemize}

We will distinguish between the volume (bulk) contribution $\Psi\om$ and its surface counterpart $\Psi\ga$, in the sense that in general\footnote{The symbol $A|_S$ stands for the restriction of the quantity $A$ to a set $S$.} $\Psi{\om}|_{\ga}\neq\Psi\ga$, that is, the restriction of the bulk quantity to the surface need not coincide with the corresponding surface quantity. This assumption corresponds to the fact that interfaces are in general zones where material properties may undergo abrupt changes and the transition layers are typically very thin. It is reasonable to treat them as $n{-}1$ dimensional manifolds, $n$ being the dimension of the ``bulk'' space, and the corresponding averaged (over the thickness of the layer) bulk quantities are then taken as the independent surface counterparts.

By taking arbitrary control volume $\cv$, using the generalized Reynolds' transport theorem, Gauss' theorem, and tools of the differential geometry, under an additional assumption of sufficient smoothness of all the involved quantities we can derive the following local form of the integral balance equations \citep[][section 1.3.2]{slattery-1990}:

\begin{itemize}
\begin{subequations}
\item{In the bulk $\mathcal{B}^+{\cup}\mathcal{B}^-$}:
    \be
        \label{eq:b1}
        \frac{\pa\Psi\om}{\pa t} + \div\left(\genericbulkfluxdensity+ \Psi\om \vv\right) - \genericbulkproductiondensity - \genericbulksupplydensity = 0\ ,
    \ee
    where ${\bf v}$ is the material velocity.

\item{At the interface $\Gamma$}:
    \be
    \label{eq:general_balance_surface}
        \frac{D\ga\Psi\ga}{Dt} + \Psi\ga\left(\div\ga\vv\gat -  2 K_M \vv\gan\right) + \div\ga\genericsurffluxdensity        - \genericsurfproductiondensity - \genericsurfsupplydensity = -\llbracket\genericbulkfluxdensity + \Psi\om(\vv - \vv\ga)\rrbracket\cdot\nn\ga\ ,
    \ee
    \end{subequations}
\end{itemize}
where  $\frac{D\ga}{Dt}$ denotes the surface material time derivative defined as
\begin{align}
       \frac{D\ga A}{Dt} \eqdef \left.\frac{\pa A}{\pa t}\right|_{{\bf X}\ga}\ ,
\end{align}
where ${\bf X}\ga$ denotes the surface material point, \citep[see][section 1.2.5]{slattery-1990}. Next, $\vv\ga$ is the surface velocity and $\vv\gat$ is its projection to the surface $\Gamma$ and $\vv\gan$ is the normal component:
\begin{align}
\label{def:tangent-normal}
\vv\gat \eqdef (\mathbb{I}{-}\nn\ga{\otimes}\nn\ga)\vv\ga\ ,\hspace{1cm}\vv\gan\eqdef \vv\ga{\cdot}\nn\ga\ ,
\end{align}
where $\mathbb{I}$ is the identity tensor, $K_M$ is the mean curvature of the surface, and $\lb z\rb$ denotes the jump of the bulk quantity $z$ across the surface $\Surface$ defined by $\lb z\rb := z^+ - z^-$ where $z^+$ and $z^-$ are the restrictions of $z|_{\mathcal{B}^+}$ and $z|_{\mathcal{B}^-}$, respectively, to $\Gamma$,
see Fig.~\ref{fig-mueller}. Note that in \eqref{eq:general_balance_surface}, we keep the surface velocity $\vv\ga$ inside the ``jump'' brackets $\llbracket\rrbracket$. This is standard notation in the literature understood in the sense that all surface quantities are tacitly taken as continuous, i.e. $\vv\ga^+=\vv\ga^-$ and thus the term on the right hand side in \eqref{eq:general_balance_surface} is interpreted as 
$
\llbracket\Psi\om(\vv - \vv\ga)\rrbracket\cdot\nn\ga 
=\llbracket\Psi\om\vv\rrbracket\cdot\nn\ga - \llbracket\Psi\om\rrbracket\vv\ga\cdot\nn\ga
$. Finally, $\div\ga$ denotes the surface divergence operator. For the definition of kinematic quantities and operators on the surfaces, see \cite{slattery-1990}(Appendix A) or the coordinate-free exposition by \cite{buscaglia-2011}.

\subsection{Local forms of the balance equations in the bulk}
The local forms of the balance equations in the bulk $\mathcal{B}^+\cup\mathcal{B}^-$ (i.e. outside the iterface $\Gamma$) for a single-component non-polar material read as follows (we omit the subscript $\om$ for bulk quantities for brevity):
\begin{itemize}
\begin{subequations}
  \item Balance of mass:
    \begin{align}
        \label{eq:bulk-mass}
        \frac{\pa\rho}{\pa t} + \div{(\rho\vv)} = 0\ ,
    \end{align}
    where $\rho$ is the density and $\vv$ is the material velocity.
  \item Balance of linear momentum:
    \begin{align}
        \label{eq:bulk-lin-mom}
        \frac{\pa(\rho\vv)}{\pa t} + \div(\rho\vv{\otimes}\vv) = \div{\stress} + \rho{\bb}\ ,
    \end{align}
    where $\stress$ is the Cauchy stress tensor and $\bb$ is the specific body force. Here and in what follows the dyadic product $\mathbf{a}\otimes\mathbf{d}$ of two vectors $\mathbf{a}$ and $\mathbf{d}$ is the second order tensor with the components $(\mathbf{a}\otimes\mathbf{d})_{ij} = a_i d_j$.
  \item Balance of angular momentum is reduced to the statement that the Cauchy stress tensor $\stress$ is symmetric, i.e.,
    \begin{align}
        \label{eq:bulk-ang-mom}
        \stress = \stress^\mathrm{T}\ ,
    \end{align}
    where $^\mathrm{T}$ denotes the transposition of a tensor.
  \item Balance of energy:
    \begin{align}
        \label{eq:bulk-total-energy}
        \frac{\pa\left(\rho(\ienergy + \frac{1}{2}|\vv|^2)\right)}{\pa t} + \div\left(\rho(\ienergy + \frac{1}{2}|\vv|^2)\vv\right)  = -\div{\heatflux} + \div{\left(\stress\vv\right)} + \rho r\ ,
    \end{align}
    where $\ienergy$ is the specific internal energy, $\heatflux$ denotes the energy flux, and $r$ is the specific energy supply. Employing the mass and momentum balances, we obtain the balance equation for the energy in the form:
    \begin{align}
        \label{eq:bulk-energy}
        \rho\frac{D\ienergy}{Dt} = -\div{\heatflux} + \stress:\DD + \rho r\ ,
    \end{align}
   where $\DD=\frac{1}{2}(\grad\vv+(\grad\vv)^\mathrm{T})$ is the symmetric part of the velocity gradient and $\frac{D}{Dt} = \frac{\pa}{\pa t}+ \vv\cdot\nabla$ denotes the material time derivative.
  \item The formulation of the second law of thermodynamics:
    \begin{align}
        \label{eq:bulk-entropy}
        \frac{\pa(\rho\eta)}{\pa t} + \div{\left(\rho\eta\vv + {\entropybulkfluxdensity}}\right) - {\entropybulksupplydensity} = {\entropybulkproductiondensity}\hspace{1cm}\text{with}\hspace{1cm} {\entropybulkproductiondensity} \geq 0\ ,
    \end{align}
    \end{subequations}
    where $\eta$ is the specific entropy, $\entropybulkfluxdensity$ is the bulk entropy flux, $\entropybulksupplydensity$ is the entropy supply
    and $\entropybulkproductiondensity$ is the entropy production, which must be non-negative according to the second law of thermodynamics.
\end{itemize}

\subsection{Local form of the balance quations at the interface}
The local forms of the balance equations at the interface $\Gamma$ for a single-component non-polar material read as follows:
\begin{itemize}
\begin{subequations}
  \item Balance of mass:
    \be
        \label{eq:surf-mass}
        \frac{D\ga\rho\ga}{Dt} + \rho\ga(\div\ga\vv\gat - 2K_M \vv\gan) = -\lb\rho(\vv-\vv\ga)\rb\cdot\nn\ga\ ,
  	\ee
  	      where $\rho\ga$ is the surface mass density.
 
	 \item Balance of linear momentum:
    \be
        \label{eq:surf-momentum}
        \frac{D\ga(\rho\ga\vv\ga)}{Dt} + \rho\ga\vv\ga(\div\ga\vv\gat - 2K_M \vv\gan)  - \div\ga\stress\ga - \rho\ga\bb\ga = -\lb\rho\vv\otimes(\vv{-}\vv\ga) - \stress\rb\cdot\nn\ga\ ,
    \ee
   where $\stress\ga$ denotes the surface Cauchy stress tensor and $\bb\ga$ is the specific surface force. 
    
  \item Balance of angular momentum (for a non-polar material):
    \be
        \label{eq:surf-angular}
        \stress\ga = \stress\ga^\mathrm{T}\ ,
    \ee
    i.e., symmetry of the surface Cauchy stress tensor.
  \item{Balance of energy:}
    \begin{align}\nonumber
        &\frac{D\ga\left(\rho\ga(\ienergy\ga + \frac{1}{2}|\vv\ga|^2)\right)}{Dt} 
        + \rho\ga(\ienergy\ga+\frac{1}{2}|\vv\ga|^2)(\div\ga\vv\gat - 2K_M\vv\gan) = -\div\ga\heatflux\ga + \div\ga\left(\stress\ga\vv\ga\right) \\\label{eq:surf-tot-energy}
        & + \rho\ga r\ga + \rho\ga\bb\ga\cdot\vv\ga + \left\lb-\rho\left(\ienergy+\frac{1}{2}|\vv|^2\right)(\vv{-}\vv\ga) + \stress\vv - \heatflux\right\rb\cdot\nn\ga\ ,
    \end{align}
    where $\ienergy\ga$ denotes the specific surface internal energy and $\heatflux\ga$ is the surface energy flux. Multiplying the surface momentum balance \eqref{eq:surf-momentum} by $\vv\ga$, one 
    can obtain the surface balance equation for kinetic energy. This can be subtracted from \eqref{eq:surf-tot-energy}, which yields the balance of surface energy in the reduced form
			    \begin{align}\nonumber
        \frac{D\ga(\rho\ga\ienergy\ga)}{Dt} + \rho\ga\ienergy\ga(\div\ga\vv\gat - 2K_M\vv\gan)
        &= - \div\ga\heatflux\ga + \stress\ga{:}\grad\ga\vv\ga + \rho\ga r\ga\\\label{eq:surf-energy_P}  & 
        + \left\lb-\rho\left(e + \frac{1}{2}|\vv\ga-\vv|^2\right)(\vv{-}\vv\ga) + \stress(\vv{-}\vv\ga) -\heatflux\right\rb\cdot\nn\ga\ , 
    \end{align}
    where $\stress_\Surface{:}\grad_\Surface\vv_{\Surface} \eqdef \mathrm{tr}\left(\stress\ga\ \nabla\ga\vv\ga\right)$. 

  \item{The formulation of the second law of thermodynamics:}
    \be
        \label{eq:surf-entropy}
        \frac{D\ga(\rho\ga\eta\ga)}{Dt} + \rho\ga\eta\ga\left(\div\ga\vv\gat{-}2 K_M\vv\gan\right)
        + \div\ga\entropysurffluxdensity + \left\lb{\entropybulkfluxdensity}{+}\rho\eta(\vv{-}\vv\ga)\right\rb\cdot\nn\ga{-}\entropysurfsupplydensity = \entropysurfproductiondensity\ ,
    \ee
    with
    \be
 \entropysurfproductiondensity\ \geq 0\ .
    \ee
  Here $\eta\ga$ denotes the surface specific entropy, $\entropysurffluxdensity$ is the surface entropy flux, $\entropysurfsupplydensity$ is the surface entropy supply and $\entropysurfproductiondensity$ is the surface entropy production, which must be non-negative in order to comply with the second law of thermodynamics.
  \end{subequations}
\end{itemize}

\section{Derivation of constitutive equations for Korteweg-type fluids in the bulk}
\label{sec:Korteweg fluids}
In this section, we recall and slightly modify the derivation of a constitutive model for a Korteweg-type fluid developed in \cite{heida.m.malek.j:on}. 
The derivation is based on imposing the following constitutive ansatz for the (specific) internal energy in the bulk:
\be
    \ienergy = \hat{\ienergy}(\eta,\rho,\nabla\rho)\ .
\ee
Assuming that $\hat{\ienergy}$ is differentiable, the thermodynamic temperature $\vartheta$ is introduced through
\be
	\label{def:vartheta}
    \vartheta \eqdef \frac{\pa\hat{\ienergy}}{\pa\eta}\ . \ee
The corresponding Helmholtz free energy $\helm$ is then obtained via the Legendre transform giving 
\begin{align}
\label{def-helm}
\helm = e - \vartheta\eta\ ,\hspace{1cm}\ \
\widehat{\helm}(\vartheta,\rho,\nabla\rho)\ &\eqdef \inf_{\eta}\left(\hat{e}(\eta,\rho,\nabla\rho)-\vartheta\eta\right) = \left.\left(\hat{e}(\eta,\rho,\nabla\rho)-\vartheta\eta\right)\right|_{\eta=\hat{\eta}(\vartheta,\rho,\nabla\rho)}\  ,
	\end{align}
where, in the last equality, we assume the invertibility of \eqref{def:vartheta} with respect to $\eta$. As a consequence of \eqref{def-helm}, we obtain the standard thermodynamic relation \be
	\label{eq:pahelm-pavartheta}
	\eta = -\frac{\pa\widehat{\helm}}{\pa\vartheta}\ .
\ee
Taking the material time derivative of \eqref{def-helm}, we obtain, after using \eqref{eq:pahelm-pavartheta}, that
\be
	\frac{\pa\widehat{\helm}}{\pa\rho}\frac{D{\rho}}{Dt} + \frac{\pa\widehat{\helm}}{\pa\nabla\rho}\cdot\frac{D{\nabla\rho}}{Dt} = \frac{De}{Dt} - \vartheta\frac{D{\eta}}{Dt}\ . \label{pepa1}
\ee
{Multiplying \eqref{pepa1} by $\rho$ and applying the energy balance (\ref{eq:bulk-energy}) we obtain}
\be
    \label{eq:1}
    \rho\left(\vartheta\frac{D{\eta}}{Dt}+\frac{p}{\rho^2}\frac{D{\rho}}{Dt}+\frac{\pa\widehat{\helm}}{\pa\nabla\rho}\cdot\frac{D{\nabla \rho}}{Dt} \right) = -\div\heatflux + \stress:\DD + \rho r\ ,
\ee
where $p$ denotes the thermodynamic pressure defined through
\be
\label{def:td-press}
p \eqdef \rho^2\frac{\pa\widehat{\helm}}{\pa\rho}\ .
\ee
Taking the gradient of (\ref{eq:bulk-mass}) we obtain
\be
    \label{eq:nabla-rho}
    \frac{D{\nabla\rho}}{Dt} + (\nabla\vv)^T(\nabla\rho) + \nabla(\rho\div\vv) = {\bf 0}\ .
\ee
Using (\ref{eq:bulk-mass}) and (\ref{eq:nabla-rho}) in (\ref{eq:1}), we arrive at
\be
    \label{eq:2}
    \rho\vartheta\frac{D{\eta}}{Dt} = (m+p)\div\vv + \stress^d:\DD^d - \div\heatflux + \rho\left(\frac{\pa\widehat{\helm}}{\pa\nabla\rho}\otimes\nabla\rho\right):(\nabla\vv)^T + \rho\frac{\pa\widehat{\helm}}{\pa\nabla\rho}\cdot\nabla(\rho\div\vv) + \rho r\ ,
\ee
where $m$ denotes the mean normal stress $m \eqdef \frac{1}{3}\mathrm{tr}(\stress)$ and $()^d \eqdef () - \frac{1}{3}\mathrm{tr}(){\Id}$ denotes the deviatoric part of a tensor. By the principle of material frame indifference \citep[see, e.g.,][]{truesdell.c.noll.w:non-linear}, the internal energy $\hat{e}$ can only depend on the magnitude of $\nabla \rho$ which immediately implies symmetry of the tensor $\rho\frac{\pa\widehat{\helm}}{\pa\nabla\rho}\otimes\nabla\rho$. Consequently, we may replace $\nabla\vv$ by its symmetric part $\DD$ in the fourth term in the right-hand side of (\ref{eq:2}). Using the identity \be\rho\frac{\pa\widehat{\helm}}{\pa\nabla\rho}\cdot\nabla(\rho\div\vv) = \div\left(\rho^2\frac{\pa\widehat{\helm}}{\pa\nabla\rho}\div\vv\right)
    -\rho\div\vv\ \div\left(\rho\frac{\pa\widehat{\helm}}{\pa\nabla\rho}\right)\ ,\ee
dividing \eqref{eq:2} by $\vartheta$ we obtain, after suitable rearrangements, the following local form of the balance equation for the bulk entropy:
\begin{align}
    \nonumber
    \rho\frac{D\eta}{Dt} &= -\div\left(\frac{\heatflux - \rho^2\frac{\pa\widehat{\helm}}{\pa\nabla\rho}\div\vv}{\vartheta}\right) + \left(\heatflux - \rho^2\frac{\pa\widehat{\helm}}{\pa\nabla\rho}\div\vv\right)\cdot\nabla\left(\frac{1}{\vartheta}\right)
    + \frac{\rho r}{\vartheta}\\\label{eq:3} &+ \frac{1}{\vartheta}\left\{\left(m + p + \frac{1}{3}\rho\frac{\pa\widehat\helm}{\pa\nabla\rho}\cdot\nabla\rho -\rho\div\left(\rho\frac{\pa\widehat{\helm}}{\pa\nabla\rho}\right)
    \right)\div\vv + \left(\stress + \rho\frac{\pa\widehat{\helm}}{\pa\nabla\rho}\otimes\nabla\rho\right)^d:\DD^d\right\}\ .
\end{align}
Recalling (\ref{eq:bulk-entropy}) and assuming that the entropy supply is given only by the corresponding energy supply term, that is, postulating that\footnote{It is also possible to split $\rho r{=}\rho r_A{+}\rho r_B$ and postulate $ \entropybulksupplydensity{=} \frac{\rho r_A}{\vartheta}$, and incorporate $\frac{\rho r_B}{\vartheta}$ among the entropy producing mechanisms. For simplicity, we do not consider this possibility here.} 
\be
   \entropybulksupplydensity = \frac{\rho r}{\vartheta}\ ,
\ee
we can identify the entropy flux in (\ref{eq:3}) as
\be
    \label{eq:entropy-flux}
   \entropybulkfluxdensity = 
   \frac{\heatflux - \rho^2\frac{\pa\widehat{\helm}}{\pa\nabla\rho}\div\vv}{\vartheta}\ .
\ee
The second and last terms in the right-hand side of (\ref{eq:3}) represent the entropy production. In accordance with the usual approach in the constitutive theory within linear irreversible thermodynamics \citep{groot.sr.mazur.p:non-equilibrium}, we want to express this term as a sum of binary products between the thermodynamic ``affinities'' (forces) and the corresponding thermodynamic ``fluxes''. Even if we pick as the set of affinities $(\div\vv,\DD^d, \nabla(\frac{1}{\vartheta}))$, the splitting into two groups is still not unique. Note, in particular, that the term $\rho^2\frac{\pa\widehat{\helm}}{\pa\nabla\rho}\cdot\nabla\left(\frac{1}{\vartheta}\right)\div\vv$ can contribute to both products with affinities $\nabla(\frac{1}{\vartheta})$ and $\div\vv$. Without knowing a-priori which choice is preferable, we split this term via a convex combination governed by a free parameter $\alpha\in\langle 0,1\rangle$ between both these terms. With such a choice, we finally arrive at the formula for the rate of entropy production,
\bea
\label{eq:bulk_entropy_production}
   \entropybulkproductiondensity &=& \frac{1}{\vartheta}\left\{\left( m + \tilde{p}_\alpha \right)\div\vv
    + \left(\stress + \stress_\rho\right)^d:\DD^d\right\}
    + \tilde{\heatflux}_\alpha\cdot\nabla\left(\frac{1}{\vartheta}\right)\ ,
\eea
where we set 
\begin{subequations}
\label{eq:aux}
\begin{align}
    \label{def:tilde-heatfux}
    \tilde{\heatflux}_\alpha &\eqdef \heatflux - \alpha\rho^2\frac{\pa\widehat{\helm}}{\pa\nabla\rho}\div\vv\ ,\\
    \label{def:tilde-p}
    \tilde{p}_\alpha &\eqdef p + \frac{1}{3}\rho\frac{\pa\widehat\helm}{\pa\nabla\rho}\cdot\nabla\rho -\rho\div\left(\rho\frac{\pa\widehat{\helm}}{\pa\nabla\rho}\right) - (1-\alpha)\vartheta\rho^2\frac{\pa\widehat{\helm}}{\pa\nabla\rho}\cdot\nabla\left(\frac{1}{\vartheta}\right)\ ,\\
    \label{tilde:stress-rho}
    \stress_\rho &\eqdef \rho\frac{\pa\widehat{\helm}}{\pa\nabla\rho}\otimes\nabla\rho\ .
\end{align}
\end{subequations}
We recognize the right-hand side of (\ref{eq:bulk_entropy_production}) as three entropy-producing mechanisms, each in the form of a product of two terms (sometimes called thermodynamic ``flux'' and thermodynamic ``affinity''), each couple describing a different physical process. Restricting ourselves here to linear relationships among the two types of terms, we arrive at the constitutive equations
\begin{subequations}
\label{eq:constitutive}
\begin{align}
    \label{eq:rheo-stress-rho}
    (\stress+\stress_\rho)^d &= 2\mu\DD^d\ , &&\mu>0\ ,\\
    \label{eq:rheo-tilde-m+p}
    m+\tilde{p}_\alpha &= \frac{2\mu+3\lambda}{3}\div\vv\ , &&2\mu + 3\lambda >0\ ,\\
    \label{eq:rheo-tilde-heatflux}
    \tilde{\heatflux}_\alpha &= \kappa\nabla\left(\frac{1}{\vartheta}\right)\ , &&\kappa>0\, .
\end{align}
\end{subequations}
If these relationships are inserted back into \eqref{eq:bulk_entropy_production}, we obtain the rate of entropy production expressed as a piece-wise quadratic function in terms of the ``affinities'',
\bea
\label{eq:entprod1}
    \entropybulkproductiondensity &=& \frac{1}{\vartheta}
    \left\{
    \frac{2\mu+3\lambda}{3}(\div\vv)^2+
    2\mu|\DD^d|^2\right\}+
    {\kappa}\left|\nabla\left(\frac{1}{\vartheta}\right)\right|^2
    ,
\eea
(here and in what follows $|\mathbb{A}|\eqdef \sqrt{\mathbb{A}_{ij}\mathbb{A}_{ij}}$), or equivalently, in terms of ``fluxes'' as follows
\bea
\label{eq:entprod2}
        \entropybulkproductiondensity &=& \frac{1}{\vartheta}
    \left\{
    \frac{3}{2\mu+3\lambda}(\tilde{m}+\tilde{p}_\alpha)^2+
    \frac{1}{2\mu}|(\stress+\stress_\rho)^d|^2\right\} +
    \frac{1}{\kappa}|\tilde{\heatflux}_\alpha|^2
    .
\eea
In both cases we see that the positivity of the coefficients together with the piece-wise quadratic form of \eqref{eq:entprod1} and \eqref{eq:entprod2} ensure that the second law of thermodynamics, i.e., the non-negativity of the rate of entropy production, is guaranteed. Inserting the formulas \eqref{eq:aux} into \eqref{eq:constitutive}, we obtain the following expressions for the Cauchy stress $\stress$, energy flux $\heatflux$, and entropy flux $\entropybulkfluxdensity$ in the bulk:
\begin{subequations}
\begin{align}
\nonumber
    \stress =& -\left(\rho^2\frac{\pa\widehat{\helm}}{\pa\rho} - \rho\div\left(\rho\frac{\pa\widehat{\helm}}{\pa\nabla\rho}\right) - (1-\alpha)\vartheta\rho^2\frac{\pa\widehat{\helm}}{\pa\nabla\rho}\cdot\nabla\left(\frac{1}{\vartheta}\right)
    \right){\Id} \\    \label{eq-bulk-stress}
 &+ \lambda\div\vv{\Id} + 2\mu\DD - \rho\left(\frac{\pa\widehat{\helm}}{\pa\nabla\rho}\otimes\nabla\rho\right)\ ,\\
    \label{eq-bulk-heat-flux}
    \heatflux =& \kappa\nabla\left(\frac{1}{\vartheta}\right) + \alpha\rho^2\frac{\pa\widehat{\helm}}{\pa\nabla\rho}\div\vv\ ,\\
    \label{eq-bulk-entropyflux}
    \entropybulkfluxdensity =& \frac{1}{\vartheta}\left(\kappa\nabla\left(\frac{1}{\vartheta}\right) - (1-\alpha) \rho^2\frac{\pa\widehat{\helm}}{\pa\nabla\rho}\div\vv\right)\ ,
\end{align}
\end{subequations}
where $\alpha \in \langle 0,1\rangle$, $2\mu + 3\lambda > 0$, $\mu > 0$, $\kappa > 0$.

\section{Derivation of boundary conditions for Korteweg-type fluids}
\label{sec:Korteweg-bcs}
\begin{figure}[h!]
\begin{center}
\includegraphics[scale=0.1]{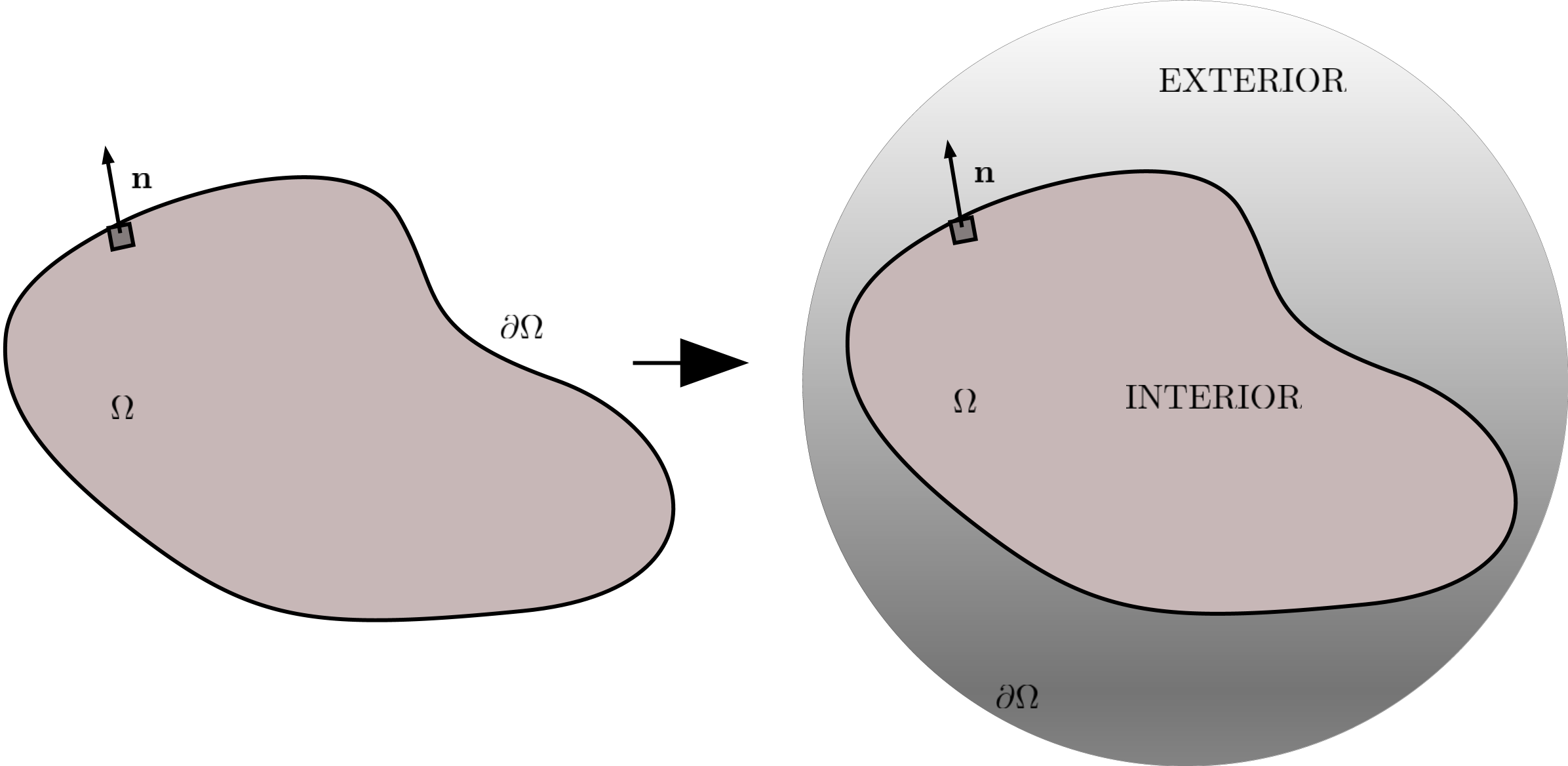} 
\caption{Visualization of the concept of application of the framework of continuum theory with singular surfaces on external boundaries. We adopt the viewpoint that the outer boundary of $\Omega$, i.e., the surface $\pa\Omega$, is an interface between $\Omega$ and its exterior.}
\label{fig-boundary}
\end{center}
\end{figure}

In this section, we extend the constitutive theory for Korteweg-type fluids presented in the previous section to dissipative processes at the boundary which will allow us to formulate thermodynamically based constitutive relations (compatible with the (local form) of the second law of thermodynamics) in the form of boundary conditions. Let us now consider a domain $\Omega$ which contains Korteweg (two-phase) fluid and let us denote its boundary $\pa\Omega$. We will look at the boundary $\pa\Omega$ as an interface between the domain $\Omega$ and its exterior; see Fig.~\ref{fig-boundary}. Adopting this viewpoint, we can employ the framework of continuum mechanics with singular surfaces introduced in Section \ref{sec:General form}. 

For Korteweg-type fluids we assume, in accordance with the physical theories of capillarity \citep[e.g.][]{rowlinson}, that the outer boundary is in fact a boundary layer with certain specific properties. This layer will be treated as infinitely thin and all the corresponding bulk quantities in this layer will be described by their surface (boundary) counterparts, obtained by space averaging over the thickness of the boundary layer. Unlike Navier-Stokes fluids,  Korteweg fluids naturally incorporate the notion of surface tension as the interfacial energy in transition regions separating the phases. With the goal of formulating boundary conditions for Korteweg-type fluids and describing phenomena such as wetting (i.e., contact angles), it appears reasonable to extend the notion of interfacial interaction and to include also interaction of the fluid with the boundary walls. To this end, we postulate the existence of boundary surface energy $\ienergy\ga$, and boundary surface entropy $\eta\ga$ expressing this fluid-boundary interaction. By postulating the different constitutive equations for $\ienergy\ga$, we will obtain a hierarchy of models of various complexity, which will result in a corresponding hierarchy of classes of boundary conditions.

For simplicity, we will investigate a model in which we ignore convective mechanisms on the boundary and we shall thus consider the boundary $\pa\Omega$ to be static by postulating zero surface velocity, i.e.,
\begin{subequations}
\label{eq:static-surface}
\be
    \label{eq:6}
    \vv\ga = {\bf 0}\ .
\ee
This condition should not be confused with the no-slip boundary condition in which $\vv^-$ vanishes on the boundary $\pa\Omega$, which is not required here. 

Next, we will also need to specify conditions on the exterior side of the boundary (denoted by a $+$ sign with the convention of exterior unit normal pointing from $-$ to $+$). We will assume that the material outside $\Omega$ is at rest, i.e.,
\be
    \label{eq:7}
    \vv^+ = {\bf 0}\ ,
\ee
implying that any mass exchange between the exterior and interior is excluded. Concerning momentum exchange, the exterior may exert force on the interior domain, but due to \eqref{eq:7}, this force does not produce any mechanical power and as such does not contribute to the energy balance. We will, however, assume that the exterior may facilitate (non-convective) energy and entropy transfer and we relate them through the standard relation of thermodynamics \citep{Coleman1963}:
\begin{align}
\label{eq:exterior-entropy-energy-rel}
{\entropybulkfluxdensity}^+ = \frac{\heatflux^+}{\vartheta^+}\ .
\end{align}

Furthermore, we assume that from the ``inside'' the boundary is impermeable, i.e.,
\be
    \label{eq:9}
    {\vv^-}\cdot\nn\ga = 0\ ,
\ee
allowing for slip of the Korteweg fluid along the boundary but no penetration.
\end{subequations}

In the following, it will be convenient to not explicitly invoke the concept of surface mass density, as we will only require the notions of surface energy and surface entropy. Thus instead of using the specific (i.e. related to unit of mass) surface energy and surface entropy, we formulate the equations directly for the products $\rho\ga\ienergy\ga$ and $\rho\ga\eta\ga$, using the following notation:
\begin{subequations}
\be
	\widetilde{\ienergy}\ga\eqdef\rho\ga\ienergy\ga\ ,\hspace{1cm}	\widetilde{\eta}\ga\eqdef\rho\ga\eta\ga\ ,
\ee
representing the surface energy and surface entropy (per unit surface), respectively. Similarly for the energy and momentum surface supply terms, we set
\be
	\widetilde{r}\ga\eqdef\rho\ga r\ga\ ,\hspace{1cm}	\widetilde{\bb}\ga\eqdef\rho\ga\bb\ga\ .
\ee
\end{subequations}

Similarly as in the bulk, the key constitutive relation representing the assumption of a local thermodynamic equlibrium takes the form
\be
\widetilde\ienergy\ga = \widehat{\ienergy}\ga(\widetilde\eta\ga,\dots)\ ,
\ee
where the dots stand for other state variables. We then define the surface thermodynamic temperature $\vartheta\ga$ through
\bea
\label{def-varthetaga}
    \vartheta\ga \eqdef \frac{\pa\widehat{\ienergy}\ga}{\pa\widetilde{\eta}\ga}\ .
\eea
We also introduce the corresponding surface Helmholtz free energy $\widetilde{\helm}\ga$ via the Legendre transform obtaining
\be
\label{def-helm-ga}
\widetilde{\helm}\ga = \widetilde{\ienergy}\ga  - \vartheta\ga\widetilde{\eta}\ga\ ,\hspace{1cm}
      \widehat{\helm}\ga(\vartheta\ga,\dots)  \eqdef \inf_{\widetilde{\eta}\ga}\left(\widehat{\ienergy}\ga(\widetilde{\eta}\ga,\dots)-\vartheta\ga\widetilde{\eta\ga}\right) = \left.\left(\widehat{\ienergy}\ga(\widetilde{\eta}\ga,\dots)-\vartheta\ga\widetilde{\eta}\ga\right)\right|_{\widetilde\eta\ga=\widehat{\eta}\ga(\vartheta\ga,\dots)}\  ,
\ee
where in the last equality we assume invertibility of \eqref{def-varthetaga} with respect to $\widetilde{\eta}\ga$. It then follows from \eqref{def-helm-ga} that
\be
\label{eq:pahelmga-pavarthetaga}
\widetilde{\eta}\ga = -\frac{\pa\widehat{\helm}\ga}{\pa\vartheta\ga}\ .
\ee
Concerning the structure of the constitutive equation for the surface Helmholtz free energy, we will consider the following three  situations:

\begin{itemize}
\begin{subequations}
    \item{{ Model A}:} \\
	Motivated by statistical physics description of the surface free energy (see \cite{rowlinson}(eq. 4.114)) we assume that 
        \bea
            \label{def: modelA}
            \widetilde{\helm}\ga = \widehat{\helm}\ga(\vartheta\ga, \rho^-)\ ,
        \eea
i.e., the surface Helmholtz free energy depends not only on the surface temperature, but also on the density of the neighboring bulk. Here only dependence on the interior density $\rho^-$ is considered.
    \vspace{0.5cm}    
    \item{{ Model B}:}
        \bea
            \label{def: modelB}
            \widetilde{\helm}\ga = \widehat{\helm}\ga(\vartheta\ga)\ ,
        \eea
            i.e., dependence of the surface Helmholtz free energy only on surface temperature.
    \vspace{0.5cm}
    \item{{ Model C}:}
        \bea
            \label{def: model3}
            \widetilde{\helm}\ga \equiv 0\ ,
        \eea
        i.e., the ``trivial'' model without any surface Helmholtz free energy, which however still provides non-trivial boundary conditions for the bulk terms.
        \end{subequations}
\end{itemize}
We will now inspect these three cases in detail. 

\subsection{{\bf Model A}}
Applying the surface material time derivative to \eqref{def-helm-ga} with $\widetilde{\helm}\ga$ given by \eqref{def: modelA}, using \eqref{eq:pahelmga-pavarthetaga}, and employing the surface energy balance in the form \eqref{eq:surf-energy_P}, we obtain the identity
\begin{align}
\nonumber
    \vartheta\ga\frac{D\ga\widetilde{\eta}\ga}{Dt} &= -\widetilde{\ienergy}\ga\left(\div\ga\vv\gat - 2 K_M \vv\gan\right) -
    \div\ga\heatflux\ga + \stress_\Surface:\grad_\Surface\vv_{\Surface } + \widetilde{r}\ga - \frac{\pa\widehat{\helm}\ga}{\pa\rho}\frac{D\ga\rho^-}{Dt}
  \\\label{eq:5} &   
    + \left\lb-\rho\left(e + \frac{1}{2}|\vv\ga-\vv|^2\right)(\vv{-}\vv\ga) + \stress(\vv{-}\vv\ga) -\heatflux\right\rb\cdot\nn\ga\ .
\end{align}
The surface being static (see the assumption \eqref{eq:6}, all terms containing $\vv\ga$ vanish and the surface material time derivative in (\ref{eq:5}) becomes just the partial time derivative, i.e. $\frac{D\ga}{Dt} = \frac{\pa }{\pa t}$. We then use (\ref{eq:bulk-mass}) and \eqref{eq:9} and obtain
\bea
    \label{eq:10}
    \vartheta\ga\frac{\pa{\widetilde{\eta}\ga}}{\pa t} &=&  -\div\ga\heatflux\ga + \widetilde{r}\ga
+ \left(\rho\frac{\pa\widehat{\helm}\ga}{\pa\rho}\div\vv + \frac{\pa\widehat{\helm}\ga}{\pa\rho}\vv\cdot\grad\rho\right)^-    
         + \left\lb\stress\vv\right\rb\cdot\nn\ga   -\lb\heatflux\rb\cdot\nn\ga \ .
\eea
With the assumptions (\ref{eq:6}--\ref{eq:9}) and neglecting for simplicity the surface ``body'' forces, i.e.,
\be
    \label{eq:bbeq0}
    \widetilde{\bb}\ga = {\bf 0}\ ,
\ee
the balance of linear momentum on the surface (\ref{eq:surf-momentum}) reads
\be
    \label{eq:11}
     - \div\ga\stress\ga = \lb\stress\rb\,\nn\ga\ .
\ee
We shall consider a membrane model\footnote{A generalization that would involve more complex structure of the surface stress tensor or non-constant surface tension is possible, but not straightforward. We will therefore not pursue this possiblity here; see also a comment in the conclusion.}, where only the surface tension $\sigma$ (which is assumed to be constant here for simplicity) constitutes the surface stress tensor, i.e., we consider
\be
    \label{eq:surfacestresstensor}
    \stress\ga = \sigma \mathbb{I}\ga\ , \quad \textrm{ where } \mathbb{I}\ga\eqdef\mathbb{I}-\nn\ga\otimes\nn\ga.
\ee
Then it holds \citep[][Appendix A]{slattery-1990} that
\be
    \div\ga\stress\ga = 2K_M\sigma\nn\ga\ ,
\ee
where $K_M$ is the mean curvature of the surface, and thus by (\ref{eq:11}), we get
\be
    \lb\stress\rb\,\nn\ga = - 2K_M\sigma\nn\ga \qquad \textrm{ leading to } \qquad   (\lb\stress\rb\,\nn\ga)_\tau = {\bf 0}\ ,
\ee
where $()_\tau$ denotes the projection of a vector to the tangent plane. Hence 
\be
   (\stress\nn\ga)^{+}_\tau = (\stress^{+}\nn\ga)_\tau = (\stress^{-}\nn\ga)_\tau =    (\stress\nn\ga)^{-}_\tau\,. \label{eq:119}
\ee
Due to \eqref{eq:7} and \eqref{eq:9} and by virtue of continuity of tangent traction (\ref{eq:119}) and the symmetry of $\stress$, it holds that
\bea
\label{eq:12}
    \left\lb\stress\vv\right\rb\cdot\nn\ga &=& \lb{\vv_\tau}\rb\cdot(\stress\nn\ga)_\tau
    = - {\vv^-_\tau}\cdot(\stress\nn\ga)^\pm_\tau\ .
\eea
Assuming Korteweg fluid inside $\Omega$, we have, by (\ref{def:tilde-heatfux}) and (\ref{eq:rheo-tilde-heatflux}), the following expression for the energy flux
\bea
    \label{eq:13}
    \heatflux^- = \left({\kappa}\nabla\left(\frac{1}{\vartheta}\right) + \alpha\rho^2\frac{\pa\widehat{\helm}}{\pa\nabla\rho}\div\vv\right)^-\ .
\eea
Inserting (\ref{eq:12}) and (\ref{eq:13}) into (\ref{eq:10}), we finally obtain 
\bea
\label{eq:15}
    \vartheta\ga\frac{\pa{\widetilde{\eta}\ga}}{\pa t} &=&  -\div\ga\heatflux\ga + \widetilde{r}\ga
         - \tt\cdot{\vv^-_\tau}
         +\left(s_\rho + \alpha s_{\nabla\rho}\right)\div\vv^-
         - \heatflux^+\cdot\nn\ga
         + \kappa^-\frac{\pa}{\pa\nn\ga}\left(\frac{1}{\vartheta^-}\right)\ ,
\eea
where we set
\begin{subequations}
\label{def:trho-srho-snablarho}
\bea
    \label{def: trho_model1}
    \tt &\eqdef& \left((\stress\nn\ga)_\tau - \frac{\pa\widehat{\helm}\ga}{\pa\rho}\nabla\ga\rho\right)^-\ ,\\
    \label{def: srho_model1}
    s_\rho &\eqdef& \left(\rho\frac{\pa\widehat{\helm}\ga}{\pa\rho}\right)^- ,\\
    s_{\nabla\rho} &\eqdef& \left(\rho^2\frac{\pa\widehat{\helm}}{\pa\nabla\rho}\cdot\nn\ga\right)^-\ ,\\
       \frac{\pa}{\pa\nn\ga}\left(\frac{1}{\vartheta^-}\right) &\eqdef&
 \nabla\left(\frac{1}{\vartheta^-}\right)\cdot\nn\ga\ .        
         \eea
\end{subequations}
In \eqref{def: trho_model1}, we used the fact that $\nabla\rho^-\cdot\vv^- = \nabla\ga\rho^-\cdot\vv^-_\tau$, due to \eqref{eq:9}. 

The next step consists of transforming \eqref{eq:15} into the form \eqref{eq:surf-entropy} where we must also incorporate all the simplifying assumptions used above, such as \eqref{eq:6}-\eqref{eq:9}. Then \eqref{eq:surf-entropy} takes the form 
\bea
    \label{eq:17a}
     \frac{\pa\widetilde{\eta}\ga}{\pa t} &=& 
-\div\ga\entropysurffluxdensity
    -  \left\lb{\entropybulkfluxdensity}\right\rb\cdot\nn\ga\ 
   + \entropysurfsupplydensity + \entropysurfproductiondensity\ \hspace{1cm}\text{with}\ \ \entropysurfproductiondensity   \geq 0\ ,
\eea
which, upon inserting \eqref{eq-bulk-entropyflux} and \eqref{eq:exterior-entropy-energy-rel}, leads to
    \bea 
    \label{eq:17}
     \frac{\pa\widetilde{\eta}\ga}{\pa t} &=& 
-\div\ga\entropysurffluxdensity
    - \frac{\heatflux^+}{\vartheta^+}\cdot\nn\ga
    + \frac{\kappa^-}{\vartheta^-}\frac{\pa}{\pa\nn\ga}\left(\frac{1}{\vartheta^-}\right)
    - (1-\alpha)\frac{s_{\nabla\rho}}{\vartheta^-}\div\vv^-
    + \entropysurfsupplydensity + \entropysurfproductiondensity\ .
\eea
We proceed in two different ways. These ways differ in the manner in which the term $\div\vv^-$ in \eqref{eq:15} and \eqref{eq:17} is treated. In the first procedure, called Model A1, $\div\vv^-$ is kept unaltered, while in the second procedure, called Model A2, we will split $\div\vv^-$ into the surface divergence and the normal derivative.
\subsubsection{Model A1}
\label{subsec-model-a1}
We first observe that the equation (\ref{eq:15}) can be rewritten in the following form of entropy balance:
\begin{align}
\nonumber
       \frac{\pa{\widetilde{\eta}\ga}}{\pa t} &=  -\div\ga\left(\frac{\heatflux\ga}{\vartheta\ga}\right)
        + \heatflux\ga\cdot\nabla\ga\left(\frac{1}{\vartheta\ga}\right) \\
        \label{eq:16}
         & + \frac{1}{\vartheta\ga}\left\{-\tt\cdot{\vv^-_\tau}
         + \left(s_\rho+\alpha s_{\nabla\rho}\right)\div\vv^-
         + \widetilde{r}\ga  - \heatflux^+\cdot\nn\ga
         + \kappa^-\frac{\pa}{\pa\nn\ga}\left(\frac{1}{\vartheta^-}\right)\right\}\ .
\end{align}
Subtracting (\ref{eq:16}) from (\ref{eq:17}) yields
\bea
    \nonumber
    0 &=& -\div\ga\left(\entropysurffluxdensity - \frac{\heatflux\ga}{\vartheta\ga}\right) -
    \heatflux\ga\cdot\nabla\ga\left(\frac{1}{\vartheta\ga}\right)
    + \frac{\tt\cdot{\vv^-_\tau}}{\vartheta\ga} - \div\vv^-\left((1-\alpha)\frac{s_{\nabla\rho}}{\vartheta^-} + \frac{s_\rho+\alpha s_{\nabla\rho}}{\vartheta\ga} \right)\\\label{eq:18} &+& \left(\entropysurfsupplydensity - \frac{\widetilde{r}\ga}{\vartheta\ga}\right)
    - \heatflux^+\cdot\nn\ga\left(\frac{1}{\vartheta^+}-\frac{1}{\vartheta\ga}\right)
    + \kappa^-\frac{\pa}{\pa\nn\ga}\left(\frac{1}{\vartheta^-}\right)\left(\frac{1}{\vartheta^-}-\frac{1}{\vartheta\ga}\right) + \entropysurfproductiondensity\ .
\eea
Since $\widetilde{r}\ga$ is the surface energy supply, it is reasonable to postulate the surface entropy supply to be
\be
    \label{eq:entropy_supply_A1}
    \entropysurfsupplydensity = \frac{\widetilde{r}\ga}{\vartheta\ga}\ .
\ee
Since $\heatflux\ga$ is the surface energy flux, classical thermodynamics together with \eqref{eq:18} suggest setting
\be
    \label{eq:entropy_flux_A1}
    \entropysurffluxdensity = \frac{\heatflux\ga}{\vartheta\ga}\ .
\ee
Consequently, \eqref{eq:18} reduces to the equation
\begin{align}
    \nonumber
    \entropysurfproductiondensity &= \heatflux\ga\cdot\nabla\ga\left(\frac{1}{\vartheta\ga}\right) -\frac{\tt\cdot{\vv^-_\tau}}{\vartheta\ga} + \div\vv^-\left((1-\alpha)\frac{s_{\nabla\rho}}{\vartheta^-} + \frac{s_\rho+\alpha s_{\nabla\rho}}{\vartheta\ga} \right)\\ & + \heatflux^+\cdot\nn\ga\left(\frac{1}{\vartheta^+}-\frac{1}{\vartheta\ga}\right)
     - \kappa^-\frac{\pa}{\pa\nn\ga}\left(\frac{1}{\vartheta^-}\right)\left(\frac{1}{\vartheta^-}-\frac{1}{\vartheta\ga}\right)\ ,
\end{align}
which identifies the entropy-producing mechanisms and which has the usual structure of a scalar product
\be
    \leftidx{^\eta}\Pi\ga = {\bf J}\cdot{\bf A}\ ,
\ee
where we choose
\begin{subequations}
\begin{align}
\label{model-A1-J}
    \mathbf{J} &= \left(\heatflux\ga, \tt, \div\vv^-, \heatflux^+\cdot\nn\ga, -\kappa^-\frac{\pa}{\pa\nn\ga}\left(\frac{1}{\vartheta^-}\right)\right)\ ,\\
\label{model-A1-A}
    \mathbf{A} &= \left(\nabla\ga\left(\frac{1}{\vartheta\ga}\right), - \frac{\vv_\tau^-}{\vartheta\ga}, (1-\alpha)\frac{s_{\nabla\rho}}{\vartheta^-} + \frac{s_\rho+\alpha s_{\nabla\rho}}{\vartheta\ga}, \frac{1}{\vartheta^+}{-}\frac{1}{\vartheta\ga}, \frac{1}{\vartheta^-}{-}\frac{1}{\vartheta\ga}\right)\ .
\end{align}
\end{subequations}
We propose linear constitutive relations between the ``fluxes'' $\mathbf{J}$ and ``affinities'' $\mathbf{A}$, following thus the framework of linear irreversible thermodynamics \citep{groot.sr.mazur.p:non-equilibrium}. We shall also consider possible cross-coupling among the vectorial quantities. The constitutive relations in such case take the form
\begin{subequations}
\label{model-A1}
\begin{align}
\label{model-A1-qga}
    \heatflux\ga &= L_{11}\nabla\ga\left(\frac{1}{\vartheta\ga}\right) + L_{12}\left(-\frac{\vv_\tau^-}{\vartheta\ga}\right)\ ,\\
\label{model-A1-t}
    \tt &=   L_{21}\nabla\ga\left(\frac{1}{\vartheta\ga}\right) +  L_{22}\left(-\frac{\vv_\tau^-}{\vartheta\ga}\right)\ ,\\
    \label{model-A1-qdivv-0}
    \div\vv^- &= L_{33}\left((1-\alpha)\frac{s_{\nabla\rho}}{\vartheta^-} + \frac{s_\rho+\alpha s_{\nabla\rho}}{\vartheta\ga}\right)\ ,\\
    \label{model-A1-q1}
    \heatflux^+\cdot\nn\ga &= L_{44} \left(\frac{1}{\vartheta^+}-\frac{1}{\vartheta\ga}\right)\ ,\\
    \label{model-A1-q2}
    -\kappa^-\frac{\pa}{\pa\nn\ga}\left(\frac{1}{\vartheta^-}\right) &= L_{55} \left(\frac{1}{\vartheta^-}-\frac{1}{\vartheta\ga}\right)\ .
\end{align}
\end{subequations}
Since the two affinities $-\frac{\vv_\tau^-}{\vartheta\ga}$, and $\nabla\ga\left(\frac{1}{\vartheta\ga}\right)$, for which cross-effect is assumed, have opposite behavior with respect to time reversal (the first one changes the sign, the other does not), the Onsager-Casimir relations (see, e.g., \citep{groot.sr.mazur.p:non-equilibrium}) imply anti-symmetry of the cross-coupling coefficient, i.e., $L_{12}=-L_{21}$. The coefficients must fulfill $L_{ii}\geq 0$, for all  $i{=}1,{\dots},5$  and $L_{11}L_{22}+(L_{12})^2\geq 0$, in order to ensure non-negativity of the rate of entropy production. We shall impose the stronger yet natural assumption that $L_{ii}{>}0$ for all  $i{=}1,{\dots},5$ in order to avoid degeneracy of the system.

Let us interpret the derived constitutive relations \eqref{model-A1}. The first relation \eqref{model-A1-qga} represents the in-surface heat conduction (Fourier law). The last two relations \eqref{model-A1-q1} and \eqref{model-A1-q2} represent heat transfer across the interface, the so-called Kapitza resistance \citep{kapitza1941}. Condition \eqref{model-A1-t} represents a generalized Navier-slip condition. Inspecting \eqref{def: trho_model1}, we can see that it is a relation among the surface traction $(\stress\nn\ga)^-$, a cross-coupling term involving the surface gradient of temperature, a term proportional to slip velocity $\vv_\tau$, i.e., traditional Navier-slip term, and finally, a term involving the surface Helmholtz free energy and tangent derivative of density. Perhaps the most interesting is the relation \eqref{model-A1-qdivv-0}, which we will later interpret as the static and dynamic contact angle condition - a condition characterizing the angle between the liquid-vapor interface and the boundary. This interpretation will be made explicit in Sections \ref{sec:van-der-Waals-Korteweg} and \ref{sec:numerical-experiments}, where we will consider a particular type of bulk and  surface Helmholtz free energy functions and support our arguments with numerical experiments. This condition in that case will relate the normal derivative of the density $\frac{\pa\rho^-}{\pa\nn\ga}$ with two effects - one due to surface tension and the other due to motion of the fluid in the vicinity of the interface. Since the latter effect vanishes when the body is in equilibrium, we will interpret that part as the {\it dynamic contact angle condition}, while the first effect persists in equilibrium and will be called the {\it static contact angle condition}. Let us note here, that in most of the literature related to Korteweg-type models, the dynamic contact angle condition is completely ignored, and the static one is simplified dramatically to $\frac{\pa\rho^-}{\pa\nn\ga}{=}0$, which corresponds to the contact angle $\frac{\pi}{2}$.

\subsubsection{Model A2}
The second approach is based on the decomposition of the term $\div\vv^-$ appearing in \eqref{eq:15} and \eqref{eq:17} by means of the following identity from differential geometry \citep[][{Appendix A}]{slattery-1990}:
\be
    \label{eq: divv}
    \div\vv^- = \frac{\pa\vv^-_{\mathrm{n}}}{\pa\nn\ga} + 2K_M\vv^-\cdot\nn\ga +\div\ga\vv^-_\tau = \frac{\pa\vv^-_\mathrm{n}}{\pa\nn\ga} +\div\ga\vv^-_\tau\ ,
\ee
where 
\begin{align}
\label{def:tangent-normal2}
\vv^-_\tau \eqdef (\mathbb{I}{-}\nn\ga{\otimes}\nn\ga)\vv^-\ ,\hspace{1cm}\vv^-_\mathrm{n}\eqdef \vv^-{\cdot}\nn\ga\ ,
\end{align}
where we used \eqref{eq:9} in the last equality in \eqref{eq: divv}. We will also employ the identity
\be
    \label{eq:identity}
   (s_\rho+\alpha s_{\nabla\rho})\div\ga\vv_\tau^- = \div\ga((s_\rho+\alpha s_{\nabla\rho})\vv_\tau^-) - \vv_\tau^-\cdot\nabla\ga(s_\rho+\alpha s_{\nabla\rho})\ .
\ee
Incorporating \eqref{eq: divv} and \eqref{eq:identity} into \eqref{eq:15}, we obtain
\begin{align}
    \nonumber
    \vartheta\ga\frac{\pa{\widetilde{\eta}\ga}}{\pa t} &= -\div\ga\left(\heatflux\ga - (s_\rho+\alpha s_{\nabla\rho})\vv^-_\tau\right) + \widetilde{r}\ga 
    - (\tt + \nabla\ga(s_\rho + \alpha s_{\nabla\rho})) \cdot{\vv^-_\tau}
    + (s_\rho+\alpha s_{\nabla\rho})\frac{\pa\vv_\mathrm{n}^-}{\pa\nn\ga}
    \\\label{eq:16b} &- \heatflux^+\cdot\nn\ga
    + \kappa^-\frac{\pa}{\pa\nn\ga}\left(\frac{1}{\vartheta^-}\right)\ ,    
\end{align}
which leads to the following balance equation for the entropy:
\begin{align}
    \nonumber
    \frac{\pa{\widetilde{\eta}\ga}}{\pa t} &=  -\div\ga\left(\frac{\heatflux\ga - (s_\rho+\alpha s_{\nabla\rho})\vv^-_\tau}{\vartheta\ga}\right) + \frac{\widetilde{r}\ga}{\vartheta\ga} 
    + (\heatflux\ga - (s_\rho+\alpha s_{\nabla\rho})\vv^-_\tau)\cdot\nabla\ga\left(\frac{1}{\vartheta\ga}\right)\\\label{eq:17b}
    &+ \frac{1}{\vartheta\ga}\left\{
      -(\tt + \nabla\ga(s_\rho+\alpha s_{\nabla\rho})) \cdot{\vv^-_\tau}
    + (s_\rho+\alpha s_{\nabla\rho})\frac{\pa\vv^-_{\mathrm{n}}}{\pa\nn\ga}
     - \heatflux^+\cdot\nn\ga
    + \kappa^-\frac{\pa}{\pa\nn\ga}\left(\frac{1}{\vartheta^-}\right)
    \right\}\ .
\end{align}
Similarly, applying \eqref{eq: divv} together with the identity
\be
   \frac{s_{\nabla\rho}}{\vartheta^-}\div\ga\vv^-_\tau
   = \div\ga\left(\frac{s_{\nabla\rho}}{\vartheta^-}\vv_\tau^-\right)
   - \vv_\tau^-\cdot\nabla\ga\left(\frac{s_{\nabla\rho}}{\vartheta^-}\right)\ 
\ee
to \eqref{eq:17}, we arrive, after rearranging the terms, at the following form of the entropy balance:
\bea
    \nonumber
     \frac{\pa\widetilde{\eta}\ga}{\pa t} &=&  -\div\ga\left(\entropysurffluxdensity + (1-\alpha)\frac{s_{\nabla\rho}}{\vartheta^-}\vv^-_\tau \right)
    -(1-\alpha)\frac{s_{\nabla\rho}}{\vartheta^-}\frac{\pa\vv^-_\mathrm{n}}{\pa\nn\ga}
    - \frac{\heatflux^+}{\vartheta^+}\cdot\nn\ga
    + \frac{\kappa^-}{\vartheta^-}\frac{\pa}{\pa\nn\ga}\left(\frac{1}{\vartheta^-}\right)\\
    \label{eq:19b}
    &+& (1-\alpha)\vv_\tau^-\cdot\nabla\ga\left(\frac{s_{\nabla\rho}}{\vartheta^-}\right)
    + \entropysurfsupplydensity + \entropysurfproductiondensity\ .
\eea
Subtracting (\ref{eq:17b}) from (\ref{eq:19b}), we obtain
\begin{align}
\nonumber
0 &= -\div\ga\left(\entropysurffluxdensity + (1-\alpha)\frac{s_{\nabla\rho}}{\vartheta^-}\vv^-_\tau - \frac{\heatflux\ga - (s_\rho+\alpha s_{\nabla\rho})\vv^-_\tau}{\vartheta\ga} \right) + \left(\entropysurfsupplydensity- \frac{\widetilde{r}\ga}{\vartheta\ga}\right)
- \heatflux^+\cdot\nn\ga\left(\frac{1}{\vartheta^+}-\frac{1}{\vartheta\ga}\right)  \\\nonumber &+ \kappa^-\frac{\pa}{\pa\nn\ga}\left(\frac{1}{\vartheta^-}\right)\left(\frac{1}{\vartheta^-}-\frac{1}{\vartheta\ga}\right)
-(\heatflux\ga - (s_\rho+\alpha s_{\nabla\rho})\vv^-_\tau)\cdot\nabla\ga\left(\frac{1}{\vartheta\ga}\right)
-\frac{\pa\vv_\mathrm{n}^-}{\pa\nn\ga}\left((1-\alpha)\frac{s_{\nabla\rho}}{\vartheta^-} + \frac{s_\rho+\alpha s_{\nabla\rho}}{\vartheta\ga}\right)\\\label{eq:modelA2-aux1} &+
\vv_\tau^-\cdot\left((1-\alpha)\nabla\ga\left(\frac{s_{\nabla\rho}}{\vartheta}\right) + \frac{\tt + \nabla\ga (s_\rho+\alpha s_{\nabla\rho})}{\vartheta\ga}\right)  +	\entropysurfproductiondensity
\ .
\end{align}
As in Model A1 we postulate the surface entropy supply to be
\be
    \entropysurfsupplydensity = \frac{\widetilde{r}\ga}{\vartheta\ga}\ ,
\ee
and \eqref{eq:modelA2-aux1} suggests setting
\be
    \entropysurffluxdensity = \frac{\heatflux\ga - (s_\rho+\alpha s_{\nabla\rho})\vv^-_\tau}{\vartheta\ga} - (1-\alpha)\left(\frac{s_{\nabla\rho}}{\vartheta^-}\right)\vv^-_\tau\ .
\ee
Consequently, \eqref{eq:modelA2-aux1} gives
\begin{align}
\nonumber
\entropysurfproductiondensity &=
(\heatflux\ga - (s_\rho+\alpha s_{\nabla\rho})\vv^-_\tau)\cdot\nabla\ga\left(\frac{1}{\vartheta\ga}\right)
-\left(\tt + \nabla\ga (s_\rho+\alpha s_{\nabla\rho}) + (1-\alpha)\vartheta\ga\nabla\ga\left(\frac{s_{\nabla\rho}}{\vartheta^-}\right) \right)\cdot\frac{\vv_\tau^-}{\vartheta\ga}\\
&+ \frac{\pa\vv_\mathrm{n}^-}{\pa\nn\ga}\left((1-\alpha)\frac{s_{\nabla\rho}}{\vartheta^-} + \frac{s_\rho+\alpha s_{\nabla\rho}}{\vartheta\ga}\right)
+ \heatflux^+\cdot\nn\ga\left(\frac{1}{\vartheta^+}-\frac{1}{\vartheta\ga}\right)
- \kappa^-\frac{\pa}{\pa\nn\ga}\left(\frac{1}{\vartheta^-}\right)\left(\frac{1}{\vartheta^-} - \frac{1}{\vartheta\ga}\right)\ ,
\end{align}
or, written again as a scalar product of two vectors,
\be
    \entropysurfproductiondensity = {\bf J}\cdot{\bf A}\ ,
\ee
where  
\begin{subequations}
\begin{align}
    {\bf J} &= \left(\heatflux\ga{-}(s_\rho{+}\alpha s_{\nabla\rho})\vv^-_\tau, \tt{+}\nabla\ga (s_\rho{+}\alpha s_{\nabla\rho}){+} (1{-}
\alpha)\vartheta\ga\nabla\ga\left(\frac{s_{\nabla\rho}}{\vartheta^-}\right), \frac{\pa\vv_\mathrm{n}^-}{\pa\nn\ga}, \heatflux^+\cdot\nn\ga, -\kappa^-\frac{\pa}{\pa\nn\ga}\left(\frac{1}{\vartheta}\right)^-\right)\ ,\\
    {\bf A} &= \left(\nabla\ga\left(\frac{1}{\vartheta\ga}\right), - \frac{\vv_\tau^-}{\vartheta\ga}, (1-\alpha)\frac{s_{\nabla\rho}}{\vartheta^-} + \frac{s_\rho+\alpha s_{\nabla\rho}}{\vartheta\ga}, \frac{1}{\vartheta^+}-\frac{1}{\vartheta\ga}, \frac{1}{\vartheta^-}-\frac{1}{\vartheta\ga}\right)\ .
\end{align}
\end{subequations}
As in Subsection \ref{subsec-model-a1}, restricting ourselves to the linear constitutive relations between the ``fluxes'' ${\bf J}$ and ``affinities'' ${\bf A}$ (with cross-coupling only among the vectorial quantities), we end up with the following set of constitutive relations
\begin{subequations}
\label{model-A2}
\begin{align}
\label{model-A2-qga}
    \heatflux\ga - (s_\rho+\alpha s_{\nabla\rho})\vv^-_\tau &= L_{11}\nabla\ga\left(\frac{1}{\vartheta\ga}\right) + L_{12}\left(-\frac{\vv_\tau^-}{\vartheta\ga}\right)\ ,
    \\ \label{model-A2-t}
    \tt + \nabla\ga (s_\rho+\alpha s_{\nabla\rho}) + (1-\alpha)\vartheta\ga\nabla\ga\left(\frac{s_{\nabla\rho}}{\vartheta^-}\right) &=   L_{21}\nabla\ga\left(\frac{1}{\vartheta\ga}\right) +  L_{22}\left(-\frac{\vv_\tau^-}{\vartheta\ga}\right)\ ,\\\label{model-A2-pavn}
    \frac{\pa\vv_\mathrm{n}^-}{\pa\nn\ga} &= L_{33}\left((1-\alpha)\frac{s_{\nabla\rho}}{\vartheta^-} + \frac{s_\rho+\alpha s_{\nabla\rho}}{\vartheta\ga}\right)\ ,\\
    \label{model-A2-q1}
    \heatflux^+\cdot\nn\ga &= L_{44} \left(\frac{1}{\vartheta^+}-\frac{1}{\vartheta\ga}\right)\ ,\\\label{model-A2-q2}
    -\kappa^-\frac{\pa}{\pa\nn\ga}\left(\frac{1}{\vartheta}\right)^- &= L_{55} \left(\frac{1}{\vartheta^-}-\frac{1}{\vartheta\ga}\right)\ .
\end{align}
\end{subequations}
Since the two ``affinities'' $-\frac{\vv_\tau^-}{\vartheta\ga}$, and $\nabla\ga\left(\frac{1}{\vartheta\ga}\right)$, for which cross-effect is assumed, have opposite behavior with respect to time reversal, the Onsager-Casimir relations suggest the requirement that the cross-coupling coefficients are anti-symmetric, i.e., $L_{12}{=}-L_{21}$. The coefficients are assumed to satisfy
$L_{ii}{>}0$, $i{=}1,{\dots},5$, and 
 $L_{11}L_{22}+(L_{12})^2\geq 0$, in order to ensure both non-negativity of the rate of entropy production and non-degeneracy of the system.

The interpretation of the constitutive relations for Model A2 is analogous to Model A1, namely the first two relations \eqref{model-A2-qga} and \eqref{model-A2-t} represent the (generalized) in-surface heat conduction (Fourier law) and generalized Navier-slip condition, respectively, together with a possible cross-coupling of the two mechanisms. Relations \eqref{model-A2-q1} and \eqref{model-A2-q2} represent heat transfer across the interface (the so-called Kapitza resistance) and relation \eqref{model-A2-pavn} is again the static and dynamic contact angle condition, as will become apparent in Sections \ref{sec:van-der-Waals-Korteweg} and \ref{sec:numerical-experiments}.

\subsection{{\bf Model B}}
The derivation of boundary conditions for Model B proceeds in an analogous way as for Model A. The only difference between these models is the absence of the term $s_\rho$, which is now identically zero. Consequently, we obtain the following two sets of boundary conditions, which again differ in the manner how the terms involving $\div\vv^-$ are treated.
\subsubsection{Model B1}
\begin{subequations}
\label{model-B1}
\begin{align}
\label{model-B1-qga}
    \heatflux\ga &= L_{11}\nabla\ga\left(\frac{1}{\vartheta\ga}\right) + L_{12}\left(-\frac{\vv_\tau^-}{\vartheta\ga}\right)\ ,\\
     \label{model-B1-t}
    (\stress\nn\ga)^-_\tau &=   L_{21}\nabla\ga\left(\frac{1}{\vartheta\ga}\right) +  L_{22}\left(-\frac{\vv_\tau^-}{\vartheta\ga}\right)\ ,\\\label{model-B1-divv}
    \div\vv^- &= L_{33}\left((1-\alpha)\frac{s_{\nabla\rho}}{\vartheta^-} + \frac{\alpha s_{\nabla\rho}}{\vartheta\ga}\right)\ ,\\ \label{model-B1-q1}
    \heatflux^+\cdot\nn\ga &= L_{44} \left(\frac{1}{\vartheta^+}-\frac{1}{\vartheta\ga}\right)\ ,\\\label{model-B1-q2}
    -\kappa^-\frac{\pa}{\pa\nn\ga}\left(\frac{1}{\vartheta}\right)^- &= L_{55} \left(\frac{1}{\vartheta^-}-\frac{1}{\vartheta\ga}\right)\ ,
\end{align}
\end{subequations}
 where $L_{12}{=}-L_{21}$ and 
where $L_{ii}{>}0$, $i{=}1,{\dots},5$ and 
 $L_{11}L_{22}{+}(L_{12})^2\geq 0$, in order to ensure both non-negativity of the rate of entropy production and non-degeneracy of the system.
 
\subsubsection{Model B2}
\begin{subequations}
\label{model-B2}
\begin{align}
\label{model-B2-qga}
    \heatflux\ga - \alpha s_{\nabla\rho}\vv^-_\tau &= L_{11}\nabla\ga\left(\frac{1}{\vartheta\ga}\right) + L_{12}\left(-\frac{\vv_\tau^-}{\vartheta\ga}\right)\ ,\\
         \label{model-B2-t}
     (\stress\nn\ga)^-_\tau + \alpha\nabla\ga s_{\nabla\rho} + (1-\alpha)\vartheta\ga\nabla\ga\left(\frac{s_{\nabla\rho}}{\vartheta^-}\right) &=   L_{21}\nabla\ga\left(\frac{1}{\vartheta\ga}\right) +  L_{22}\left(-\frac{\vv_\tau^-}{\vartheta\ga}\right)\ ,\\\label{model-B2-pavn}
    \frac{\pa\vv_\mathrm{n}^-}{\pa\nn\ga} &= L_{33}\left((1-\alpha)\frac{s_{\nabla\rho}}{\vartheta^-} + \frac{\alpha s_{\nabla\rho}}{\vartheta\ga}\right)\ ,\\\label{model-B2-q1}
    \heatflux^+\cdot\nn\ga &= L_{44} \left(\frac{1}{\vartheta^+}-\frac{1}{\vartheta\ga}\right)\ ,\\\label{model-B2-q2}
    -\kappa^-\frac{\pa}{\pa\nn\ga}\left(\frac{1}{\vartheta}\right)^- &= L_{55} \left(\frac{1}{\vartheta^-}-\frac{1}{\vartheta\ga}\right)\ ,
\end{align}
\end{subequations}
 where $L_{12}{=}-L_{21}$ and 
where $L_{ii}{>}0$, $i{=}1,{\dots},5$ and 
 $L_{11}L_{22}{+}(L_{12})^2\geq 0$, in order to ensure both non-negativity of the rate of entropy production and non-degeneracy of the system.

The only difference between Models A and B is due to the absence of terms  $s_\rho$, its main implication being that conditions \eqref{model-B1-divv} and \eqref{model-B2-pavn} represent solely dynamic angle conditions, with static (equilibrium) contact angle (equal to $\frac{\pi}{2}$ for the Korteweg - van der Waals fluid studied in Section \ref{sec:van-der-Waals-Korteweg}).

\subsection{{\bf Model C}}
Assuming that the surface Helmholtz free energy, and consequently also both the surface internal energy and surface entropy are identically equal to zero, it also makes sense to assume the same for the corresponding energy and entropy surface fluxes. Therefore we set
\begin{subequations}
\label{eq:modelC-assumptions}
\be
    \widetilde{\helm}\ga \equiv 0\ ,\widetilde{\ienergy}\ga \equiv 0,\ \widetilde{\eta}\ga \equiv 0\ ,
    {\heatflux\ga} \equiv {\bf 0},\ {\entropysurffluxdensity} \equiv {\bf 0}\ .
\ee
Furthermore, we assume that neither surface energy supply nor entropy supply are present, i.e., 
\bea
    \widetilde{r}\ga \equiv 0,\ \entropysurfsupplydensity \equiv 0\ .
\eea
\end{subequations}
Employing also the assumptions on the velocity field (\ref{eq:6})--(\ref{eq:9}), the absence of the surface body forces (\ref{eq:bbeq0}), and the character of the surface stress tensor (\ref{eq:surfacestresstensor}), the surface energy balance (\ref{eq:surf-energy_P}) reduces to the following form of a jump condition:
\bea
    \label{eq:energyC}
    0 &=&   (\stress\nn\ga)^-_\tau\cdot\vv_\tau^- + \lb\heatflux\rb\cdot\nn\ga\ .
\eea
Following the same arguments, the surface entropy balance (\ref{eq:surf-entropy}) reduces, with the use of \eqref{eq-bulk-heat-flux}, \eqref{eq-bulk-entropyflux}, \eqref{eq:modelC-assumptions}, and \eqref{eq:6}--\eqref{eq:9}, to

\bea
    \label{eq:entropyC}
    \entropysurfproductiondensity &=& \left\lb\frac{\heatflux}{\vartheta}\right\rb\cdot\nn\ga + \frac{s_{\nabla\rho}}{\vartheta^-}\div\vv^-\ .
\eea
Employing the identity for the jump of a product of two fields $a,b$
\be \left\lb a b\right\rb = \lb a\rb\langle b\rangle + \langle a\rangle\lb b\rb\ ,\ee
where $\langle a \rangle$ denotes the average of $a^+$ and $a^-$, i.e., $\langle a\rangle = \frac{1}{2}\left(a^+ + a^-\right)$, applying this identity to $\left\lb\frac{\heatflux}{\vartheta}\right\rb$ in (\ref{eq:entropyC}), and inserting its result into \eqref{eq:energyC} instead of $\lb\heatflux\rb$, we obtain the equation for the rate of surface entropy production
\be
    \entropysurfproductiondensity = -\left\langle\frac{1}{\vartheta}\right\rangle(\stress\nn\ga)^-_\tau\cdot\vv_\tau^- + \frac{s_{\nabla\rho}}{\vartheta^-}\div\vv^- + \left\lb\frac{1}{\vartheta}\right\rb\langle\heatflux\rangle\cdot\nn\ga\ .
\ee
This expression again takes the form of a scalar product
\be
    \leftidx{^\eta}\Pi\ga = {\bf J}\cdot{\bf A}\ ,
\ee
where we choose
\begin{subequations}
\bea
    {\bf J} &=& \left((\stress\nn\ga)^-_\tau, \div\vv^-, \langle\heatflux\rangle\cdot\nn\ga\right)\ ,\\
    {\bf A} &=& \left(-\left\langle\frac{1}{\vartheta}\right\rangle\vv_\tau^-, \frac{s_{\nabla\rho}}{\vartheta^-}, \left\lb\frac{1}{\vartheta}\right\rb\right)\ .
\eea
\end{subequations}
The linear relations between the ``fluxes'' ${\bf J}$ and ``affinities'' {\bf A} yield the following relations:
\begin{subequations}
\label{model-C}
\begin{align}
\label{model-C-t}
    (\stress\nn\ga)^-_\tau &= -L_{11}\left\langle\frac{1}{\vartheta}\right\rangle\vv_\tau^-\ ,\\
    \label{model-C-divv}
    \div\vv^- &= L_{22}\frac{s_{\nabla\rho}}{\vartheta^-}\ ,\\
    \label{model-C-q}
    \langle\heatflux\rangle\cdot\nn\ga &= L_{33}\left\lb\frac{1}{\vartheta}\right\rb\ .
\end{align}
\end{subequations}
Assuming that $L_{11}{>}0$, $L_{22}{>}0$, $L_{33}{>}0$, the rate of entropy production is non-negative. The boundary condition \eqref{model-C-t} represents the Navier-slip, 
\eqref{model-C-divv} describes the dynamic contact angle condition, and 
\eqref{model-C-q} stands for the heat transfer across the interface (Kapitza resistance), respectively, see Sections \ref{sec:van-der-Waals-Korteweg} and \ref{sec:numerical-experiments} for further details.
\\
\\
\\
\noindent{\bf Isothermal process}\\
\\
In the following, we will study a variant of the above models in which the temperature is continuous across the interface. This can in particular be achieved if we consider an isothermal process and assume that \be\vartheta^+{=}\vartheta^-{=}\vartheta\ga{=} \mathrm{const.}\ee If we also ignore the cross-coupling effects for simplicity and absorb the (constant) temperature into the coefficients, we obtain the following sets of reduced boundary conditions (recall that $\tt$, $s_\rho$ and $s_{\nabla\rho}$ are defined in \eqref{def:trho-srho-snablarho}):
\begin{itemize}
	\item { Model A1:}
\begin{subequations}
\label{model-A1i}
\begin{align}
  \label{model-A1i-t}
    \tt &= -L_{22}\vv_\tau^-\ ,\\
    \label{model-A1i-qdivv}
    \div\vv^- &= L_{33}\left(s_{\nabla\rho} + s_\rho\right)\ ;
\end{align}
\end{subequations}

	\item { Model A2:}
\begin{subequations}
\label{model-A2i}
\begin{align}
  \label{model-A2i-t}
    \tt + \nabla\ga\left(s_\rho + s_{\nabla\rho}\right) &=   -L_{22}\vv_\tau^-\ ,\\
    \label{model-A2-qdivv}
   \frac{\pa\vv_\mathrm{n}^-}{\pa\nn\ga} &= L_{33}\left(s_{\nabla\rho} + s_\rho\right)\ ;
\end{align}
\end{subequations}
\item {Model B1 $\&$ Model C:}
\begin{subequations}
\label{model-B1i}
\begin{align}
  \label{model-B1i-t}
    (\stress\nn\ga)^-_\tau &=   -L_{22}\vv_\tau^-\ ,\\
    \label{model-B1i-qdivv}
    \div\vv^- &= L_{33}s_{\nabla\rho}\ .
\end{align}
\end{subequations}
\item {Model B2}
\begin{subequations}
\label{model-B2i}
\begin{align}
  \label{model-B2i-t}
    (\stress\nn\ga)^-_\tau + \nabla\ga s_{\nabla\rho}&=   -L_{22}\vv_\tau^-\ ,\\
    \label{model-B2i-pavn}
   \frac{\pa\vv_\mathrm{n}^-}{\pa\nn\ga} &= L_{33}s_{\nabla\rho}\ ,
\end{align}
\end{subequations}
where $L_{22}{>}0$ $L_{33}{>}0$ for all sets.
Each set consists of a (generalized) Navier-slip condition and a contact angle condition, which is either solely dynamic or combines static and dynamic terms, as clarified in the following sections.
\end{itemize}

\section{Particular example of Korteweg fluid model and boundary conditions}
\label{sec:van-der-Waals-Korteweg}
Let us now consider a particular Helmholtz free energy $\widehat\helm$ of the form
\be
\label{helm-Kvdw}
	\widehat{\helm}(\vartheta,\rho,\nabla\rho) = \widehat{\helm_{vdW}}(\vartheta,\rho) + \frac{\sigma}{2\rho}|\nabla\rho|^2\ ,
\ee
where the term $\widehat{\helm}_{vdW}$ corresponds to van der Waals fluid  
\citep{van-der-Waals-1983,landau-statistical,diehl2007} and takes the following form:
\be
\label{helm-vdW}
	\widehat{\helm_{vdW}}(\vartheta,\rho) = -a'\rho + \ell\vartheta\ln\left(\frac{\rho}{b'-\rho}\right) - c\vartheta\ln\left(\frac{\vartheta}{\vartheta_0}\right) - d\vartheta + e'\ ,
\ee
where $a'$, $b'$, $c$, $d$, $e$, $\ell$ and $\sigma$ are constant parameters and $\vartheta_0$ is some reference temperature. The correspondence of \eqref{helm-vdW} to the van der Waals model is revealed by identifying the equation of state for thermodynamic pressure associated with $\widehat{\helm_{vdW}}$, which we show next.

Using the standard thermodynamic definition of the thermodynamic pressure \eqref{def:td-press}, we obtain the expression
\begin{align}
	\label{eq:vdw-eos1}
	p_{vdW} = \widehat{p_{vdW}}(\vartheta,\rho) \eqdef \rho^2\frac{\pa\widehat{\helm_{vdW}}}{\pa\rho} = -a'\rho^2 + \ell b'\vartheta\frac{\rho}{b'-\rho}\ .
\end{align}
Defining $a\eqdef a' M_m^2$, $b\eqdef \frac{M_m}{b'}$ and $R \eqdef \ell M_m$, where $M_m$ is the molar mass of the molecules of the considered gas-liquid system, and considering a homogeneous system with $n$ moles in volume $V$, which means $\rho=\frac{n M_m}{V}$, we can recast \eqref{eq:vdw-eos1} into the standard form 
\be
\label{pvdw}
	\left(p_{vdW} + \frac{n^2 a}{V^2}\right)\left(V-nb\right) = nR\vartheta\ ,
\ee
which is the traditional van der Waals equation of state \citep[e.g.][]{callen}, provided we suitably interpret the parameters $a$, $b$, and $R$. 

The chemical potential for a single-component fluid is simply the Gibbs free energy $g$ as follows from the Euler relation \citep[e.g.][]{callen}. We thus obtain
\begin{align}
\label{muvdw}
	\mu_{vdW} = g_{vdW} \eqdef \helm_{vdW} - \frac{p_{vdW}}{\rho} = \frac{\pa(\rho\widehat{\psi_{vdW}})}{\pa\rho}\ .
\end{align}

\begin{figure}[h!]
\begin{center}
\includegraphics[scale=0.4]{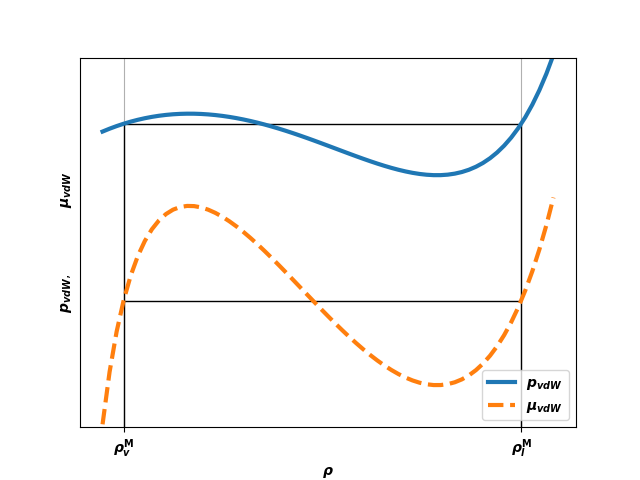} 
\caption{A sketch of thermodynamic pressure and chemical potential for a dimensionless van der Waals fluid with the corresponding Maxwell states $\rho_v^M$ and $\rho_l^M$ for a subcritical temperature $\vartheta=0.85\vartheta_c$.}
\label{fig-maxwell}
\end{center}
\end{figure}
The pressure $p_{vdW}$ and the chemical potential $\mu_{vdW}$ are sketched in Fig.~\ref{fig-maxwell}. For temperatures above a critical temperature $\vartheta_{c}$, both $p_{vdW}$ and $\mu_{vdW}$ are increasing functions of density, but for a temperature below the critical temperature $\vartheta_c$, both functions have two increasing branches separated by a region where these functions are decreasing. The critical temperature as well as the critical pressure are found by identifying the inflection point of the $\widehat{p_{vdW}}$ equation of state, i.e., by finding $(\vartheta_c,\rho_c)$ such that 
\be
	\frac{\pa \widehat{p_{vdW}}}{\pa\rho}(\vartheta_c,\rho_c) = 0,\hspace{1cm}
	\frac{\pa^2 \widehat{p_{vdW}}}{\pa\rho^2}(\vartheta_c,\rho_c) = 0\ ,
\ee
which for \eqref{helm-vdW} yields
\begin{align}
\label{def:vdW-parameters}
	\vartheta_c = \frac{8a'b'}{27 \ell}\ ,\hspace{1cm}\rho_c = \frac{b'}{3}\ ,\hspace{1cm} p_c \eqdef p_{vdW} (\vartheta_c,\rho_c) = \frac{a'{b'}^2}{27}\ .
\end{align}
In the subcritical region, i.e., for each $\vartheta{<}\vartheta_c$, there are two  states (called Maxwell states) $\rho^M_v$ (vapor) and $\rho^M_l$ (liquid) defined by two phase-coexistence equilibrium 
relations expressing the equality of pressures and chemical potentials (see Fig.~\ref{fig-maxwell}),
\begin{subequations}
\label{def:maxwell-states}
\begin{align}
	\widehat{p_{vdW}}(\rho^M_v(\vartheta),\vartheta) &= \widehat{p_{vdW}}(\rho^M_l(\vartheta),\vartheta)\ ,\\
	\widehat{\mu_{vdW}}(\rho^M_v(\vartheta),\vartheta) &= \widehat{\mu_{vdW}}(\rho^M_l(\vartheta),\vartheta)\ .
\end{align}
\end{subequations}  

Let us now consider an isothermal setting below the critical temperature (meaning the temperature $\vartheta$ is uniform and equals the constant $\vartheta_0$, satisfying $\vartheta_0{<}\vartheta_c$) and let us set up the system of governing equations and boundary conditions for such a Korteweg - van der Waals fluid in the bulk. Using the definition of the Helmholtz free energy \eqref{helm-Kvdw}, the Cauchy stress $\stress$ reads according to \eqref{eq-bulk-stress} as
\be
	\label{Cauchy-stress-KvdW}
	\stress = -p_{vdW}\Id + \sigma\left(\rho\Delta\rho + \frac{1}{2}|\nabla\rho|^2\right)\Id - \sigma \nabla\rho\otimes\nabla\rho + \lambda \div\vv\Id + 2\mu\DD\ .
\ee
Using the identity 
\be
	\div\left(\left(\rho\Delta\rho + \frac{1}{2}|\nabla\rho|^2\right)\Id - \nabla\rho\otimes\nabla\rho\right) = \rho\nabla\Delta\rho\ ,
\ee
the governing equations (balances of mass and momentum) in the bulk for the Korteweg - van der Waals fluids read
\begin{subequations}
\label{eq:mass-and-momentum-kvdw}
\begin{align}	
\frac{\pa\rho}{\pa t} + \div(\rho\vv) &= 0\ ,\\
\frac{\pa(\rho\vv)}{\pa t} +\div(\rho\vv\otimes\vv) &= \nabla \left(- p_{vdW} + \lambda\div\vv\right) + \div\left(2\mu\DD\right) + \sigma\rho\nabla\Delta\rho +\rho\bb\ . 
\end{align}
\end{subequations}

Boundary conditions corresponding to Models A, B, and C, (see \eqref{model-A1i}--\eqref{model-B2i}) can now be expressed in more explicit forms since for the Korteweg - van der Waals model we can explicitly evaluate the term
\be
	s_{\nabla\rho} = \left(\rho^2\frac{\pa\widehat\helm}{\pa\nabla\rho}\right)^-\cdot\nn\ga = \sigma\rho^-\nabla\rho^-\cdot\nn\ga\ = \sigma\rho^-\frac{\pa\rho^-}{\pa\nn\ga}\ .
\ee
Employing the definition of $\tt$ \eqref{def: trho_model1}, we obtain the following set of boundary conditions (depending on the form $\widehat{\helm}\ga$ and the way how $\div{\vv}^-
$ is treated on $\Gamma$):
\begin{itemize}
	\item {Model A1:}
\begin{subequations}
\label{model-A1i-KvdW}
\begin{align}
  \label{model-A1i-t-KvdW}
    (\stress\nn\ga)_\tau^- &= \left(\frac{\pa\widehat{\helm}\ga}{\pa\rho^-}\right)\nabla\ga\rho^- -\alpha\vv_\tau^-\ ,\\
    \label{model-A1-qdivv-KvdW}
    \beta\div\vv^- &= \left(\frac{\pa\widehat{\helm}\ga}{\pa\rho^-} + \sigma\frac{\pa\rho^-}{\pa\nn\ga}\right)\ ;
\end{align}
\end{subequations}
	\item {Model A2:}
\begin{subequations}
\label{model-A2i-KvdW}
\begin{align}
  \label{model-A2i-t-KvdW}
      (\stress\nn\ga)_\tau^- &= \left(\frac{\pa\widehat{\helm}\ga}{\pa\rho^-}\right)\nabla\ga\rho^- - \nabla\ga\left(\rho^-\frac{\pa\widehat{\helm}\ga}{\pa\rho^-} + \sigma\rho^-\frac{\pa\rho^-}{\pa\nn\ga} \right)-\alpha\vv_\tau^-\ ,\\
    \label{model-A2-qdivv-KvdW}
   \beta\frac{\pa\vv_\mathrm{n}^-}{\pa\nn\ga} &= \left(\frac{\pa\widehat{\helm}\ga}{\pa\rho^-} + \sigma\frac{\pa\rho^-}{\pa\nn\ga}\right)\ ;
\end{align}
\end{subequations}
\item {Model B1 $\&$ Model C:}
\begin{subequations}
\label{model-B1i-KvdW}
\begin{align}
  \label{model-B1i-t-KvdW}
    (\stress\nn\ga)^-_\tau &=   -\alpha\vv_\tau^-\ ,\\
    \label{model-B1-qdivv-KvdW}
    \beta\div\vv^- &= \sigma\frac{\pa\rho^-}{\pa\nn\ga}\ ;
\end{align}
\end{subequations}
\item {Model B2:}
\begin{subequations}
\label{model-B2i-kvdW}
\begin{align}
  \label{model-B2i-t-KvdW}
    (\stress\nn\ga)^-_\tau &=  - \nabla\ga\left(\sigma\rho^-\frac{\pa\rho^-}{\pa\nn\ga} \right)-\alpha\vv_\tau^-\  ,\\
    \label{model-B2-pavn-KvdW}
   \beta\frac{\pa\vv_\mathrm{n}^-}{\pa\nn\ga} &= \sigma\frac{\pa\rho^-}{\pa\nn\ga}\ ,
\end{align}
\end{subequations}
where $\alpha$, $\beta$ are some non-negative parameters ($\alpha{=}L_{22}$, $\beta{=}\frac{L_{33}}{\rho^-}$), possibly depending on $\vartheta^-$, $\vartheta\ga$, $\rho^-$. 
\end{itemize}

It will be convenient to incorporate the constitutive relation for the Cauchy stress \eqref{Cauchy-stress-KvdW} in the above conditions. Since, by \eqref{Cauchy-stress-KvdW},
\be
	(\stress\nn\ga)_\tau^- = -\sigma\nabla\ga\rho^- \frac{\pa\rho^-}{\pa\nn\ga} + (2\mu\DD\nn\ga)^-_\tau\ ,
\ee 
we obtain by simple manipulation (in particular substituting from the second equation into the first) the following conditions:
\begin{itemize}
	\item {Model A1:}
\begin{subequations}
\label{model-A1i-KvdW3}
\begin{align}
  \label{model-A1i-t-KvdW3}
    (2\mu\DD\nn\ga)_\tau^- &= \beta\div\vv^-\nabla\ga\rho^- -\alpha\vv_\tau^-\ ,\\
    \label{model-A1-qdivv-KvdW3}
    \beta\div\vv^- &= \left(\frac{\pa\widehat{\helm}\ga}{\pa\rho^-} + \sigma\frac{\pa\rho^-}{\pa\nn\ga}\right)\ ;
\end{align}
\end{subequations}
	\item {Model A2:}
\begin{subequations}
\label{model-A2i-KvdW3}
\begin{align}
  \label{model-A2i-t-KvdW3}
      (2\mu\DD\nn\ga)_\tau^- &= -\rho^-\nabla\ga\left(\beta\frac{\pa\vv_\mathrm{n}^-}{\pa\nn\ga}\right)  -\alpha\vv_\tau^-\ ,\\
    \label{model-A2-qdivv-KvdW3}
  \beta\frac{\pa\vv_\mathrm{n}^-}{\pa\nn\ga} &= \left(\frac{\pa\widehat{\helm}\ga}{\pa\rho^-} + \sigma\frac{\pa\rho^-}{\pa\nn\ga}\right)\ ;
\end{align}
\end{subequations}
\item {Model B1 $\&$ Model C:}
\begin{subequations}
\label{model-B1i3}
\begin{align}
  \label{model-B1i-t3}
    (2\mu\DD\nn\ga)^-_\tau &= \beta\div\vv^-\nabla\ga\rho^- - \alpha\vv_\tau^-\ ,\\
    \label{model-B1-qdivv3}
    \beta\div\vv^- &= \sigma\frac{\pa\rho^-}{\pa\nn\ga}\ ;
\end{align}
\end{subequations}
\item {Model B2:}
\begin{subequations}
\label{model-B2i3}
\begin{align}
  \label{model-B2i-t3}
    (2\mu\DD\nn\ga)^-_\tau  &=  - \rho^-\nabla\ga\left(\beta\frac{\pa\vv_\mathrm{n}^-}{\pa\nn\ga}\right)-\alpha\vv_\tau^-\  ,\\
    \label{model-B2-pavn3}
   \beta\frac{\pa\vv_\mathrm{n}^-}{\pa\nn\ga} &= \sigma\frac{\pa\rho^-}{\pa\nn\ga}\ ,
\end{align}
\end{subequations}
where $\alpha$, $\beta$ are non-negative parameters, possibly depending on $\vartheta^-$, $\vartheta\ga$, $\rho^-$.  
\end{itemize}

In the above sets of boundary conditions, the right-hand side of the second equation represents the static contact angle condition (static in the sense that it does not depend explicitly on $\vv$). Let us note that in the class of diffuse interface methods, to which the Korteweg model presented here belongs (as well as other models including Cahn-Hilliard and Allen-Cahn models or numerous variants of the level set method), a standard way to impose a given static contact angle $\varphi$ is expressed through the formula (imposed on the boundary)
\be
\label{eq:standard-contact}
\frac{\nabla \rho^-}{|\nabla\rho^-|}\cdot\nn\ga = \cos{\varphi}\ \hspace{1cm}\mathrm{i.e.,}\hspace{1cm}  \frac{\pa\rho^-}{\pa\nn\ga} = |\nabla\rho^-|  \cos{\varphi}\ ,
\ee
see, e.g., \cite{brackbill1992}. Such a formula, despite its apparent simplicity, is problematic from several points of view. First, it cannot be incorporated into the framework developed above as this would require that the surface Helmholtz free energy $\widehat{\helm}\ga$ depends also on $|\nabla\rho^-|$ instead of just $\rho^-$ (and temperature). Second, the term $\frac{\pa\rho^-}{\pa\nn\ga}$ appears naturally in the weak formulation of associated initial boundary value problems (see Section~\ref{sec:weak-formulation}, eq.~\eqref{def-B-pa-Omega}) and from  \eqref{eq:standard-contact}$_2$, we can see that it is a non-linear function of the density gradient. This, however, represents quite a severe constraint on the regularity of the density field in terms of mathematical well-posedness. Last, but not least, in order to apply the formula \eqref{eq:standard-contact} in numerical calculations, a sufficiently accurate numerical approximation of the term $|\nabla\rho^-|$ on the boundary is required. Interestingly, all these problems can be circumvented in the framework developed here. In particular, it is possible to replace \eqref{eq:standard-contact} by a relation (not involving $\nabla\rho$ at all) of the form
\be
\label{eq:contact-simple}
\frac{\pa\rho^-}{\pa\nn\ga} = \gamma(\varphi)P(\rho^-)\ ,
\ee
where $P(\rho^-)$ is a low-order polynomial and $\gamma(\varphi)$ is a function depending only on the imposed contact angle $\varphi$. Towards this goal, let us follow the so-called energy-based approach \citep[see e.g.][]{Jacqmin2000} and postulate the surface Helmholtz free energy as follows: 
\begin{align}
\label{def:psi-ga-our}
\widehat{\helm}\ga(\rho^-) = \helm\ga^0 + (\sigmalw - \sigmavw)\frac{\int_{\rho_v^M}^{\rho^-}(x-\rho_v^M)(\rho_l^M-x)\,dx}{\int_{\rho_v^M}^{\rho^M_l}(x-\rho_v^M)(\rho_l^M-x)\,dx}\ , 
\end{align}
where $\sigmavw$ and $\sigmalw$ are the vapor-wall and liquid-wall surface tensions, respectively, and $\helm\ga^0$ is a constant. In this form, the surface Helmholtz free energy $\widehat{\helm}\ga$ is constant in the boundary regions that are in contact with the pure bulk phases characterized by the Maxwell states $\rho^M_v$ or $\rho^M_l$. The value of the constant differs for the two cases by $(\sigmalw{-}\sigmavw)$, and this jump takes place across the boundary ``contact line'', i.e., across the region on the boundary where the phases change from one to another. Let us substitute into \eqref{def:psi-ga-our} the standard contact-angle formula (Young equation)
\begin{align}
\sigmalv\cos{\varphi} = \sigmavw-\sigmalw\ ,
\end{align}
relating the liquid-vapor surface tension $\sigmalv$ with $\sigmalw$ and $\sigmavw$ through the cosine of the wetting angle $\varphi$. Here the wetting angle $\varphi$ denotes the contact angle of the liquid-vapor interface with respect to the wall measured inside the liquid domain. Employing the ansatz for the surface Helmholtz free energy \eqref{def:psi-ga-our}, the static part (i.e., corresponding to $\vv{=}{\bf 0}$) of the boundary conditions \eqref{model-A1-qdivv-KvdW3} and \eqref{model-A2-qdivv-KvdW3} becomes
\be
\label{eq:contact-our}
\frac{\pa\rho^-}{\pa\nn\ga} = \gamma_0\cos{\varphi}\ (\rho^-{-}\rho_v^M)(\rho_l^M{-}\rho^-)\ ,\hspace{1cm}\text{where}\hspace{1cm}\gamma_0 = \frac{\sigmalv}{\sigma}\left(\int_{\rho_v^M}^{\rho^M_l}(x-\rho_v^M)(\rho_l^M-x)\,dx\right)^{-1}\ ,
\ee
which is of the desired form \eqref{eq:contact-simple}. We will test this formula in the numerical simulations in Section \ref{sec:numerical-experiments}, where we also provide the specific value for the parameter $\gamma_0$. Let us only note here that due to the temperature dependence of the Maxwell states (see \eqref{def:maxwell-states}), $\gamma_0$ depends on temperature even in the current setting with constant surface tensions.


Let us summarize the conditions \eqref{model-A1i-KvdW3} -- \eqref{model-B2i3} corresponding to the ansatz for the surface Helmholtz free energy $\widehat{\psi}\ga$ of the form \eqref{def:psi-ga-our}:

\begin{itemize}
\item{Model A1:}
\begin{subequations}
\label{eq:bcs-final}
\begin{align}
    (2\mu\DD\nn\ga)_\tau^- &= \beta\div\vv^-\nabla\ga\rho^- -\alpha\vv_\tau^-\ ,\\
\label{eq:bc1}
    \frac{\pa\rho^-}{\pa\nn\ga}  &= \gamma_0\cos{\varphi}\ (\rho^-{-}\rho_v^M)(\rho_l^M{-}\rho^-) + \frac{\beta}{\sigma}\div\vv^-\ ; 
        \end{align}
\end{subequations}
 \item {Model A2:}
\begin{subequations}
    \begin{align}
      (2\mu\DD\nn\ga)_\tau^- &= -\rho^-\nabla\ga\left(\beta\frac{\pa\vv_\mathrm{n}^-}{\pa\nn\ga}\right)  -\alpha\vv_\tau^-\ ,\\
\label{eq:bc2}
    \frac{\pa\rho^-}{\pa\nn\ga}  &= \gamma_0\cos{\varphi}\ (\rho^-{-}\rho_v^M)(\rho_l^M{-}\rho^-) + \frac{\beta}{\sigma}\frac{\pa\vv_\mathrm{n}^-}{\pa\nn\ga}\ ;
    \end{align}
   \end{subequations} 
    \item {Models B1 and C}:
    \begin{subequations}
\begin{align}    
    (2\mu\DD\nn\ga)^-_\tau &= \beta\div\vv^-\nabla\ga\rho^- - \alpha\vv_\tau^-\ ,\\
   \frac{\pa\rho^-}{\pa\nn\ga} &= \frac{\beta}{\sigma}\div\vv^-\ ;
   \end{align}
   \end{subequations}
   \item{Model B2:}
   \begin{subequations}
   \begin{align}
    (2\mu\DD\nn\ga)^-_\tau  &=  - \rho^-\nabla\ga\left(\beta\frac{\pa\vv_\mathrm{n}^-}{\pa\nn\ga}\right)-\alpha\vv_\tau^-\  ,\\
   \frac{\pa\rho^-}{\pa\nn\ga} &= \frac{\beta}{\sigma}\frac{\pa\vv_\mathrm{n}^-}{\pa\nn\ga}\ .
\end{align}

\end{subequations}
\end{itemize}

As will be documented in the following numerical simulations, by explicitly evaluating the parameter $\gamma_0$ from \eqref{eq:contact-our} in Models A1 and A2, the value $\varphi$ equals the equilibrium (static) contact angle for the Korteweg - van der Waals fluid while the (positive) value of parameter $\beta$ governs the dynamic relaxation to this equilibrium. Clearly, Models B and C admit only homogeneous Neumann boundary conditions in equilibrium $\frac{\pa\rho^-}{\pa\nn\ga}{=}0$, i.e., the static contact angle is $\frac{\pi}{2}$.

\section{Numerical experiments focused on contact angle phenomena}
\label{sec:numerical-experiments}

Numerical experiments presented below are focused on the qualitative understanding of phenomena connected with the novel boundary conditions \eqref{eq:bcs-final}. We validate our interpretation of the static and dynamic parts of the contact-angle conditions by  demonstrating that the first term on the right-hand side of \eqref{eq:bc1} and \eqref{eq:bc2} determines the equilibrium contact angle, while the remaining terms on the right-hand side cause a dynamic delay in the attainment of this equilibrium contact angle value, see Experiments 1 and 2 below. Finally, we show that the dynamic terms have the potential to describe the phenomenon called dynamic contact angle hysteresis, see Experiment 3 below.

This section is structured in the following way. We first provide a dimensionless form of the governing equations. Then we proceed with identifying the corresponding continuous weak form and its discrete counterpart based on the Galerkin discretization. Finally, we briefly describe the numerical method and show the results of the three numerical experiments.

\subsection{Dimensionless formulation}
\label{sec-dimensionless-formulation}
We introduce the same scaling as in \cite{gomez2010}. Each field  quantity is expressed as $\varphi{=}[\varphi]\tilde{\varphi}$, where $[\varphi]$ denotes the scale of the quantity and $\tilde{\varphi}$ denotes the dimensionless counterpart. 
We introduce a spatial scale $[{\bf x}]{=}L_0$, and consider the scaling of spatial differential operators\footnote{This spatial scaling is clearly not optimal in the interfacial regions where another length scale corresponding to the thickness of the interfacial zone should probably be introduced. However, since we do not perform any scaling-based simplifications and the scaling only serves to provide dimensionless formulation, this issue can be ignored.} $[\nabla]{=}L_0^{-1}$ and $[\div]{=}L_0^{-1}$. We scale the density by $[\rho]{=}b'$, where $b'$ occurs as a parameter in the van der Waals model \eqref{def:vdW-parameters}. The temperature is scaled by the critical temperature, i.e., $[\vartheta]{=}\vartheta_c$ (see \eqref{helm-vdW}) and the pressure $p^{vdW}$ is scaled by $[p^{vdW}]{=}a'(b')^2$. For time, we pick the scale $[t]{=}L_0/\sqrt{a'b'}$ and the velocity is thus scaled by $[\vv]{=}[{\bf x}]/[t] =\sqrt{a'b'}$. We introduce the dimensionless numbers
\begin{center}

\begin{tabular}{ccccccccc}
	$Re_\lambda$ & $\eqdef$ & $\frac{L_0\sqrt{a'b'}b'}{[\lambda]}$ & Reynolds number 1 &\hspace{2cm} & $Ca$ & $\eqdef$ & $\frac{1}{L_0}\sqrt{\frac{\sigma}{a'}}$ & capillary number\\
	$Re_\mu$ & $\eqdef$ & $\frac{L_0\sqrt{a'b'}b'}{[\mu]}$ & Reynolds number 2 & & $G$ & $\eqdef$ & $\frac{[\bb] L_0}{a'b'}$ & 
\end{tabular}
\end{center}
\noindent
and consequently, we can rewrite the system of balance equations \eqref{eq:mass-and-momentum-kvdw} as
\begin{subequations}
\label{eq:mass-and-momentum-kvdw-dimless}
\begin{align}
\label{eq:mass-dimless}	
\frac{\pa\tilde{\rho}}{\pa\tilde{t}} + \widetilde{\mathrm{div}}(\tilde{\rho}\tilde{\vv}) &= 0\ ,\\
\label{eq:momentum-dimless}
\frac{\pa(\tilde{\rho}\tilde{\vv})}{\pa\tilde{t}} +\widetilde{\mathrm{div}}(\tilde{\rho}\tilde{\vv}\otimes\tilde{\vv}) &= -\tilde{\nabla}\widetilde{p_{vdW}} + \frac{1}{Re_\lambda}\tilde{\nabla}(\widetilde{\lambda}\widetilde{\mathrm{div}}\tilde{\vv}) + \frac{1}{Re_\mu}\widetilde{\mathrm{div}}\left(2\widetilde{\mu}\widetilde{\DD}\right) + (Ca)^2\tilde{\rho}\tilde{\nabla}\tilde{\Delta}\tilde{\rho} + G\tilde{\rho}\tilde{\bb}\ , 
\end{align}
where
\begin{align}
	\widetilde{p^{vdW}} = \frac{8}{27}\frac{\tilde{\vartheta}\tilde{\rho}}{1-\tilde{\rho}} - \tilde{\rho}^2\ .
\end{align}
\end{subequations}
Next, we introduce the dimensionless numbers
\begin{center}
\begin{tabular}{ccc}
	\hspace{1cm}$\mathcal{A} = \frac{[\alpha]L_0}{[\mu]} $\ , & \hspace{1cm}$\mathcal{B} = \frac{b'[\beta]}{L_0[\mu]}$\ , & \hspace{1cm}$\mathcal{D}=\frac{[\beta]}{\sigma}\sqrt{\frac{a'}{b'}}$\ ,
\end{tabular}
\end{center}
and a dimensionless function $\mathcal{C}(\tilde{\vartheta})$ defined in \eqref{def-C} in the Appendix.
Then the boundary conditions \eqref{eq:bcs-final} read as follows:
\begin{subequations}
\label{eq:boundary-terms-dimless}
\begin{itemize}
\item Model A1:
\begin{align}
(2\tilde{\mu}\widetilde{\DD}\tilde\nn\ga)^-_\tau  &= -\mathcal{A}\tilde{\alpha}\tilde{\vv}^-_\tau + \mathcal{B}\tilde{\beta}\widetilde{\mathrm{div}}\tilde{\vv}^- \tilde{\nabla}\ga\tilde{\rho}^-\ , \\
\frac{\pa\tilde{\rho}^-}{\pa\tilde{\nn}\ga} &= \mathcal{C}(\tilde{\vartheta})\,\cos{\varphi}\ (\tilde{\rho}^-{-}\tilde{\rho}_v^M)(\tilde{\rho}_l^M{-}\tilde{\rho}^-) + \mathcal{D}\tilde{\beta}\widetilde{\mathrm{div}}\tilde{\vv}^-\ ;
\end{align}
\item Model A2:
\begin{align}
(2\tilde{\mu}\widetilde{\DD}\tilde\nn\ga)^-_\tau  &= -\mathcal{A}\tilde{\alpha}\tilde{\vv}^-_\tau - \mathcal{B}
\tilde\rho^-\tilde{\nabla}\ga\left(\tilde{\beta}\frac{\pa\tilde{\vv}_\mathrm{n}^-}{\pa\tilde\nn\ga}\right)\ , \\
\frac{\pa\tilde{\rho}^-}{\pa\tilde{\nn}\ga} &= \mathcal{C}(\tilde{\vartheta})\,\cos{\varphi}\ (\tilde{\rho}^-{-}\tilde{\rho}_v^M)(\tilde{\rho}_l^M{-}\tilde{\rho}^-) + \mathcal{D}\tilde{\beta}\frac{\pa\tilde{\vv}_\mathrm{n}^-}{\pa\tilde\nn\ga}\ ;
\end{align}

\item Model B1 and C:
\begin{align}
(2\tilde{\mu}\widetilde{\DD}\tilde\nn\ga)^-_\tau  &= -\mathcal{A}\tilde{\alpha}\tilde{\vv}^-_\tau + \mathcal{B}\tilde{\beta}\widetilde{\mathrm{div}}\tilde{\vv} \tilde{\nabla}\ga\tilde{\rho}^-\ , \\
\frac{\pa\tilde{\rho}^-}{\pa\tilde{\nn}\ga} &= \mathcal{D}\tilde{\beta}\widetilde{\mathrm{div}}\tilde{\vv}^-\ ;
\end{align}
\item Model B2:
\begin{align}
(2\tilde{\mu}\widetilde{\DD}\tilde\nn\ga)^-_\tau  &= -\mathcal{A}\tilde{\alpha}\tilde{\vv}^-_\tau - \mathcal{B}
\tilde\rho^-\tilde{\nabla}\ga\left(\tilde{\beta}\frac{\pa\tilde{\vv}_\mathrm{n}^-}{\pa\tilde\nn\ga}\right)\ , \\
\frac{\pa\tilde{\rho}^-}{\pa\tilde{\nn}\ga} &= \mathcal{D}\tilde{\beta}\frac{\pa\tilde{\vv}_\mathrm{n}^-}{\pa\tilde\nn\ga}\ .
\end{align}

\end{itemize}
\end{subequations}
\subsection{Weak formulations of the initial and boundary value problems and their numerical discretization}
\label{sec:weak-formulation}
In this subsection, we first introduce the weak formulations to the system of governing equations \eqref{eq:mass-and-momentum-kvdw-dimless}--\eqref{eq:boundary-terms-dimless} and then we present its discretization. For simplicity, we avoid using tildes in the dimensionless formulations \eqref{eq:mass-and-momentum-kvdw-dimless} and \eqref{eq:boundary-terms-dimless}. Since with respect to density, the strong form of the momentum balance \eqref{eq:momentum-dimless} involves the third derivative, we employ a mixed formulation by introducing 
\be
\label{eq:z}
z\eqdef\Delta\rho\ \ee
 as a new variable.
We also assume that both the bulk and the shear viscosities are constant, meaning that $\mu{=}[\mu]$, $\lambda{=}[\lambda]$ (implying that $\tilde{\mu}{=}\tilde{\lambda}{=}1$).

In order to specify a weak solution to \eqref{eq:mass-and-momentum-kvdw-dimless}--\eqref{eq:z}, we first introduce several standard function spaces: the Lebesgue space $(L^2(\Omega),(\cdot,\cdot)_\Omega)$, the Sobolev space
$W^{1,2}(\Omega)$, and its
dual $\left(W^{1,2}(\Omega)\right)^*$ with the 
corresponding duality pairing $\langle \cdot,\cdot\rangle_\Omega$. We also set the space 
$$W^{1,2}_\mathrm{n}(\Omega) \eqdef \left\{ \ww=(w_1,w_2,w_3)\in W^{1,2}(\Omega)\times W^{1,2}(\Omega)\times W^{1,2}(\Omega); \ww\cdot\nn\ga{=} 0\ \text{at}\ \pa\Omega \right\}$$ 
and introduce the spaces
\begin{align}
 X\eqdef  W^{1,2}(\Omega)\times W_\mathrm{n}^{1,2}(\Omega)\times\left(W^{1,2}(\Omega)\right)^*\ ,\hspace{1cm}
 Y\eqdef W^{1,2}(\Omega)\times W_\mathrm{n}^{1,2}(\Omega)\times W^{1,2}(\Omega)\ .
\end{align}
We say that $\uu\eqdef(\rho,\vv,z)\in X$ is a weak solution to \eqref{eq:mass-and-momentum-kvdw-dimless}--\eqref{eq:z}
if
\begin{subequations}
\label{eq-continuous-weak-form}
\begin{align}
\left(\frac{\pa\rho}{\pa t},\rhotest\right)_\Omega + \left(\frac{\pa(\rho\vv)}{\pa t},\vvtest\right)_\Omega + \langle z,\pitest\rangle_\Omega + B_\Omega(\uu,\uutest) + B_{\pa\Omega}(\uu^-,\uutest) = 0\hspace{0.5cm}\text{for all}\ \ \uutest = (\rhotest,\vvtest,\pitest)\in Y\ , 
\end{align}
holds, where the bulk and boundary forms $B_\Omega$ and $B_{\pa\Omega}$ are defined as
\begin{align}
\nonumber
B_\Omega(\uu,\uutest) &\eqdef (\div{(\rho\vv)},\rhotest)_\Omega - \left(\rho\vv\otimes\vv,\nabla	\vvtest\right)_\Omega 
- \left(p_{vdW},\div\vvtest\right)_\Omega
+ \frac{1}{Re_\lambda}(\div\vv,\div\vvtest)_\Omega\\ &+ \frac{2}{Re_\mu}(\DD,\nabla\vvtest)_\Omega + (Ca)^2(z,\div(\rho\vvtest))_\Omega - G(\rho\bb,\vvtest)_\Omega + (\nabla\rho,\nabla\pitest)_\Omega\ ,
\\
\label{def-B-pa-Omega}
B_{\pa\Omega}(\uu^-,\uutest) &\eqdef -\frac{2}{Re_\mu }\int_{\pa\Omega}(\DD\nn\ga)^-_\tau\cdot{\vvtest}_\tau dS - \int_{\pa\Omega}\frac{\pa\rho^-}{\pa\nn\ga}\pitest dS\ .
\end{align}
\end{subequations}
Next, we replace the integrands $(2\DD\nn\ga)^-_\tau$ and $\frac{\pa\rho^-}{\pa\nn\ga}$ 
in $B_{\pa\Omega}$ by means of \eqref{eq:boundary-terms-dimless} and employ 
the identity $\frac{\mathcal{B}}{Re_\mu}{=} \mathcal{D}(Ca)^2$. This will generate four different forms:
\begin{subequations}
\label{eq:boundary-integrals}
\begin{align}
\nonumber
B^{(A1)}_{\pa\Omega}(\uu^-,\uutest) &= \frac{\mathcal{A}}{Re_\mu}\int_{\pa\Omega}\alpha \vv_\tau^-\cdot(\vvtest)_\tau\,dS -\mathcal{D}(Ca)^2\int_{\pa\Omega}\beta\div\vv^-\nabla\ga\rho^-\cdot(\vvtest)_\tau\,dS\ \\\label{bi1} &- \int_{\pa\Omega}\left\{\mathcal{C}\, \cos{\varphi}\,({\rho^-}{-}{\rho}_v^M)({\rho}_l^M{-}{\rho^-}){+}\mathcal{D}{\beta}{\mathrm{div}}{\vv^-}\right\}\pitest\,dS\ ,
\\
\nonumber
B^{(A2)}_{\pa\Omega}(\uu^-,\uutest) &= \frac{\mathcal{A}}{Re_\mu}\int_{\pa\Omega}\alpha \vv_\tau^-\cdot(\vvtest)_\tau\,dS-\mathcal{D}(Ca)^2\int_{\pa\Omega}\beta\left(\frac{\pa\vv^-_\mathrm{n}}{\pa\nn\ga}\right)\div\ga(\rho^-(\vvtest)_\tau)\,dS \\\label{bi2}
&{-} \int_{\pa\Omega} \left\{\mathcal{C}\, \cos{\varphi}\,({\rho^-}{-}{\rho}_v^M)({\rho}_l^M{-}{\rho^-}) + \mathcal{D}{\beta}\frac{\pa\vv^-_\mathrm{n}}{\pa\nn\ga}\right\}\pitest\,dS\ ,\\
\nonumber
B^{(B1\&C)}_{\pa\Omega}(\uu^-,\uutest) &= \frac{\mathcal{A}}{Re_\mu}\int_{\pa\Omega}\alpha \vv_\tau^-\cdot(\vvtest)_\tau\,dS - \mathcal{D}(Ca)^2\int_{\pa\Omega} \beta\div\vv^-\nabla\ga\rho^-\cdot(\vvtest)_\tau\,dS\\\label{bi3} &- \int_{\pa\Omega}\mathcal{D}{\beta}{\mathrm{div}}{\vv^-}\pitest\,dS\ ,
\\
\nonumber
B^{(B2)}_{\pa\Omega}(\uu^-,\uutest) &=  \frac{\mathcal{A}}{Re_\mu}\int_{\pa\Omega}\alpha \vv_\tau^-\cdot(\vvtest)_\tau\,dS - \mathcal{D}(Ca)^2\int_{\pa\Omega}\beta\left(\frac{\pa\vv^-_\mathrm{n}}{\pa\nn\ga}\right)\div\ga(\rho(\vvtest)_\tau)\,dS\\ \label{bi4} &- \int_{\pa\Omega} \mathcal{D}{\beta}\frac{\pa\vv^-_\mathrm{n}}{\pa\nn\ga}\pitest\,dS\ .
\end{align}
\end{subequations}
Let us note that in \eqref{bi2} and \eqref{bi4}, we applied the integration by parts in the second terms on the right-hand sides. The system is supplemented with initial conditions for the density and the velocity:
$$
	\left.\rho(t,{\bf x})\right|_{t=0} = \rho_0(\bf{x})\ ,\hspace{0.5cm}	\left.\vv(t,{\bf x})\right|_{t=0} = \vv_0(\bf{x})\ ,\hspace{0.5cm} {\bf x}\in \Omega\ .
$$

The weak formulation \eqref{eq-continuous-weak-form} and \eqref{eq:boundary-integrals} is discretized 
in time by a simple $\Theta$-scheme and in space by the Galerkin method. Denoting (finite-element) discrete subspaces of $X$ and $Y$ by $X_h$ and $Y_h$, respectively, we define the discrete solution at the $n$-th time level as $\uu_h^n\eqdef(\rho_h^n,\vv_h^n,z_h^n)\in X_h$ satisfying
\begin{subequations}
\label{eq-discrete-weak-form}
\begin{align}
\nonumber
& \left(\frac{\rho_h^{n+1}{-}\rho_h^n}{\delta t},\rhotesth\right)_{\Omega_h} + \left(\frac{\rho_h^{n+1}\vv_h^{n+1}{-}\rho_h^n\vv_h^n}{\delta t},\vvtesth\right)_{\Omega_h} + (z_h^{n+1},\pitesth)_{\Omega_h} + \Theta B_{\Omega_h}(\uu^{n+1}_h,\uutesth) + (1-\Theta) B_{\Omega_h}(\uu^{n}_h,\uutesth)\\ &+ \Theta B_{\pa{\Omega_h}}(\uu_h^{n+1},\uutest) +  (1-\Theta) B_{\pa{\Omega_h}}(\uu_h^{n},\uutest) = 0, \hspace{1cm}\text{for all }\uutesth\in Y_h\ , 
\end{align}
where $B_{\Omega_h}$ and $B_{\pa\Omega_h}$ differ from $B_\Omega$ and $B_{\pa\Omega}$ only by the integration domains - 
here $\Omega_h$ and $\pa{\Omega_h}$ denote the (finite element) approximations of $\Omega$ and $\pa\Omega$, respectively. The initial values are taken as $\rho_h^{0} = (\rho_0)_h$ and $\vv_h^{0} = (\vv_0)_h$. Finally, $\Theta{\in}\langle 0,1\rangle$, where the value $\Theta{=}1$ yields a fully implicit time discretization and $\Theta{=}0.5$ corresponds to the Crank-Nicolson scheme.
\end{subequations}

\subsection{Numerical solution}
The discrete system \eqref{eq-discrete-weak-form} is implemented by a finite-element method in the software package FEniCS \citep{alnaes2015} and for the discrete spaces, we choose continuous piece-wise polynomial approximations $X_h = P_3{\times}(P_2)^3{\times}P_3$, where $P_N$ denotes polynomials of order $N$. We apply a structured mesh to a two-dimensional domain $\Omega$ with the aspect height-to-length ratio $1{:}3$. The mesh consists of $N_1{\times}N_2$ squares, each divided into 4 regular triangles. We apply a scaling of the capillary number $Ca$ based on the refinement methodology proposed by \cite{gomez2010}. \cite{gomez2010} argue that since the realistic resolution of the diffuse interface zone in the Korteweg models is out of the scope of macroscopic models, it is reasonable to treat the capillary number in such cases as a regularizing parameter; it's adjustment is based on the given spatial resolution of the model in such a way that the diffuse interface remains reasonably resolved. Based on this idea \cite{gomez2010} introduce the parameterization $Ca{=}\frac{h}{L_0}$, where $h$ is the characteristic length scale of the spatial mesh, here defined as $h{=}\frac{L_0}{2\sqrt{N_1 N_2}}$, with the length scale $L_0{=}1$ and $N_1{=}90$, $N_2{=}30$. Moreover, we also adopt the scaling of the Reynolds numbers from \cite{gomez2010}, setting 
$Re_\mu{=}Re_\lambda{=}2 Ca^{-1}$. Being interested only in qualitative properties of the model, we set all but one of the remaining dimensionless numbers equal to one, i.e., we assign $\mathcal{A}{=}\mathcal{B}{=}\mathcal{D}{=}1$. The exception is the parameter $\mathcal{C}$ (depending on temperature), which governs the equilibrium contact angle, which we want to control quantitatively. For temperature $\vartheta{=}0.85\vartheta_c$, we get $\mathcal{C}(0.85){\doteq}\frac{25.5}{9\sqrt{3}Ca}$, see \eqref{eq:magic-value} in the Appendix.
  
Since the boundary conditions corresponding to Models B1, B2, and C represent a subclass of the boundary conditions for Models A1 and A2, we only consider the latter two models. The dimensionless boundary conditions in the considered setting simplify to (we omit tilde symbols for brevity): 
\begin{subequations}
\label{eq:boundary-terms-dimless2}
\begin{itemize}
\item Model A1:
\begin{align}
(2{\mu}{\DD}\nn\ga)^-_\tau  &=  -\alpha\vv^-_\tau + \beta{\mathrm{div}}{\vv}^- \nabla\ga{\rho}^-\ , \\
\frac{\pa{\rho}^-}{\pa{\nn}\ga} &= \, \mathcal{C}\cos{\varphi}\,({\rho}^-{-}{\rho}_v^M)({\rho}_l^M{-}{\rho}^-) + {\beta}{\mathrm{div}}{\vv}^-\ ;
\end{align}
\item Model A2:
\begin{align}
(2{\mu}{\DD}\nn\ga)^-_\tau  &= -\alpha\vv^-_\tau - \rho^-{\nabla}\ga\left({\beta}\frac{\pa{\vv}_\mathrm{n}^-}{\pa\nn\ga}\right)\ , \\
\frac{\pa{\rho}^-}{\pa{\nn}\ga} &= \, \mathcal{C}\cos{\varphi}\,({\rho}^-{-}{\rho}_v^M)({\rho}_l^M{-}{\rho}^-) + {\beta}\frac{\pa{\vv}_\mathrm{n}^-}{\pa\nn\ga}\ .
\end{align}
\end{itemize}
\end{subequations}
\noindent {\bf Experiment 1}\\
\noindent In this numerical experiment, we study the evolution of the Korteweg - van der Waals fluid in a two-dimensional container $\Omega$ in the absence of body forces, meaning that $\bb{=}{\bf 0}$ in \eqref{eq-continuous-weak-form}. The system is initially at rest ($\vv_0{=}{\bf 0}$). Consequently, the only driving mechanisms for its evolution are the boundary conditions \eqref{eq:boundary-terms-dimless2} provided that $\varphi{\neq}\frac{\pi}{2}$ on the upper or lower part of the boundary. In order to isolate the effect of the novel contact angle condition and the generalized Navier slip condition from the traditional Navier slip boundary condition, we set $\alpha{=}0$.

\begin{figure}[h!]
\begin{tabular}{ccc}
\includegraphics[scale=0.32]{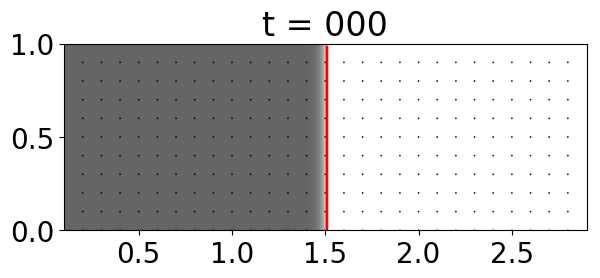}   & 
\includegraphics[scale=0.32]{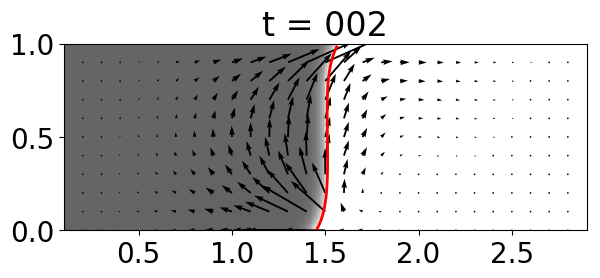}   
&
\includegraphics[scale=0.32]{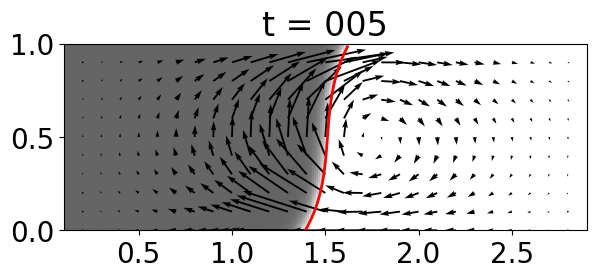}   \\
\includegraphics[scale=0.32]{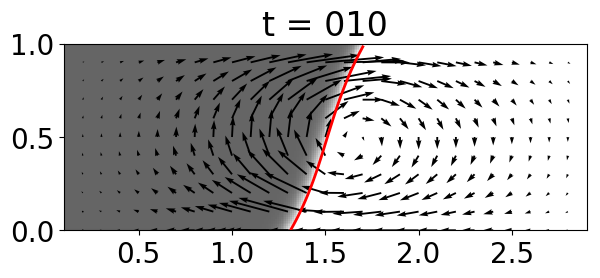}   & 
\includegraphics[scale=0.32]{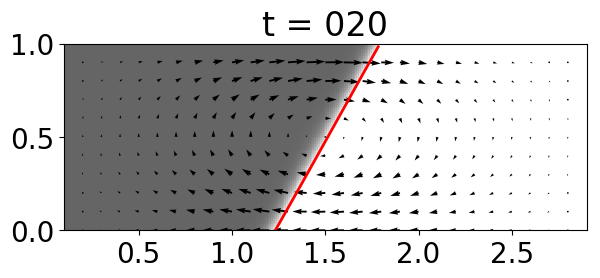}   
&
\includegraphics[scale=0.32]{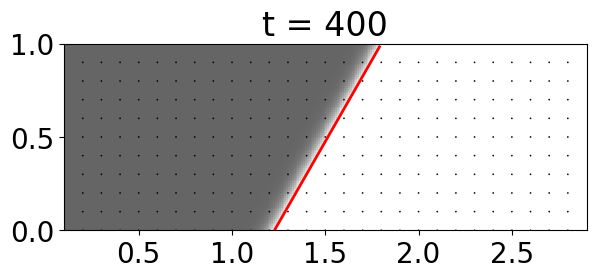}   
\end{tabular}
\caption{Evolution of the density distribution in a Korteweg - van der Waals model from the initial condition  (top left) given by vapour in the Maxwell state $\rho_v^M$ in the right part of the domain (white) and liquid in the Maxwell state with density $\rho_l^M$ in the left part of the panel (grey) to the equilibrium given by a static contact angle $\frac{\pi}{3}$ (bottom right). The solid red contour denotes the interface between the liquid and water phases defined here by the density value $\frac{\rho_v^M+\rho_l^M}{2}$. The arrows represent the velocity field.}
\label{Fig1-experiment1}
\end{figure}

\begin{figure}[h!]
\begin{tabular}{ccc}
\includegraphics[scale=0.32]{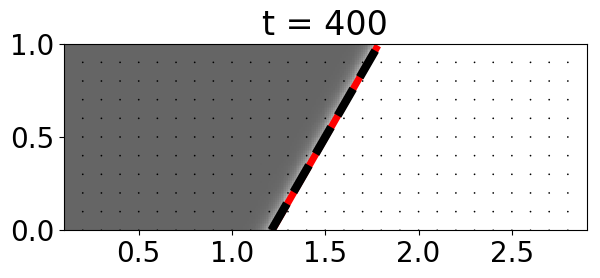}   & 
\includegraphics[scale=0.32]{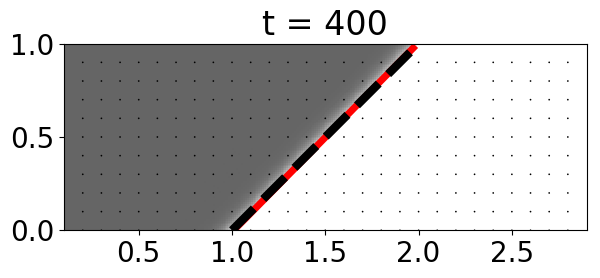}   
&
\includegraphics[scale=0.32]{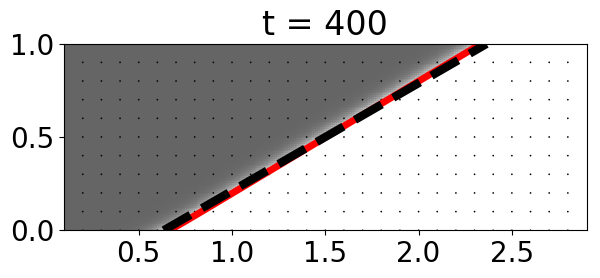}   
\end{tabular}
\caption{Equilibrium states of simulations as in Fig.~\ref{Fig1-experiment1}, which differ by the values of the equilibrium contact angle $\varphi{=}\frac{\pi}{3}$,(left) $\frac{\pi}{4}$ (middle) and $\frac{\pi}{6}$ (right). The solid red contour denotes the interface between the liquid and water phases defined here by the density value $\frac{\rho_v^M+\rho_l^M}{2}$ and the dashed black line is a linear function with the slope given by $\varphi$ passing through the center of the domain.
}
\label{Fig1-experiment1-role-of-gamma}
\end{figure}

In Fig.~\ref{Fig1-experiment1}, we depict the evolution of the density distribution from the initial condition (top left). The system consists of a vapour in the Maxwell state $\rho_v^M$ in the right part of the domain (white) and liquid in the Maxwell state with density $\rho_l^M$ in the left part of the panel (grey) separated by a flat interface perpendicular to the boundary. In order to demonstrate that with the value of $\mathcal{C}$ as in \eqref{eq:magic-value}, the parameter $\varphi$ controls (and equals) the equilibrium value of the contact angle of the fluid-vapor interface, we prescribe in all simulations in Experiment 1 the value of the static contact angle $\varphi$ on the top boundary and $\pi{-}\varphi$ on the bottom boundary. The reason is that in this case the equilibrium interfaces are particularly simple, being linear. In the snaphots of the simulation shown in Fig.~\ref{Fig1-experiment1}, we employ only the equilibrium (static) part of the contact angle-condition, which means we consider $\beta{=}0$. The red contour denotes the interface between the liquid and water phases defined here by the density value $\frac{\rho_v^M+\rho_l^M}{2}$. The arrows depict the velocity field. All quantities are dimensionless and since (with the exception of the equilibrium contact angle) we are interested only in qualitative behavior of the model, we do not show any scales.  In Fig.~\ref{Fig1-experiment1-role-of-gamma} we show the final states of three simulations, which differ only in the value of $\varphi$, considering $\varphi{=}\frac{\pi}{3}$, $\frac{\pi}{4}$, and $\frac{\pi}{6}$. For comparison, we plot also black dashed lines with the slope corresponding to the prescribed $\varphi$ and we observe very good agreement.
 
\begin{figure}[h!]
\begin{tabular}{cccccc}
\includegraphics[scale=0.15]{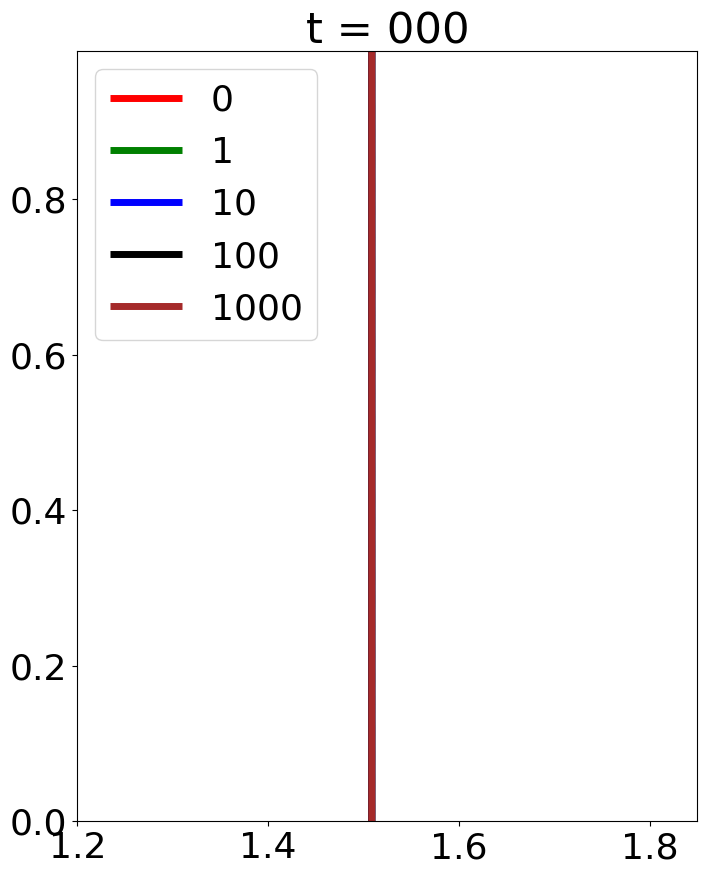}
&
\includegraphics[scale=0.15]{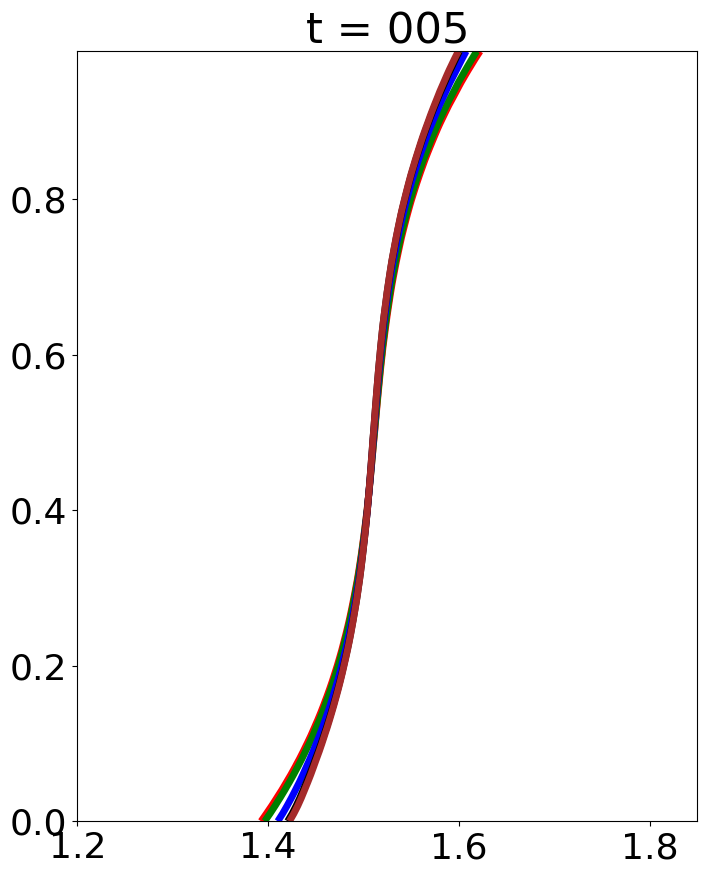}
&
\includegraphics[scale=0.15]{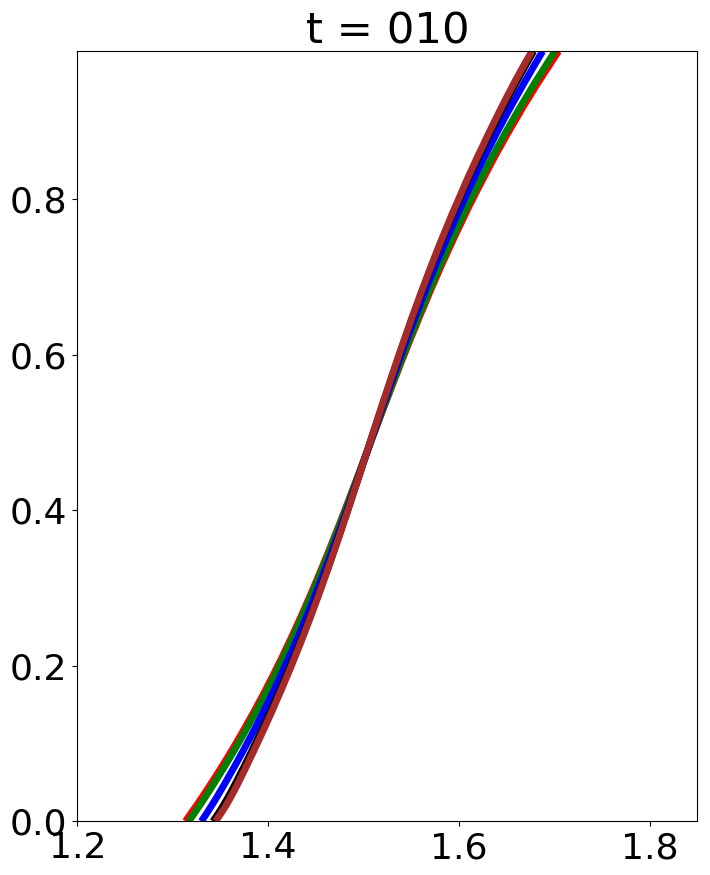}
&
\includegraphics[scale=0.15]{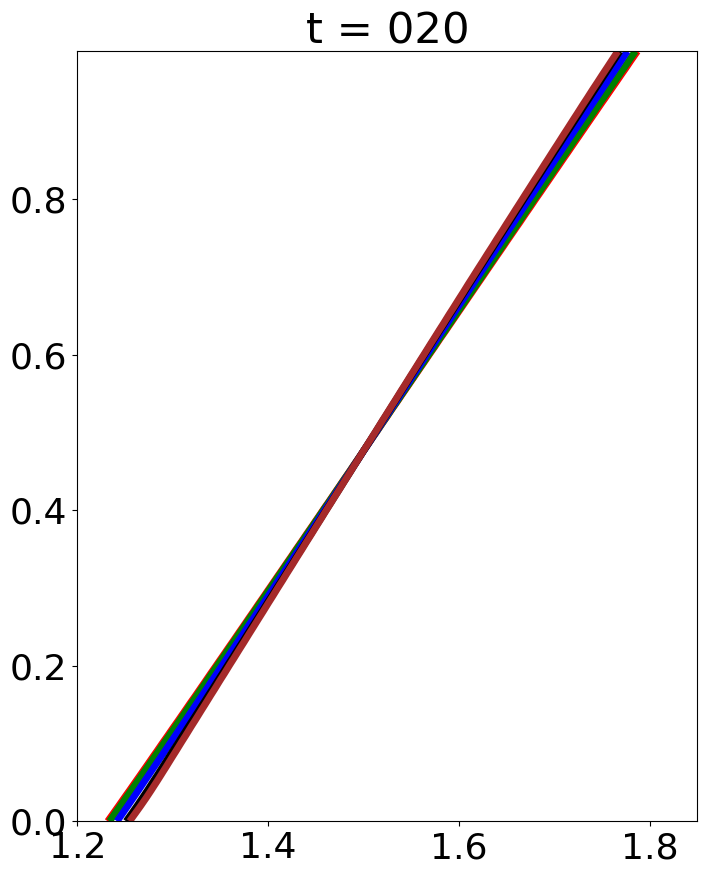}
&
\includegraphics[scale=0.15]{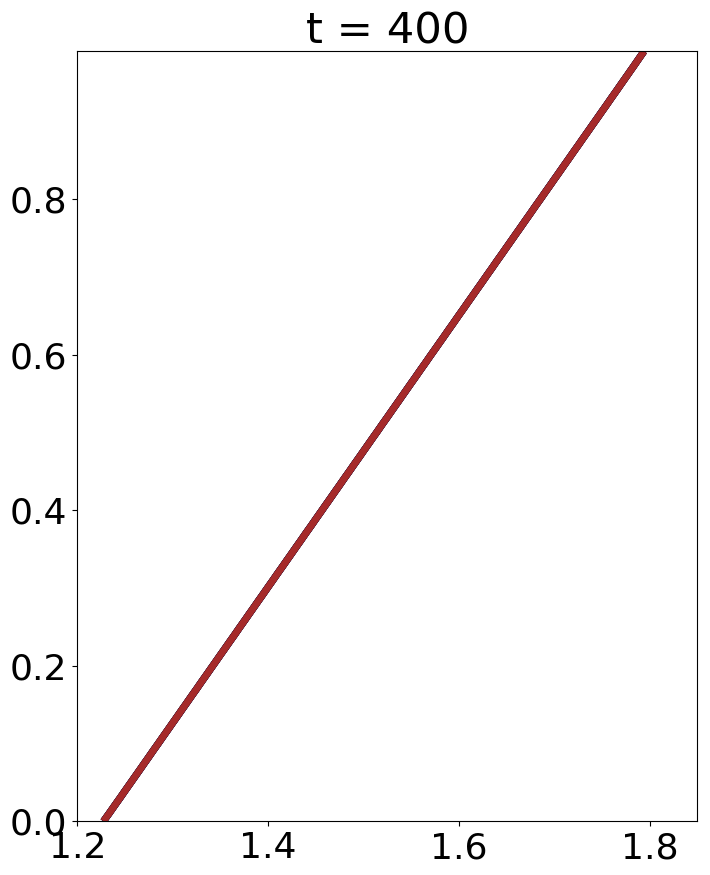}
\\
\includegraphics[scale=0.15]{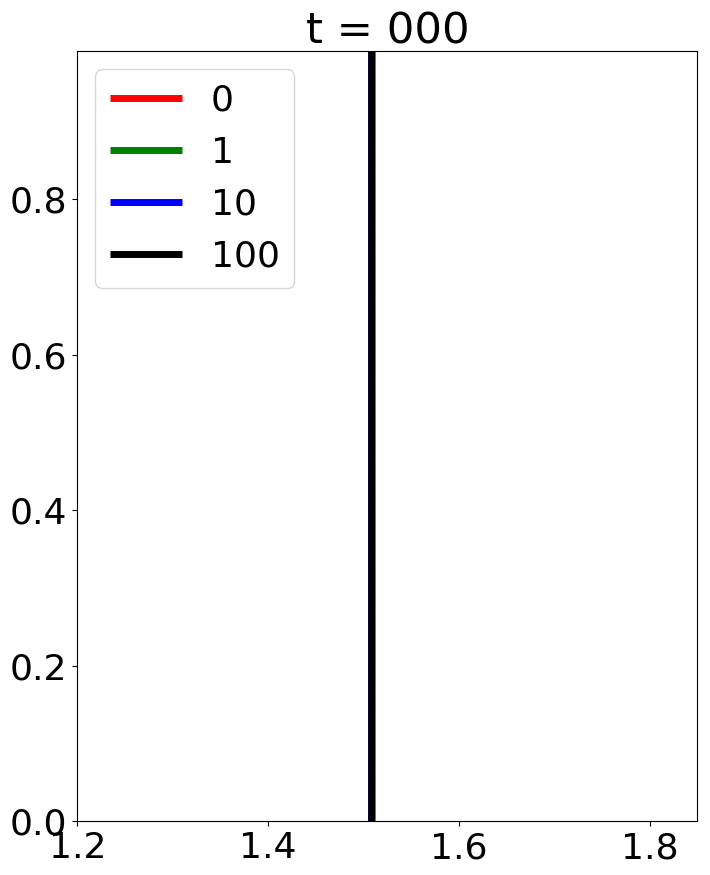}
&
\includegraphics[scale=0.15]{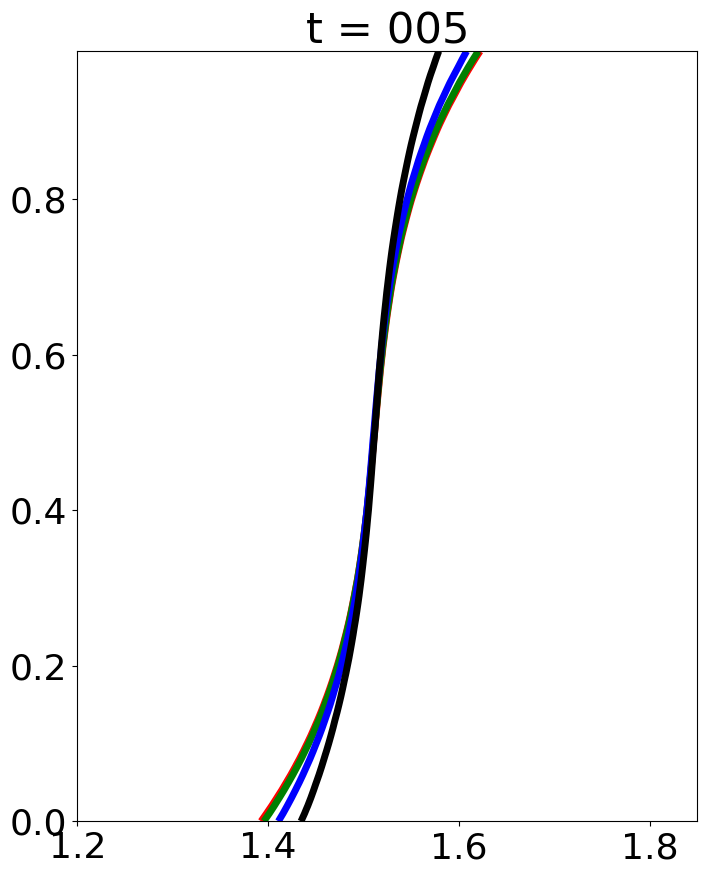}
&
\includegraphics[scale=0.15]{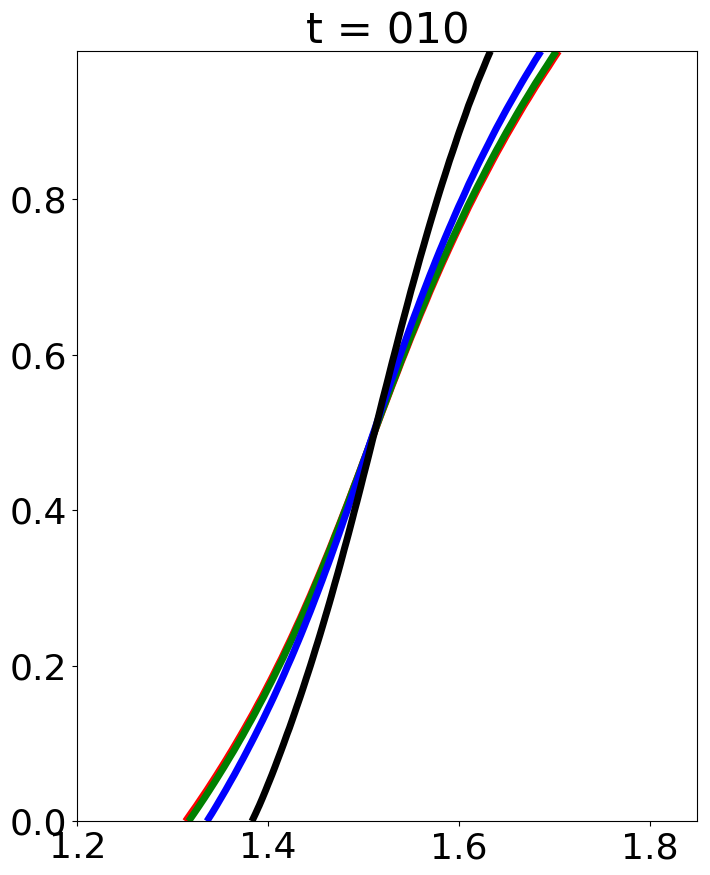}
&
\includegraphics[scale=0.15]{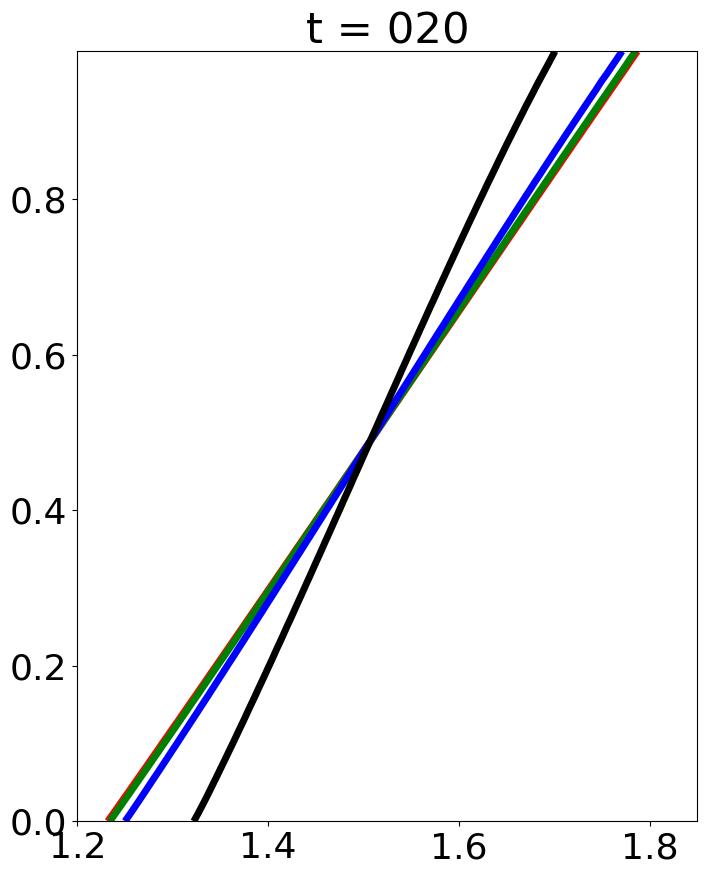}
&
\includegraphics[scale=0.15]{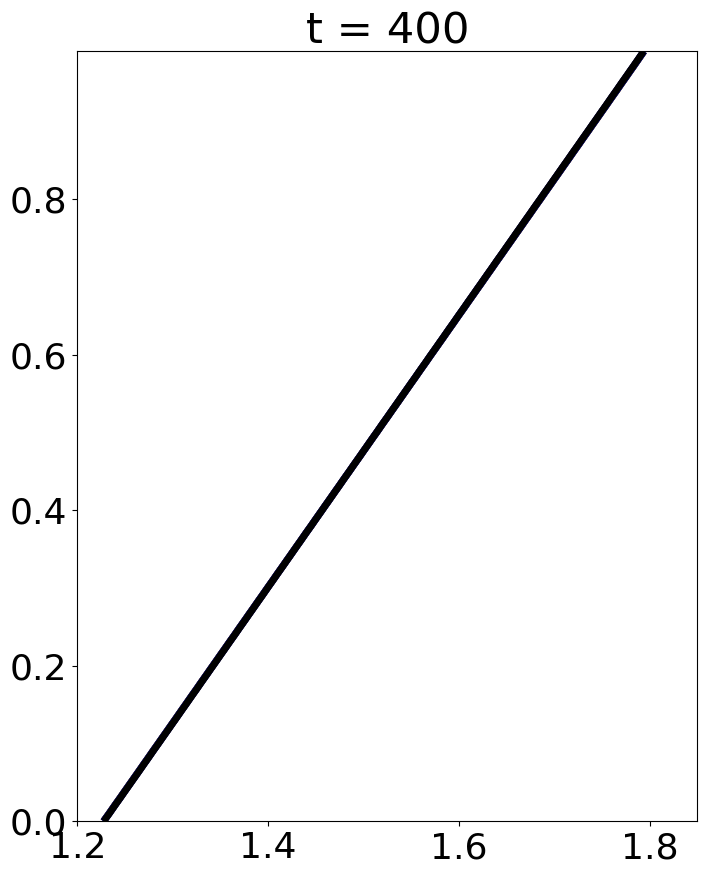}
\end{tabular}
\caption{Zoomed-in evolution of the interface between the two Maxwell states for different values of $\beta$ parameter for Model A1 (top row) and for Model A2 (bottom row).
}
\label{Fig2-experiment1}
\end{figure}

In order to study the effect of the dynamic part of the contact angle condition, in Fig.~\ref{Fig2-experiment1}, we depict the time evolution of the interface based on the value of the parameter $\beta$ for Models A1 (top row) and A2 (bottom row) for a given static contact angle $\varphi{=}\frac{\pi}{3}$. Note that the parameter $\beta$ appears both in the contact angle condition and in the generalized  Navier slip condition; see eq.~\eqref{eq:boundary-terms-dimless2}. The case $\beta{=}0$ corresponds to the solely static contact angle condition while for $\beta{>}0$ additional dissipative surface mechanism is present. In the second case, the evolution of the interface and motion of the contact points lags behind the case with the static contact angle and this dynamic effect is stronger for Model A2 than for Model A1 and depends in both cases on the values of $\beta$. While for Model A1 there appears to be a saturation of the dynamic effect with respect to increasing values of $\beta$, for Model A2 the bigger the value of $\beta$, the stronger the dynamic effect. It is important to note that for all non-zero values of $\beta$, the final equilibrium configuration matches the case with $\beta{=}0$ as expected since the additional terms are of non-equilibrium nature and must vanish in the final equilibrium state.

\vspace{0.5cm}

\begin{figure}[h!]
\begin{tabular}{ccc}
\includegraphics[scale=0.32]{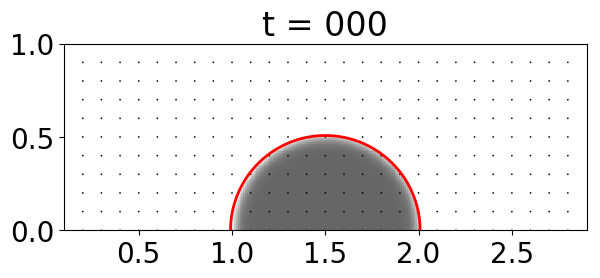}   & 
\includegraphics[scale=0.32]{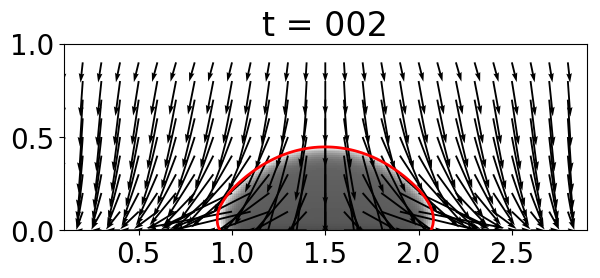}   &
\includegraphics[scale=0.32]{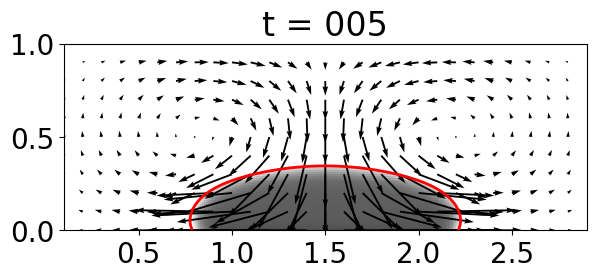}   \\
\includegraphics[scale=0.32]{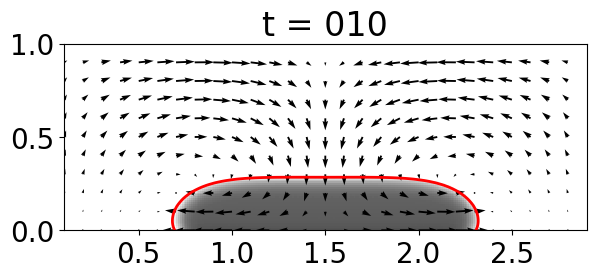}   & 
\includegraphics[scale=0.32]{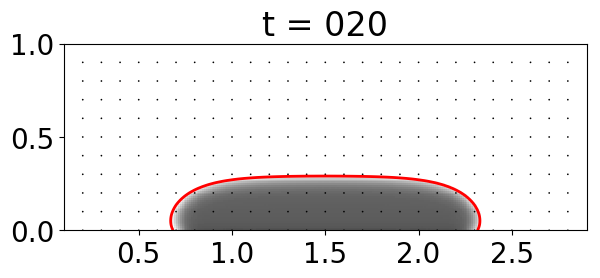}   &
\includegraphics[scale=0.32]{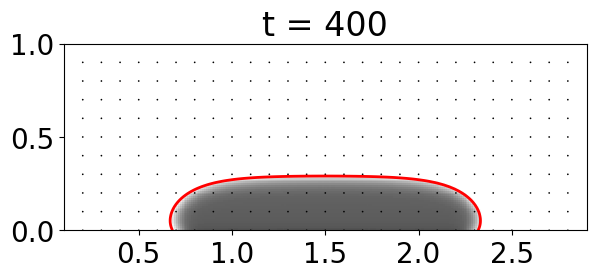}   
\end{tabular}
\caption{Evolution of a semicircular droplet of a Korteweg - van der Waals liquid with density $\rho_l^M$ surrounded by vapour with density $\rho_v^M$ from the initial condition (top left) to an equilibrium given by static contact angle $\frac{\pi}{3}$ (bottom right). The red contour denotes the interface between the liquid and vapour phases defined here by the density value $\frac{\rho_v^M+\rho_l^M}{2}$. The arrows denote the velocity field.}
\label{Fig1-experiment2}
\end{figure}

\begin{figure}[h!]
\begin{tabular}{ccccccc}
\includegraphics[scale=0.15]{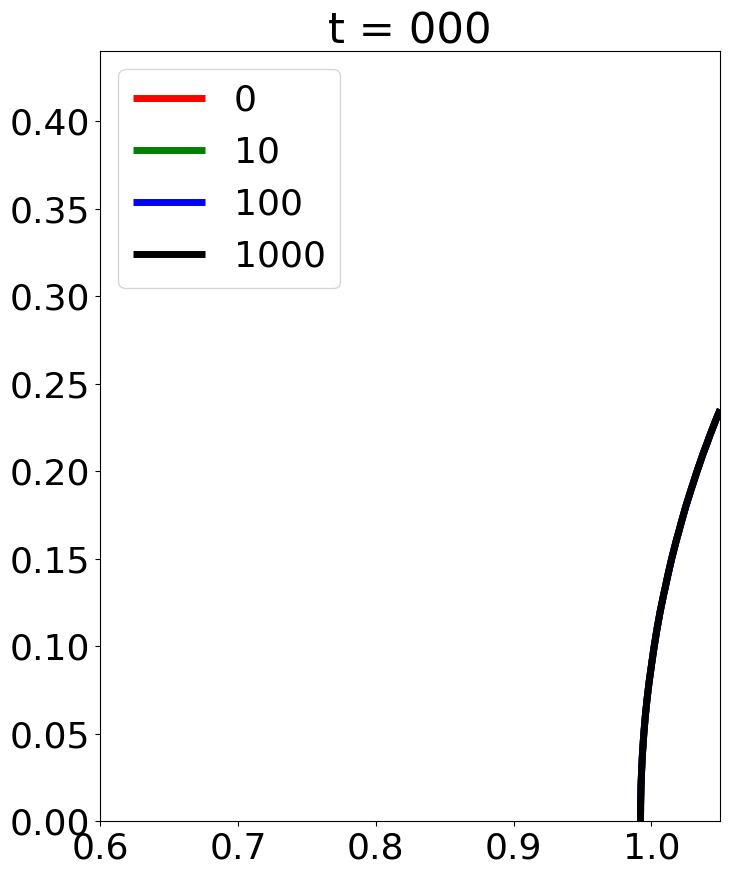}
&
\includegraphics[scale=0.15]{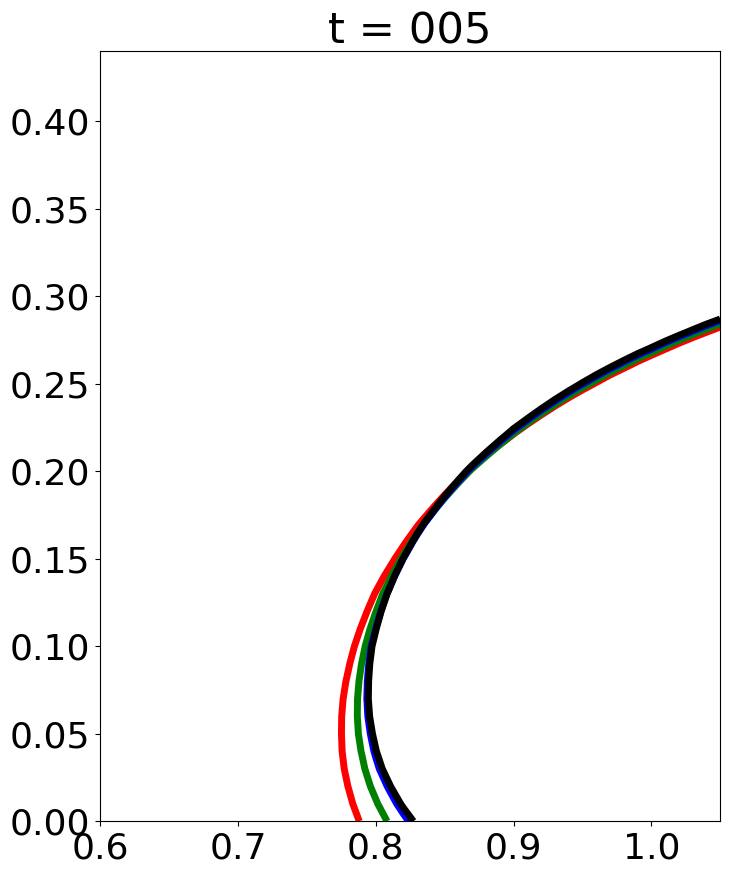}
&
\includegraphics[scale=0.15]{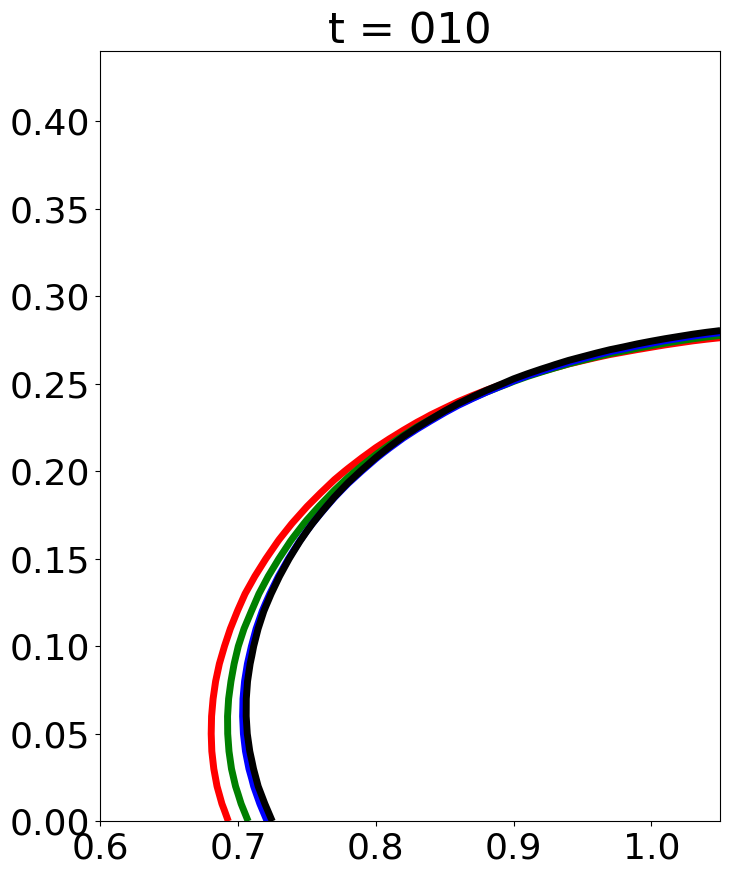}
&
\includegraphics[scale=0.15]{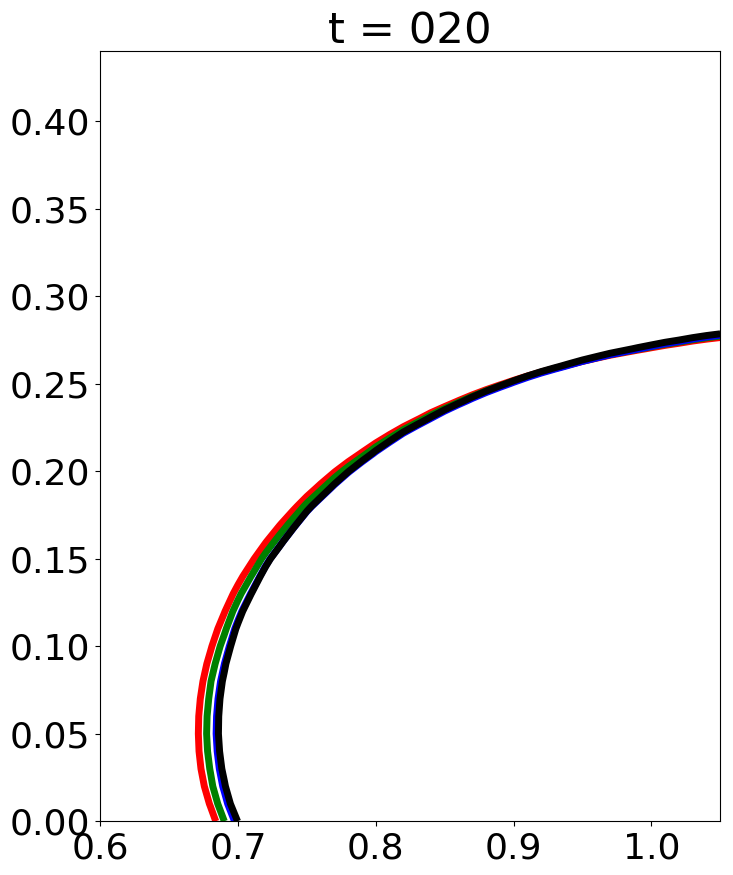}
&
\includegraphics[scale=0.15]{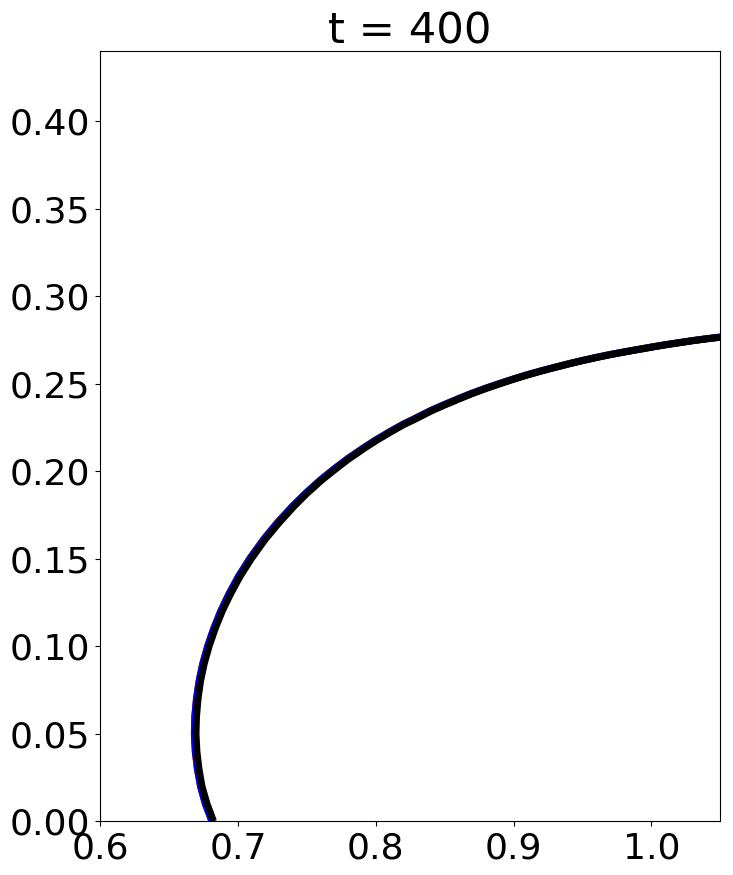}
\\
\includegraphics[scale=0.15]{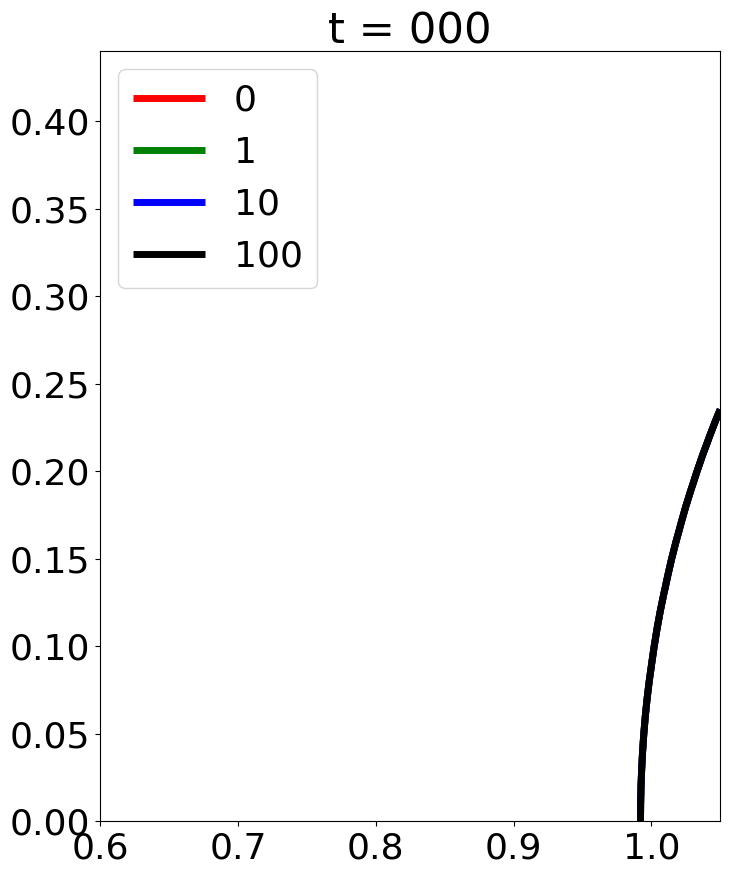}
&
\includegraphics[scale=0.15]{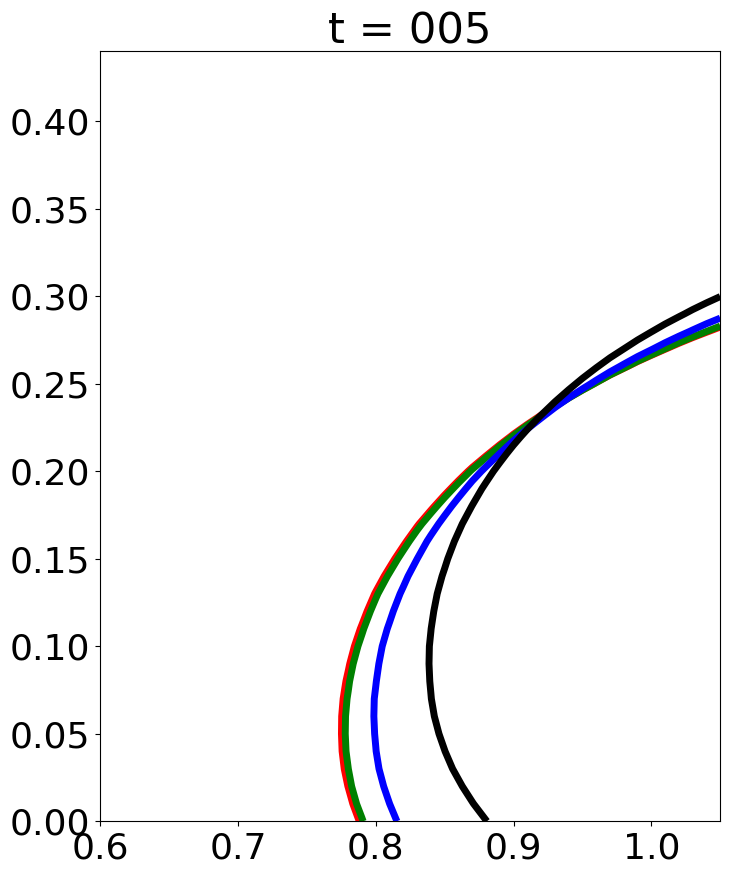}
&
\includegraphics[scale=0.15]{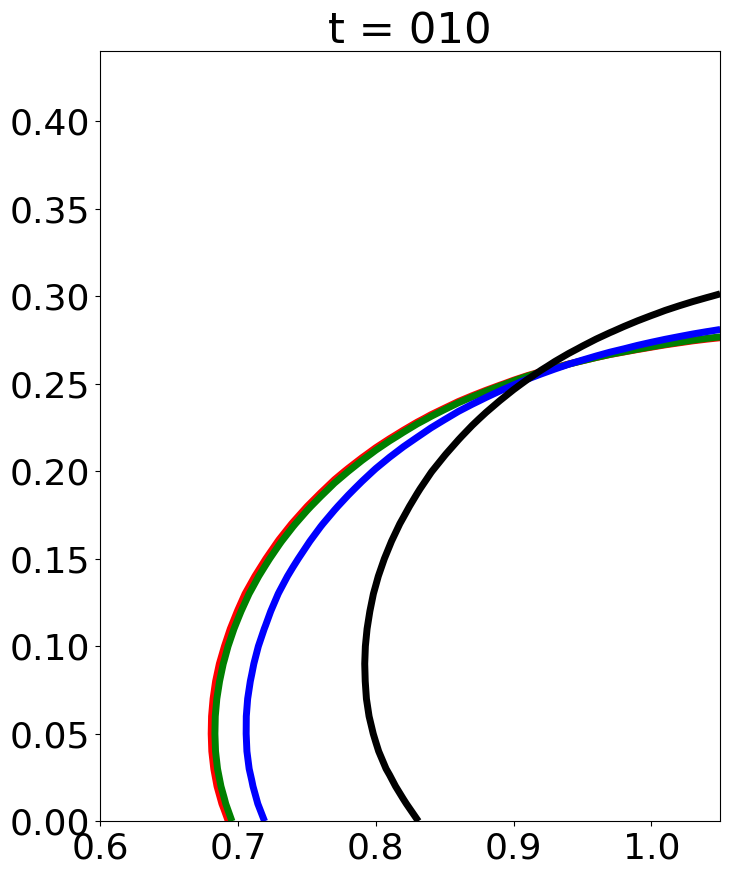}
&
\includegraphics[scale=0.15]{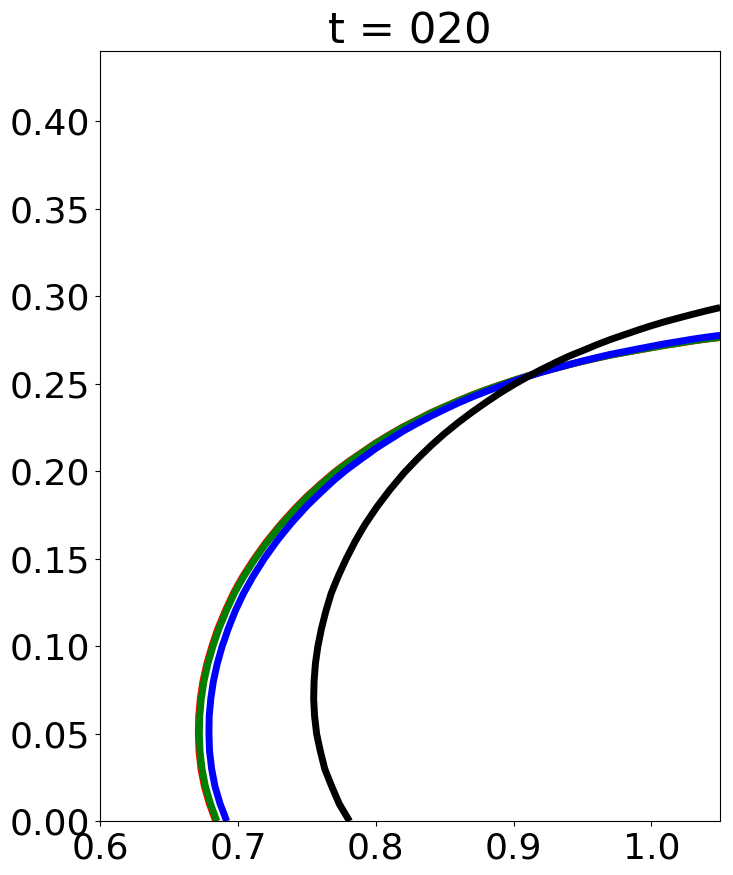}
&
\includegraphics[scale=0.15]{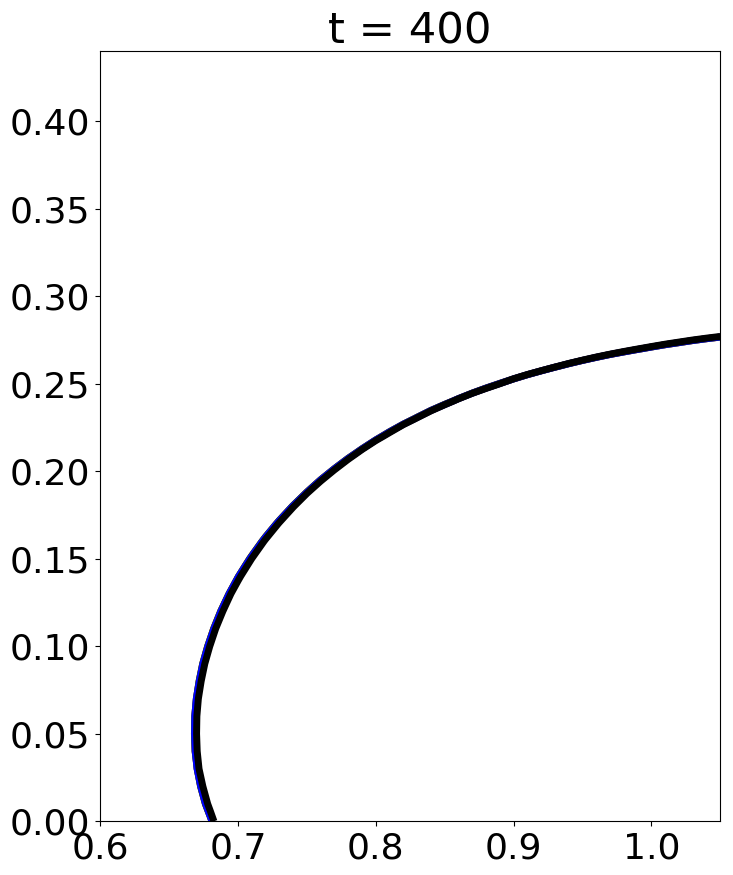}
\end{tabular}
\caption{Zoomed-in evolution of the interface between the two Maxwell states  in the vicinity of the contact point with the wall for different values of $\beta$ for Model A1 (top row) and for Model A2 (bottom row).}
\label{Fig2-experiment2}
\end{figure}

\noindent {\bf Experiment 2}\\
In the second numerical experiment, we study the spreading of a droplet of a Korteweg - van der Waals liquid \eqref{helm-Kvdw} in contact with a wall under the action of gravity (i.e. for $\bb{=}-{g}{\bf e}_z$; $g$ being the gravity acceleration and ${\bf e}_z$ the unit vector in the vertical direction). We again set $\alpha{=}0$. We consider the same set of dynamic and static boundary conditions as is the first experiment and we plot the same quantities. In particular, in Fig.~\ref{Fig1-experiment2}, we show the time evolution in the case of a solely static contact angle and in Fig.~\ref{Fig2-experiment2}, we depict the zoomed-in evolution of the interface between the liquid and vapor phases for different values of $\beta$ for Models A1 (top row) and A2 (bottom row), respectively. As in Experiment 1, we see that the introduction of the dynamic contact angle condition and associated surface traction term in the momentum balance leads to a delay in the evolution of the interface and the contact point when compared with solely static contact angle conditions. The equilibria are again the same for all models. Again, the dynamic effect is weaker for Model A1 compared to Model A2, and the dependence on the value of $\beta$ is the same as in Experiment 1 - we observe a saturation of the dynamic effect for higher values of $\beta$ for Model A1, while the effect appears to monotonously increase with the increasing value of $\beta$ for Model A2.

\begin{figure}[h!]
\begin{tabular}{ccc}
\includegraphics[scale=0.32]{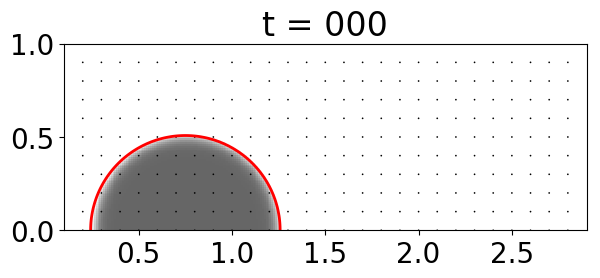}   & 
\includegraphics[scale=0.32]{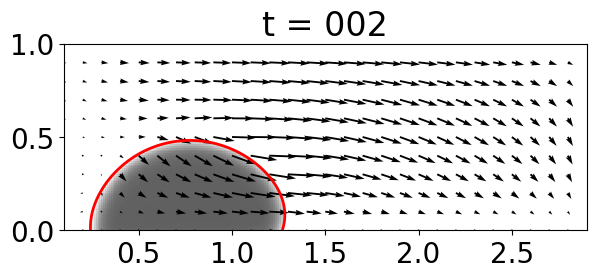}   &
\includegraphics[scale=0.32]{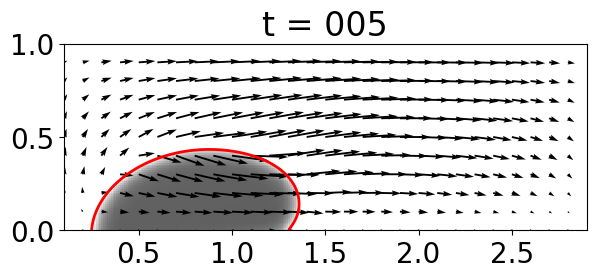}   \\
\includegraphics[scale=0.32]{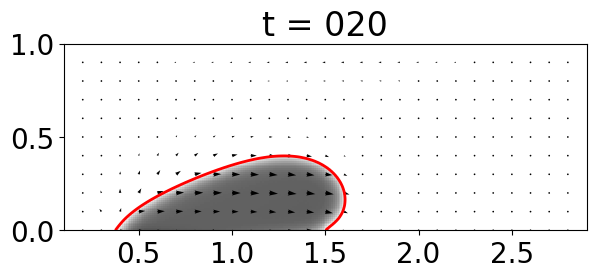}   & 
\includegraphics[scale=0.32]{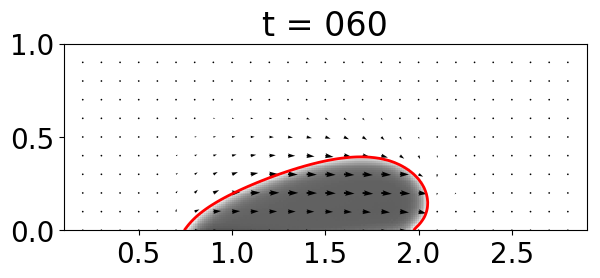}   &
\includegraphics[scale=0.32]{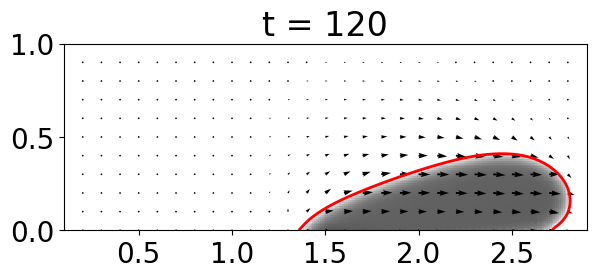}   
\end{tabular}
\caption{Sliding of an originally semicircular droplet of a Korteweg - van der Waals liquid with density $\rho_l^M$ surrounded by vapour with density $\rho_v^M$ from the initial condition (top left) over an inclined slope (inclination $30^\circ$) under the action of gravity. The equilibrium static contact angle is $\frac{\pi}{2}$. The slope is rotated such that the horizontal axis is aligned with the slope. The red contour denotes the interface between the liquid and vapour phases defined here by the density value $\frac{\rho_v^M+\rho_l^M}{2}$. The arrows represent the velocity field.}
\label{Fig1-experiment3}
\end{figure}

\vspace{0.5cm}
\noindent{\bf Experiment 3}\\
In the last numerical experiment, we study the combined effect of the dynamic contact angle condition and generalized Navier-slip at the boundary. We consider the same geometry as in Experiment 2, only the initial condition is such that the droplet is positioned more to the left. The body force is prescribed as
\be
	\bb = (g\sin{\Phi},-g\cos\Phi)\ ,
\ee
i.e., we consider a droplet sliding down an inclined slope (with an inclination $\Phi{=}\frac{\pi}{6}$), viewed from a coordinate system rotated such that its horizontal axis is aligned with the slope. The (dimensionless) friction parameter is set to $\alpha{=}50$ in all experiments. In Fig.~\ref{Fig1-experiment3}, we plot several time snapshots of the evolution (for Model A2 and $\beta{=}100$). Interestingly, we observe a difference between the values of the contact angles between the advancing side (right) and the receding side (left)
of the droplet.

\begin{figure}[h!]
\begin{tabular}{ccc}
\includegraphics[scale=0.33]{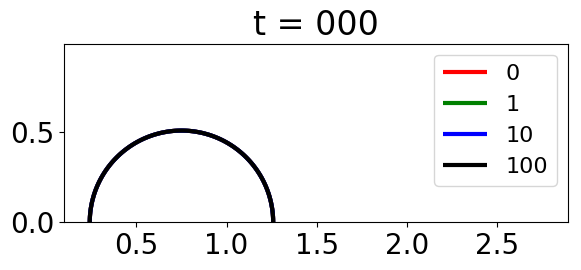}
&
\includegraphics[scale=0.33]{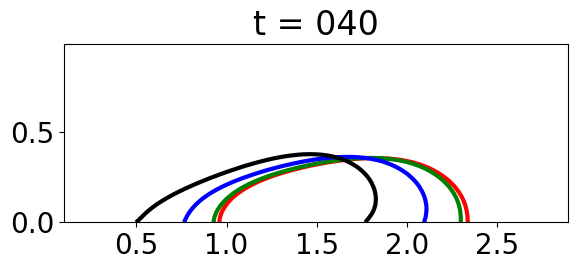}
&
\includegraphics[scale=0.33]{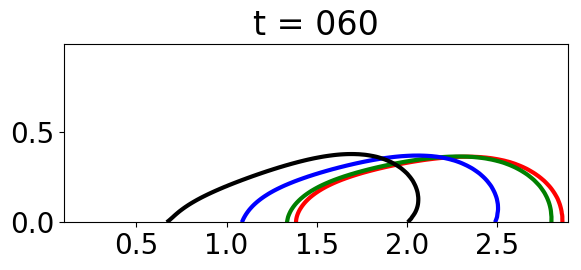}
\\
\includegraphics[scale=0.33]{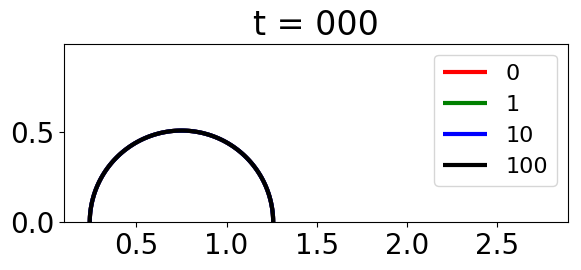}
&
\includegraphics[scale=0.33]{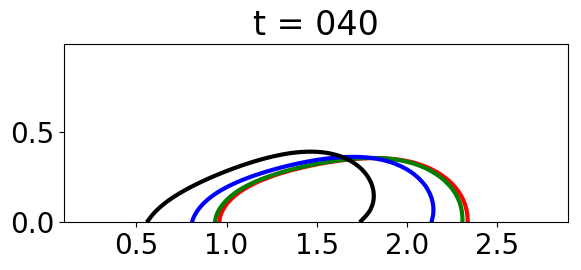}
&
\includegraphics[scale=0.33]{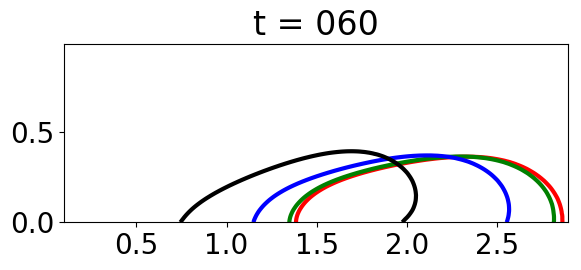}
\end{tabular}
\caption{Evolution of the interface between the two Maxwell states for different values of $\beta$ for Model A1 (top row) and for Model A2 (bottom row).}
\label{Fig2-experiment3}
\end{figure}

In Fig.~\ref{Fig2-experiment3} we show how this effect depends on the values of the dynamic coefficient $\beta$ for the two Models A1 and A2. We can see that the observed phenomenon is clearly governed by the $\beta$ parameter and is rather insensitive to the type of the Model (A1 vs. A2). The bigger the value of $\beta$, the more pronounced the effect. These results are satisfactory in the sense that they provide a possible explanation of the dynamic contact angle hysteresis observed in nature \citep[see e.g.][]{bormashenko2013}, often attributed to pinning of the contact line. Here it results from dissipative processes within the interfacial zone between the phases; such an explanation corresponds to the ideas suggested recently in \cite{makkonen2017}.

\section{Conclusions}
\label{sec:Conclusions}
In this paper we have developed a thermodynamical framework to identify the boundary conditions for a class of Korteweg-type fluids. We exploited the tools of continuum thermodynamics stemming from the balance equations both in the bulk and at the boundary of the domain, which was treated as an interface between the body and its surroundings. Assuming the constitutive equations for the Helmholtz free energy in the bulk and at the boundary, we identified the surface and bulk entropy production mechanisms giving us a starting point for the formulation of the constitutive equations in the bulk and at the boundary.

For three types of surface Helmholtz free energy of various complexity, we derived a hierarchy of corresponding constitutive equations at the boundary. While some of the constitutive relations on the boundary took standard forms, in particular the in-surface Fourier heat flux, and the heat transmission conditions across the surface (Kapitza conditions), we obtained also two novel boundary conditions mutually coupled by a common parameter. The first one represents a nontrivial generalization of the Navier slip condition, relating the traction force at the boundary with the slip velocity and a novel dynamic term - either the trace of the divergence of the bulk velocity field, or the normal derivative of the normal velocity component. The second novel boundary condition was interpreted as a contact angle boundary condition for the Korteweg fluid model and it relates the normal derivative of density with two types of terms. The static terms arise from the surface Helmhotz free energy and characterize the value of the equilibrium contact angle attained after cessation of all motion in the fluid. The other terms are dynamic and dissipative in nature and involve either the trace of the divergence of the bulk velocity or the normal derivative of its normal component. These terms do not affect the equilibrium value of the contact angle and are only active in the dynamic situation when the fluid is flowing. It should be noted that in the literature it is possible to find alternative dynamic contact angle conditions involving the term $\frac{\pa\rho^-}{\pa t}$; see \cite{Jacqmin2000}. From the perpective of the derivation presented here, such boundary conditions would be recovered provided that we do not replace the term $\frac{D\ga\rho^-}{Dt}$ in \eqref{eq:5} using the mass balance in the bulk. The approach presented here thus represents a possible generalization of such models.

Considering isothermal processes at a subcritical temperature admitting coexistence of liquid and gaseous phases, we then made the model explicit. We assumed that the Helmholtz free energy in the bulk corresponds to the Korteweg - van der Waals fluid, and the surface Helmholtz free energy reflects a simple characterization of the static contact angle. For this model, we derived explicit formulae for the contact angle condition and for the generalized Navier slip. The resulting model was implemented in the finite-element software package FEniCS. We studied the qualitative behavior of the Korteweg fluid model with the derived boundary conditions in three numerical experiments.
The first two experiments confirmed the interpretation of the novel boundary conditions, namely we observed a time lag in the attainment of the static (equilibrium) value of the contact angle and also a lag in the motion of the contact line with an increasing amplitude of the novel dynamic terms.
In the third experiment, we studied the sliding of a liquid droplet over an inclined plane, and we observed the so-called contact angle hysteresis, that is, a difference of the contact angle between the advancing and the receding side of the droplet. This phenomenon is often attributed to pinning of the contact line to irregularities on the surface; in our model it results from a dissipative process localized in the contact zone.

It should be noted that when constructing the constitutive relations, we constrained ourselves to linear relations for simplicity. A nonlinear generalization of our approach is possible. Here one could follow various thermodynamic approaches, such as the construction based on the maximization of the rate of entropy production \citep[][]{rajagopal.kr.srinivasa.ar:2004} or by defining a suitable convex dissipation potential and deriving the constitutive response accordingly in the context of the so-called generalized standard materials \citep[e.g.][]{halphen1975}. It is also worth noting that while we considered just
one particular member of the rich family of the so-called diffuse interface models - a Korteweg fluid - we are positive that the developed methodology could also be applied to other members of this class. The concept of a diffuse interface between two distinct subregions has been used since its origin at the turn of the 19th century for instance in the context of multicomponent materials \citep{cahn-hilliard1958,cahn-hilliard1959}, in the classical theory of superconductivity \citep{landau-ginzburg1965}, and, more recently in the modeling of various natural phenomena such as foams \citep{fonseca2007}, solidification \citep{kobayashi1994}, phase transitions in solids \citep{fried1994}, and glass formation \citep{rehor2017}, to name just a few of the plethora of applications. In all of these applications, generalizations of the boundary conditions and in particular, the dynamic contact angle conditions expressed as conditions for the normal derivative of the particular order parameter, should be possible following the methodology developed in this manuscript. In particular, this approach might play a key role in identification of suitable boundary conditions for viscoelastic rate type models with stress diffusion \citep{malek2018}. Yet another generalization of our models could be obtained by relaxing the assumptions made on the structure of the surface Cauchy stress tensor; here we assumed  that it is spherical (membrane model) and that the surface tension is constant. We conjecture that relaxing these assumptions would lead to the appearance of additional dynamic terms and admitting surface tension gradient would allow one to capture phenomena such as the Marangoni effect \citep{Marangoni1871}.

\appendix
\section{Evaluation of the static angle function $\gamma_0(\vartheta)$}

We provide an explicit evaluation of the parameter $\gamma_0$, which governs the static (equilibrium) part of the contact angle condition for Models A1 and A2, see \eqref{eq:contact-our}. Let us recall the definition of $\gamma_0$:
\begin{align}
\gamma_0 = \frac{\sigmalv}{\sigma}\left(\int_{\rho_v^M}^{\rho^M_l}(x-\rho_v^M)(\rho_l^M-x)\,dx\right)^{-1}\ .
\end{align}
The crucial step is to evaluate the fraction $\frac{\sigmalv}{\sigma}$, i.e. to find the relation between the surface tension of the liquid-vapor interface and the parameter $\sigma$ appearing at the gradient term in the bulk Helmholtz free energy of the Korteweg fluid (see \eqref{helm-Kvdw}).
Based on \cite{diehl2007} and \cite{dreyer_kraus_2010}, for the model described by the Helmholtz free energy \eqref{helm-Kvdw} and \eqref{helm-vdW}, it holds
\begin{align}
	\sigmalv = \sqrt{\sigma} c_0\ ,\hspace{1cm}\text{where}\hspace{1cm} c_0(\vartheta) \eqdef \sqrt{2}\int_{\rho^M_v(\vartheta)}^{\rho^M_l(\vartheta)}\sqrt{\rho \helm_{vdW}(\vartheta,\rho) - \rho\mu_{vdW}(\vartheta,\rho_v^M) + p_{vdW}(\vartheta,\rho_v^M)}\,d\rho\ ,
\end{align}
with the thermodynamic pressure $p_{vdW}$ and the chemical potential $\mu_{vdW}$ given by \eqref{pvdw} and \eqref{muvdw}, respectively. We introduce dimensionless function $\tilde{c_0}(\tilde{\vartheta})$ as in \cite{diehl2007} through
\begin{align}
c_0(\vartheta) = \rho_c\sqrt{p_c}\ \tilde{c_0}(\tilde{\vartheta})\ = \frac{\sqrt{a'}(b')^2}{9\sqrt{3}} \tilde{c_0}(\tilde{\vartheta})\ ,
\end{align}
where $\rho_c$, $p_c$ and $\vartheta_c$ are the critical density, pressure and temperature, respectively, given by \eqref{def:vdW-parameters}, and $\tilde{\vartheta}{=}\frac{\vartheta}{\vartheta_c}$ is the dimensionless temperature. The function $\tilde{c_0}(\tilde{\vartheta})$ can be approximated by the following expression \citep[][p.37]{diehl2007}:
\begin{align}
\label{eq:c0-analytic}
\tilde{c_0}(\tilde{\vartheta}) \doteq \sqrt{2}\sqrt{1-\tilde{\vartheta}}\left(6.4(1-\tilde{\vartheta}) -0.7(1-\tilde{\vartheta})^2\right)\ , 
\end{align}
which provides a good fit for $\tilde{\vartheta}{\in}\langle0.6, 1\rangle$. Applying the scaling and the definition of the capilary number from Section \ref{sec-dimensionless-formulation},   we rewrite finally \eqref{eq:contact-our}$_1$ as follows
\begin{align}
\label{eq-contact-angle-appendix}
\frac{\pa\tilde{\rho}}{\pa\nn\ga} = \mathcal{C}(\tilde{\vartheta})\cos{\varphi}\ (\tilde{\rho}^--\tilde{\rho}_v^M)(\tilde{\rho}^M_l-\tilde{\rho}^-)\ ,
\end{align}
where we introduced the dimensionless function $C(\tilde{\vartheta})$ as follows
\begin{align}
\label{def-C}
\mathcal{C}(\tilde{\vartheta}) \eqdef \frac{1}{9\sqrt{3}Ca}\frac{\tilde{c_0}(\tilde{\vartheta})}{\tilde{r}(\tilde{\vartheta})}\ \hspace{1cm}\text{with}\hspace{1cm}\tilde{r}(\tilde{\vartheta}) \eqdef \int_{\tilde{\rho}^M_v}^{\tilde{\rho}_v^M}(x-{\tilde{\rho}_v^M})({\tilde{\rho}_l^M}-x)\,dx\ ,
\end{align}
Finally, for the value $\tilde{\vartheta}{=}0.85$, considered in our numerical simulations, we evaluate the Maxwell states numerically by solving \eqref{def:maxwell-states}: $\rho_l^M{=}  0.6024$, $\rho_v^M{\doteq}0.1066$, and, consequently, from \eqref{eq:c0-analytic} and \eqref{eq-contact-angle-appendix}, we obtain 
$\tilde{r}(0.85){\doteq}0.0203 $, $\tilde{c}_0(0.85) {\doteq}0.5172$. This yields the value of $\mathcal{C}$ used in the numerical simulations in Section~\ref{sec:numerical-experiments}:
\begin{align}
\label{eq:magic-value}
\mathcal{C}(0.85)\doteq \frac{25.5}{9\sqrt{3}Ca}\ .
\end{align} 

\section*{Acknowledgements}
\label{Ack}
J. M\'{a}lek and O. Sou\v{c}ek 
acknowledge support of the project 18-12719S financed by the Czech Science Foundation. M. Heida is financed by Deutsche Forschungsgemeinschaft (DFG) through Grant
CRC 1114
``Scaling Cascades in Complex Systems'', Project C05 {\em Effective models for
materials and interfaces with multiple scales}.

%% file: Korteweg-BC.bbl
\begin{thebibliography}{38}
\expandafter\ifx\csname natexlab\endcsname\relax\def\natexlab#1{#1}\fi
\providecommand{\url}[1]{\texttt{#1}}
\providecommand{\href}[2]{#2}
\providecommand{\path}[1]{#1}
\providecommand{\DOIprefix}{doi:}
\providecommand{\ArXivprefix}{arXiv:}
\providecommand{\URLprefix}{URL: }
\providecommand{\Pubmedprefix}{pmid:}
\providecommand{\doi}[1]{\href{http://dx.doi.org/#1}{\path{#1}}}
\providecommand{\Pubmed}[1]{\href{pmid:#1}{\path{#1}}}
\providecommand{\bibinfo}[2]{#2}
\ifx\xfnm\relax \def\xfnm[#1]{\unskip,\space#1}\fi
\bibitem[{Alnaes et~al.(2015)Alnaes, Blechta, Hake, Johansson, Kehlet, Logg,
  Richardson, Ring, Rognes and Wells}]{alnaes2015}
\bibinfo{author}{Alnaes, M.S.}, \bibinfo{author}{Blechta, J.},
  \bibinfo{author}{Hake, J.}, \bibinfo{author}{Johansson, J.},
  \bibinfo{author}{Kehlet, B.}, \bibinfo{author}{Logg, A.},
  \bibinfo{author}{Richardson, C.}, \bibinfo{author}{Ring, J.},
  \bibinfo{author}{Rognes, M.E.}, \bibinfo{author}{Wells, G.N.},
  \bibinfo{year}{2015}.
\newblock \bibinfo{title}{The {FEniCS Project Version}~1.5}.
\newblock \bibinfo{journal}{Archive of Numerical Software} \bibinfo{volume}{3},
  \bibinfo{pages}{9--23}.
\bibitem[{Anderson et~al.(1998)Anderson, McFadden and Wheeler}]{anderson1998}
\bibinfo{author}{Anderson, D.M.}, \bibinfo{author}{McFadden, G.B.},
  \bibinfo{author}{Wheeler, A.A.}, \bibinfo{year}{1998}.
\newblock \bibinfo{title}{Diffuse-interface methods in fluid mechanics}.
\newblock \bibinfo{journal}{Annual Review of Fluid Mechanics}
  \bibinfo{volume}{30}, \bibinfo{pages}{139--165}.
\bibitem[{Bonn et~al.(2009)Bonn, Eggers, Indekeu, Meunier and
  Rolley}]{bonn2009}
\bibinfo{author}{Bonn, D.}, \bibinfo{author}{Eggers, J.},
  \bibinfo{author}{Indekeu, J.}, \bibinfo{author}{Meunier, J.},
  \bibinfo{author}{Rolley, E.}, \bibinfo{year}{2009}.
\newblock \bibinfo{title}{Wetting and spreading}.
\newblock \bibinfo{journal}{Rev. Mod. Phys.} \bibinfo{volume}{81},
  \bibinfo{pages}{739--805}.
\bibitem[{Bormashenko(2013)}]{bormashenko2013}
\bibinfo{author}{Bormashenko, E.}, \bibinfo{year}{2013}.
\newblock \bibinfo{title}{Wetting of Real Surfaces}.
\newblock \bibinfo{publisher}{De Gruyter}, \bibinfo{address}{Berlin, Boston}.
\bibitem[{Brackbill et~al.(1992)Brackbill, Kothe and Zemach}]{brackbill1992}
\bibinfo{author}{Brackbill, J.}, \bibinfo{author}{Kothe, D.},
  \bibinfo{author}{Zemach, C.}, \bibinfo{year}{1992}.
\newblock \bibinfo{title}{A continuum method for modeling surface tension}.
\newblock \bibinfo{journal}{Journal of Computational Physics}
  \bibinfo{volume}{100}, \bibinfo{pages}{335 -- 354}.
\bibitem[{{Buscaglia} and {Ausas}(2011)}]{buscaglia-2011}
\bibinfo{author}{{Buscaglia}, G.C.}, \bibinfo{author}{{Ausas}, R.F.},
  \bibinfo{year}{2011}.
\newblock \bibinfo{title}{{Variational formulations for surface tension,
  capillarity and wetting}}.
\newblock \bibinfo{journal}{Computer Methods in Applied Mechanics and
  Engineering} \bibinfo{volume}{200}, \bibinfo{pages}{3011--3025}.
\bibitem[{Cahn and Hilliard(1958)}]{cahn-hilliard1958}
\bibinfo{author}{Cahn, J.}, \bibinfo{author}{Hilliard, J.},
  \bibinfo{year}{1958}.
\newblock \bibinfo{title}{Free energy of a non-uniform system. {I}.
  {I}nterfacial free energy}.
\newblock \bibinfo{journal}{J. Chem. Phys.} \bibinfo{volume}{28},
  \bibinfo{pages}{258--267}.
\bibitem[{Cahn and Hilliard(1959)}]{cahn-hilliard1959}
\bibinfo{author}{Cahn, J.}, \bibinfo{author}{Hilliard, J.},
  \bibinfo{year}{1959}.
\newblock \bibinfo{title}{Free energy of a non-uniform system. {III}.
  {N}ucleation in a two-component incompressible fluid}.
\newblock \bibinfo{journal}{J. Chem. Phys.} \bibinfo{volume}{31},
  \bibinfo{pages}{688--699}.
\bibitem[{Callen(1985)}]{callen}
\bibinfo{author}{Callen, H.B.}, \bibinfo{year}{1985}.
\newblock \bibinfo{title}{Thermodynamics and an introduction to
  thermostatistics}.
\newblock \bibinfo{edition}{Revised} ed., \bibinfo{publisher}{John Wiley \&
  Sons}.
\bibitem[{Coleman and Noll(1963)}]{Coleman1963}
\bibinfo{author}{Coleman, B.D.}, \bibinfo{author}{Noll, W.},
  \bibinfo{year}{1963}.
\newblock \bibinfo{title}{The thermodynamics of elastic materials with heat
  conduction and viscosity}.
\newblock \bibinfo{journal}{Archive for Rational Mechanics and Analysis}
  \bibinfo{volume}{13}, \bibinfo{pages}{167--178}.
\bibitem[{Diehl(2007)}]{diehl2007}
\bibinfo{author}{Diehl, D.}, \bibinfo{year}{2007}.
\newblock \bibinfo{title}{Higher order schemes for simulation of compressible
  liquid-vapor flows with phase change}.
\newblock \bibinfo{publisher}{Doctoral dissertation},
  \bibinfo{address}{Freiburg im Breisgau}.
\bibitem[{Dreyer and Kraus(2010)}]{dreyer_kraus_2010}
\bibinfo{author}{Dreyer, W.}, \bibinfo{author}{Kraus, C.},
  \bibinfo{year}{2010}.
\newblock \bibinfo{title}{On the van der waals–cahn–hilliard phase-field
  model and its equilibria conditions in the sharp interface limit}.
\newblock \bibinfo{journal}{Proceedings of the Royal Society of Edinburgh:
  Section A Mathematics} \bibinfo{volume}{140}, \bibinfo{pages}{1161--1186}.
\bibitem[{Dunn and Serrin(1986)}]{dunn-serrin-1986}
\bibinfo{author}{Dunn, J.E.}, \bibinfo{author}{Serrin, J.},
  \bibinfo{year}{1986}.
\newblock \bibinfo{title}{On the thermomechanics of interstitial working}, in:
  \bibinfo{booktitle}{The Breadth and Depth of Continuum Mechanics},
  \bibinfo{publisher}{Springer Berlin Heidelberg}, \bibinfo{address}{Berlin,
  Heidelberg}. pp. \bibinfo{pages}{705--743}.
\bibitem[{Fonseca et~al.(2007)Fonseca, Morini and Slastikov}]{fonseca2007}
\bibinfo{author}{Fonseca, I.}, \bibinfo{author}{Morini, M.},
  \bibinfo{author}{Slastikov, V.}, \bibinfo{year}{2007}.
\newblock \bibinfo{title}{Surfactants in foam stability: A phase-field model}.
\newblock \bibinfo{journal}{Archive for Rational Mechanics and Analysis}
  \bibinfo{volume}{183}, \bibinfo{pages}{411--456}.
\bibitem[{Fried and Gurtin(1994)}]{fried1994}
\bibinfo{author}{Fried, E.}, \bibinfo{author}{Gurtin, M.},
  \bibinfo{year}{1994}.
\newblock \bibinfo{title}{Dynamic solid-solid transitions with phase
  characterized by an order parameter}.
\newblock \bibinfo{journal}{Physica D: Nonlinear Phenomena}
  \bibinfo{volume}{72}, \bibinfo{pages}{287 -- 308}.
\bibitem[{Gibbs(1928)}]{gibbs-selected-vol1}
\bibinfo{author}{Gibbs, J.}, \bibinfo{year}{1928}.
\newblock \bibinfo{title}{The collected works of J. Willard Gibbs, vol. 1}.
\newblock \bibinfo{publisher}{Yale University Press, New Haven}.
\bibitem[{Gomez et~al.(2010)Gomez, Hughes, Nogueira and Calo}]{gomez2010}
\bibinfo{author}{Gomez, H.}, \bibinfo{author}{Hughes, T.J.},
  \bibinfo{author}{Nogueira, X.}, \bibinfo{author}{Calo, V.M.},
  \bibinfo{year}{2010}.
\newblock \bibinfo{title}{Isogeometric analysis of the isothermal
  {N}avier-{S}tokes-{K}orteweg equations}.
\newblock \bibinfo{journal}{Computer Methods in Applied Mechanics and
  Engineering} \bibinfo{volume}{199}, \bibinfo{pages}{1828 -- 1840}.
\bibitem[{de~Groot and Mazur(1984)}]{groot.sr.mazur.p:non-equilibrium}
\bibinfo{author}{de~Groot, S.R.}, \bibinfo{author}{Mazur, P.},
  \bibinfo{year}{1984}.
\newblock \bibinfo{title}{Non-equilibrium thermodynamics}.
\newblock \bibinfo{publisher}{Dover Publications}, \bibinfo{address}{New York}.
\newblock \bibinfo{note}{Reprint of the 1962 original}.
\bibitem[{Halphen and Son~Nguyen(1975)}]{halphen1975}
\bibinfo{author}{Halphen, B.}, \bibinfo{author}{Son~Nguyen, Q.},
  \bibinfo{year}{1975}.
\newblock \bibinfo{title}{Sur les mat{\'e}riaux standard
  g{\'e}n{\'e}ralis{\'e}s}.
\newblock \bibinfo{journal}{Journal de M{\'e}canique} \bibinfo{volume}{14},
  \bibinfo{pages}{39--63}.
\bibitem[{Heida(2013)}]{heida2013}
\bibinfo{author}{Heida, M.}, \bibinfo{year}{2013}.
\newblock \bibinfo{title}{On the derivation of thermodynamically consistent
  boundary conditions for the {C}ahn-{H}illiard-{N}avier-{S}tokes system}.
\newblock \bibinfo{journal}{Int. J. Eng. Sci.} \bibinfo{volume}{62},
  \bibinfo{pages}{126 -- 156}.
\bibitem[{Heida and M\'alek(2010)}]{heida.m.malek.j:on}
\bibinfo{author}{Heida, M.}, \bibinfo{author}{M\'alek, J.},
  \bibinfo{year}{2010}.
\newblock \bibinfo{title}{On compressible {K}orteweg fluid-like materials}.
\newblock \bibinfo{journal}{Int. J. Eng. Sci.} \bibinfo{volume}{48},
  \bibinfo{pages}{1313--1324}.
\bibitem[{Hutter and Rajagopal(1994)}]{hutter1994}
\bibinfo{author}{Hutter, K.}, \bibinfo{author}{Rajagopal, K.R.},
  \bibinfo{year}{1994}.
\newblock \bibinfo{title}{On flows of granular materials}.
\newblock \bibinfo{journal}{Continuum Mechanics and Thermodynamics}
  \bibinfo{volume}{6}, \bibinfo{pages}{81 -- 139}.
\bibitem[{Jacqmin(2000)}]{Jacqmin2000}
\bibinfo{author}{Jacqmin, D.}, \bibinfo{year}{2000}.
\newblock \bibinfo{title}{{Contact-line dynamics of a diffuse fluid
  interface}}.
\newblock \bibinfo{journal}{Journal of Fluid Mechanics} \bibinfo{volume}{402},
  \bibinfo{pages}{57--88}.
\bibitem[{Kapitza(1941)}]{kapitza1941}
\bibinfo{author}{Kapitza, P.L.}, \bibinfo{year}{1941}.
\newblock \bibinfo{title}{Heat transfer and superfluidity of {H}elium {II}}.
\newblock \bibinfo{journal}{Phys. Rev.} \bibinfo{volume}{60},
  \bibinfo{pages}{354--355}.
\bibitem[{Kobayashi(1994)}]{kobayashi1994}
\bibinfo{author}{Kobayashi, R.}, \bibinfo{year}{1994}.
\newblock \bibinfo{title}{A numerical approach to three-dimensional dendritic
  solidification}.
\newblock \bibinfo{journal}{Experiment. Math.} \bibinfo{volume}{3},
  \bibinfo{pages}{59--81}.
\bibitem[{Korteweg(1901)}]{korteweg1901}
\bibinfo{author}{Korteweg, D.J.}, \bibinfo{year}{1901}.
\newblock \bibinfo{title}{Sur la forme que prennent les \'{e}quations du
  mouvement des fluides si l’on tient compte des forces capillaires
  caus\'{e}es par des variations de densit\'{e} consid\'{e}rables mais
  continues et sur la th\'{e}orie de la capillarit\'{e} dans l’hypoth\'{e}se
  d’une variation continue de la densit\'{e}.}
\newblock \bibinfo{journal}{Archives N\'{e}erlandaises des sciences exactes et
  naturelles.} \bibinfo{volume}{2}, \bibinfo{pages}{1--24}.
\bibitem[{Landau and Ginzburg(1965)}]{landau-ginzburg1965}
\bibinfo{author}{Landau, L.}, \bibinfo{author}{Ginzburg, V.},
  \bibinfo{year}{1965}.
\newblock \bibinfo{title}{On the theory of superconductivity}, in:
  \bibinfo{editor}{ter Haar, D.} (Ed.), \bibinfo{booktitle}{Collected papers of
  L.D. Landau}. \bibinfo{publisher}{Pergamon Oxford}, p.
  \bibinfo{pages}{626–633}.
\bibitem[{Landau and Lifshitz(1980)}]{landau-statistical}
\bibinfo{author}{Landau, L.D.}, \bibinfo{author}{Lifshitz, E.M.},
  \bibinfo{year}{1980}.
\newblock \bibinfo{title}{Statistical Physics (3rd Edition)}.
\newblock \bibinfo{publisher}{Butterworth-Heinemann},
  \bibinfo{address}{Oxford}.
\bibitem[{Makkonen(2017)}]{makkonen2017}
\bibinfo{author}{Makkonen, L.}, \bibinfo{year}{2017}.
\newblock \bibinfo{title}{A thermodynamic model of contact angle hysteresis}.
\newblock \bibinfo{journal}{The Journal of Chemical Physics}
  \bibinfo{volume}{147}, \bibinfo{pages}{064703}.
\bibitem[{M\'{a}lek et~al.(2018)M\'{a}lek, Pr\r{u}\v{s}a, Sk\v{r}ivan and
  S\"{u}li}]{malek2018}
\bibinfo{author}{M\'{a}lek, J.}, \bibinfo{author}{Pr\r{u}\v{s}a, V.},
  \bibinfo{author}{Sk\v{r}ivan, T.}, \bibinfo{author}{S\"{u}li, E.},
  \bibinfo{year}{2018}.
\newblock \bibinfo{title}{Thermodynamics of viscoelastic rate-type fluids with
  stress diffusion}.
\newblock \bibinfo{journal}{Physics of Fluids} \bibinfo{volume}{30},
  \bibinfo{pages}{023101}.
\bibitem[{Marangoni(1871)}]{Marangoni1871}
\bibinfo{author}{Marangoni, C.}, \bibinfo{year}{1871}.
\newblock \bibinfo{title}{Sul principio della viscosita' superficiale dei
  liquidi stabilito dalsig. j. plateau}.
\newblock \bibinfo{journal}{Il Nuovo Cimento (1869-1876)} \bibinfo{volume}{5},
  \bibinfo{pages}{239--273}.
\bibitem[{Rajagopal and Srinivasa(2004)}]{rajagopal.kr.srinivasa.ar:2004}
\bibinfo{author}{Rajagopal, K.R.}, \bibinfo{author}{Srinivasa, A.R.},
  \bibinfo{year}{2004}.
\newblock \bibinfo{title}{On thermomechanical restrictions of continua}.
\newblock \bibinfo{journal}{Proc. R. Soc. Lond., Ser. A, Math. Phys. Eng. Sci.}
  \bibinfo{volume}{460}, \bibinfo{pages}{631--651}.
\bibitem[{{\v{R}}eho{\v{r}} et~al.(2017){\v{R}}eho{\v{r}}, Blechta and
  Sou{\v{c}}ek}]{rehor2017}
\bibinfo{author}{{\v{R}}eho{\v{r}}, M.}, \bibinfo{author}{Blechta, J.},
  \bibinfo{author}{Sou{\v{c}}ek, O.}, \bibinfo{year}{2017}.
\newblock \bibinfo{title}{On some practical issues concerning the
  implementation of {C}ahn-{H}illiard-{N}avier-{S}tokes type models}.
\newblock \bibinfo{journal}{International Journal of Advances in Engineering
  Sciences and Applied Mathematics} \bibinfo{volume}{9},
  \bibinfo{pages}{30--39}.
\bibitem[{Rohde(2018)}]{Rohde2018}
\bibinfo{author}{Rohde, C.}, \bibinfo{year}{2018}.
\newblock \bibinfo{title}{Fully resolved compressible two-phase flow:
  modelling, analytical and numerical issues}, in: \bibinfo{booktitle}{New
  trends and results in mathematical description of fluid flows}.
  \bibinfo{publisher}{Birkh\"{a}user/Springer, Cham}. Ne\v{c}as Center Ser.,
  pp. \bibinfo{pages}{115 -- 181}.
\bibitem[{Rowlinson and Widom(1989)}]{rowlinson}
\bibinfo{author}{Rowlinson, J.S.}, \bibinfo{author}{Widom, B.},
  \bibinfo{year}{1989}.
\newblock \bibinfo{title}{Molecular Theory of Capillarity}.
\newblock \bibinfo{publisher}{Dover Publications, New York}.
\bibitem[{Slattery(1990)}]{slattery-1990}
\bibinfo{author}{Slattery, J.C.}, \bibinfo{year}{1990}.
\newblock \bibinfo{title}{Interfacial transport phenomena}.
\newblock \bibinfo{publisher}{Springer-Verlag}, \bibinfo{address}{New York}.
\bibitem[{Truesdell and Noll(1965)}]{truesdell.c.noll.w:non-linear}
\bibinfo{author}{Truesdell, C.}, \bibinfo{author}{Noll, W.},
  \bibinfo{year}{1965}.
\newblock \bibinfo{title}{The non-linear field theories of mechanics}, in:
  \bibinfo{editor}{Fl\"uge, S.} (Ed.), \bibinfo{booktitle}{Handbuch der
  Physik}. \bibinfo{publisher}{Springer}, \bibinfo{address}{Berlin}. volume
  \bibinfo{volume}{III/3}.
\bibitem[{van~der Waals(1893)}]{van-der-Waals-1983}
\bibinfo{author}{van~der Waals, J.D.}, \bibinfo{year}{1893}.
\newblock \bibinfo{title}{Thermodynamische theorie der capillariteit in de
  onderstelling van continue dichtheidsverandering}.
\newblock \bibinfo{publisher}{Amsterdam, J. M\"{u}ller}.

\end{thebibliography}
